\documentclass[aps,prd,floatfix,showpacs,superscriptaddress,nofootinbib,10pt]{revtex4-2}
\usepackage{amsmath}
\usepackage{amssymb}
\usepackage{graphicx}
\usepackage{braket}
\usepackage{bbm}
\usepackage{xcolor}

\usepackage[caption=false]{subfig}

\usepackage{tikz}
\usetikzlibrary{shapes}
\usepackage{bbm}
\usepackage{siunitx,pslatex}

\begin{document}

\title{Anisotropic and frame dependent chaos of suspended strings from a dynamical holographic QCD model with magnetic field}
\author{Bhaskar Shukla}\email{519ph1003@nitrkl.ac.in}\affiliation{Department of Physics and Astronomy, National Institute of Technology Rourkela, Rourkela - 769008, India}
\author{David Dudal}\email{david.dudal@kuleuven.be}\affiliation{KU Leuven Campus Kortrijk -- Kulak, Department of Physics Etienne Sabbelaan 53 bus 7657, 8500 Kortrijk, Belgium}\affiliation{
	Ghent University, Department of Physics and Astronomy, Krijgslaan 281-S9, 9000 Gent, Belgium}
\author{Subhash Mahapatra}\email{mahapatrasub@nitrkl.ac.in}\affiliation{Department of Physics and Astronomy, National Institute of Technology Rourkela, Rourkela - 769008, India}

\date{}

\begin{abstract}
We investigate both from a qualitative as well as quantitative perspective the emergence of chaos in the QCD confining string in a magnetic field from a holographic viewpoint.  We use an earlier developed bottom-up solution of the Einstein-Maxwell-Dilaton action that mimics QCD and its thermodynamics quite well.  Surprisingly, our predictions depend on the used frame: the magnetic field tends to suppress the chaos in both perpendicular and parallel directions relative to the magnetic field in the string frame whilst in the Einstein frame, the chaos suppression only happens in the perpendicular direction, with an enhanced chaos along the magnetic field. The amount of suppression/enhancement in both frames does depend on the relative orientation of the string and magnetic field.
\end{abstract}

\maketitle

\section{Introduction}
Chaos theory is a fundamental tool in modern science that has numerous applications across a wide range of fields, including meteorology, astronomy, ecology, population biology, chemistry, social psychology, economics, and many more \cite{Shen2021Jan,Fang2015Dec,Hastings1993,Philippe1993Aug,Hobbs1997Jun,Carver1997Apr,Fister2015Feb,Kelsey1988Mar}. It allows us to understand and predict the behavior of complex systems, whether they are natural or man-made, and it has had a profound impact on our understanding of the world. In recent years, chaos theory has seen an increased interest in quantum systems, also in the realm of high energy physics and quantum chromodynamics (QCD). A plethora of work has been done in this direction \cite{Pullirsch1998May,Markum1998Apr,Markum1999Mar,Bittner2001Apr,Hashimoto2016Nov} and steady progress has been made. For example, a possible explanation of quantum chaos in the hadronic spectrum  by a holographic approach can be found in \cite{PandoZayas2013Apr}. Probing the phase diagram of QCD chaos is a challenging task, an option effectively explored in \cite{Akutagawa2018Jun}.

The Maldacena-Shenker-Stanford (MSS) bound \cite{Maldacena2016Aug}, a global upper bound on chaos for the Lyapunov exponent of out-of-time-ordered correlators (OTOCs) in thermal quantum field theories, is one significant advancement in this field. The (largest) Lyapunov exponent, according to this bound, is constrained by the temperature as follows \cite{Maldacena2016Aug}
\begin{equation}\label{1}
	\lambda \leq 2 \pi T
\end{equation} with $\hbar=1$ and $k_{B}=1$. This MSS bound has been extensively tested and has had a significant impact on the understanding of the chaotic dynamics of quantum systems \cite{Shenker2014Mar,Shenker2014Dec,Leichenauer2014Aug,Shenker2015May,Jackson2015Dec,Polchinski2015May,deBoer2018May,Dalui2019Jan,Dalui2020Dec,Ageev2021May}.

The AdS/CFT correspondence \cite{Maldacena1999Apr}, also known as gauge/gravity duality, provides a way to study the behavior of quantum systems and in a dual, lower-dimensional description. This correspondence asserts that a strongly coupled gauge theory in four-dimensional spacetime can be equivalent to a classical gravity theory in five-dimensional anti-de Sitter (AdS) spacetime at zero temperature, with the field theory ``living'' on the boundary. A deconfined gauge theory at finite temperature is dual to a gravitational theory in an AdS black hole, which behaves as a thermal system in itself. For more details about the AdS/CFT correspondence, let us refer to the seminal papers \cite{Maldacena1998Jun,Maldacena1999Apr,Witten1998Feb,Gubser1998May} or the tome \cite{ammon2015gauge}. By using the AdS/CFT correspondence, many important features of QCD have been modelled and attempts to connect to experiment are made, see e.g.~\cite{Gursoy2011Mar,Gursoy2021Jul} and references therein.

In curved geometries, examples of chaotic dynamics analyses of strings have been presented in \cite{Basu2011May,Basu2011Jun,Stepanchuk2013Mar,Giataganas2014Jan,Bai2016Mar,Asano2015Aug,Panigrahi2016Oct,Basu2017Mar,Asano2016Sep,Ishii2017Mar,Rigatos:2020hlq,Giataganas:2017guj}.
In many papers detailed investigations of the quark-antiquark pair have been made by means of gauge/gravity duality where chaos is qualitatively studied using the Poincaré sections and quantified by means of $\lambda$, the Lyapunov exponent \cite{Avramis2007Apr,Arias2010Jan,Nunez2010Apr,Bellantuono2017Aug,Hashimoto2018Oct,Akutagawa2019Aug,Colangelo2020Oct}.

Strong magnetic fields are produced during the early stages of non-central heavy ion collisions and are expected to remain sufficiently large during the formation of quark-gluon plasma (QGP) \cite{Skokov2009Dec,Bzdak2012Mar,D'Elia2010Sep,D'Elia2021Dec,Deng2012Apr,Tuchin2013Jul,Voronyuk2011May}. These magnetic fields can have an impact on many QCD observables and have been the subject of much research in recent years \cite{Kharzeev:2013jha,costa2022topical}. In addition, slightly less intense magnetic fields probably exist inside neutron stars \cite{Duncan1992Jun}, and very strong magnetic fields may have been present during the formation of the early universe \cite{Vachaspati1991Aug}. The role of magnetic fields in these different contexts is thus important and has been studied extensively in the last few years \cite{Miransky2015Apr,Iwasaki2021Jul,Bohra2020Feb,Critelli:2016cvq,Aref'eva2022Mar,Aref'eva2021Apr,Jena2022Apr,Jain:2022hxl,Dudal:2021jav,Dudal:2016joz}.

However, in holographic model computations of QCD observables, the choice of Einstein or String frame can significantly impact the results. A key example is the computation of the string tension of the confining string between a static (heavy) quark-antiquark pair, for which the String frame should be used \cite{Kiritsis:2009hu}. Indeed, using the Einstein frame would not even yield a confining linear potential, \cite{Critelli:2016cvq}, a fact confirmed in \cite{Dudal2017Dec} when studying other features derivable from the quark-antiquark pair.

In a different context, scalar-tensor theories of gravity, which describe the gravitational force as being mediated by a scalar field, the dilaton in addition to the metric tensor, can also be formulated in different conformal frames. These frames, such as the String or Einstein frame, are related by a conformal transformation and can lead to different predictions for the behavior of physical systems. The debate over which frame is more suitable for describing physical processes in scalar-tensor theories of gravity is ongoing, as the choice of frame can have significant consequences for the interpretation of observations and experiments \cite{Faraoni1999Jan,Casadio1999Jun,Cho1992May,Banerjee2016Mar,Sk2017Aug,Capozziello2010May,Corda2011Jan,Dick1998Mar,Quiros2013Feb,Bhadra2007Feb,Jarv2007Nov,Nojiri2001Jul,S.J.2021Jan,Magnano1994Oct,Capozziello1997Dec,Faraoni1998Nov,Quiros2018Dec,Galaverni2022Apr,Macias2001May,Faraoni1998Jan,Faraoni2007Jan}.

In a previous work, the role of magnetic field on the chaotic dynamics of the string has been scrutinised in \cite{Colangelo2022Apr}. In our paper, we build upon their work to continue the study of a magnetic field on the chaotic dynamics of a hanging string in two different frames i.e.~String vs.~Einstein frame, to check if there is an anisotropy present for different string orientations and to verify the MSS bound by using the underlying Einstein-Maxwell-dilaton (EMD) model. Given that this happens also at the level of the connecting string, one might expect a dependence of the chaos observables on the chosen frame.

The paper is divided into two parts. In the first part, we deal with the model in the String frame, where in Sec.~\ref{sec:2.1}, we discuss the Methodology, mainly the bottom-up magnetized EMD model of \cite{Bohra2020Feb,Bohra2021Apr}. We then discuss the hanging string profile in the gravitational background in Sec.~\ref{sec:2.2} and focus on the unstable string solutions. The effect of adding a perturbation to the static string is discussed in Sec.~\ref{sec:2.3}. The chaotic dynamics, first qualitatively by means of Poincaré sections, is presented in Sec.~\ref{sec:2.4} and quantitatively via Lyapunov exponents in Sec.~\ref{sec:2.5}. We also check the MSS bound in Sec.~\ref{sec:2.6}. In the second part, we consider the Einstein frame in Sec.~\ref{sec:3}. Finally, we conclude our paper by highlighting the main results with some discussions in Sec.~\ref{sec:4}.

\section{String Frame}\label{sec:2}

\subsection{Magnetized EMD Gravity Model}\label{sec:2.1}
We consider a five-dimensional Einstein-Maxwell-Dilaton (EMD) gravity system of \cite{Bohra2020Feb},
\begin{equation}\label{2}
	S_{EM}=-\frac{1}{16\pi G_5}\int d^5x \sqrt{-g}[R-\frac{f(\phi)}{4}F_{MN}F^{MN}-\frac{1}{2}\partial_M \phi \partial^M \phi -V(\phi)] \,,
\end{equation}
where $F_{MN}$ is the field strength tensor for the $U(1)$ gauge field, $\phi$ is the dilaton field, $f(\phi)$ is the gauge kinetic function representing the coupling between the $U(1)$ gauge field and the dilaton field respectively. $V(\phi)$ is the potential of the dilaton field and $G_{5}$ is the Newton constant in five dimensions. Interestingly, using the following Ans\"atze for the metric $g_{MN}$, field strength tensor $F_{MN}$, and dilaton field $\phi$,
\begin{eqnarray}\label{3}
	ds^2&=& \frac{L^2}{z^2} e^{2 A(z)} \left[-g(z) dt^2+\frac{dz^2}{g(z)}+dx_1^2+e^{B^2 z^2} \left(dx_2^2+dx_3^2\right)\right], \nonumber\\
	F_{MN}&=&Bdx_{2} \wedge dx_{3} \,,
\end{eqnarray}
and the boundary conditions
\begin{eqnarray}\label{3a}
& & g(z=z_h)=0,~~~ g(z=0) =1,~~~A(z=0) = 0  \,,
\end{eqnarray}
the Einstein, Maxwell, and dilaton field equations can be completely solved in closed form in terms of a single parameter $a$,
\begin{eqnarray}
A(z)&=& -a z^2 \,,
\label{asol} \\
g(z) &=& 1-\frac{e^{z^2 \left(3 a-B^2\right)} \left(3 a z^2-B^2 z^2-1\right)+1}{e^{z_h^2 \left(3
   a-B^2\right)} \left(3 a z_h^2-B^2 z_h^2-1\right)+1} \,,
\label{gsol} \\
\phi(z) &=& \frac{\left(9 a-B^2\right) \log \left(\sqrt{6 a^2-B^4} \sqrt{z^2 \left(6 a^2-B^4\right)+9 a-B^2}+6 a^2
   z -B^4 z \right)}{\sqrt{6 a^2-B^4}} \nonumber\\
  & & + z \sqrt{z^2 \left(6 a^2-B^4\right)+9 a-B^2} -\frac{\left(9 a-B^2\right) \log \left(\sqrt{9 a-B^2} \sqrt{6 a^2-B^4}\right)}{\sqrt{6 a^2-B^4}}\,,
\label{phisol} \\
f(z) &=& g(z)e^{2 A(z)+2 B^2 z^2} \left(-\frac{6 A'(z)}{z}-4 B^2+\frac{4}{z^2}\right)-\frac{2 e^{2 A(z)+2 B^2 z^2} g'(z)}{z} \,,
\label{fsol} \\
V(z) &=& g'(z) \left(-3 z^2 A'(z)-B^2 z^3+3 z\right) e^{-2 A(z)} - g(z)\left(12 + 9 B^2 z^3 A'(z) \right) e^{-2 A(z)} \nonumber \\
& & +g(z) \left(-9 z^2 A'(z)^2-3 z^2 A''(z)+18 z
   A'(z)-2 B^4 z^4+8 B^2 z^2\right)e^{-2 A(z)} \,,
\label{Vsol}
\end{eqnarray}
wherein the AdS radius $L$ has been set to one and $z$ is the usual holographic radial coordinate. The above solution corresponds to a black hole having horizon at $z=z_h$. Therefore, the $z$-coordinates runs from $z=z_h$ to $z=0$ (asymptotic boundary). In most of the calculations below regarding the dynamics of the string, we will work in the radial coordinate $r=1/z$. This holographic coordinate runs from $r=r_h=1/z_h$ to $r=\infty$. One can explicitly check that this spacetime solution asymptotes to AdS at the boundary. Notice also that in the above Ans\"atze the magnetic field is chosen in the $x_1$ direction which breaks the $SO(3)$ invariance of the boundary spatial coordinates ($x_{1},x_{2},x_{3}$). Let us also mention the temperature and entropy of the black hole
\begin{eqnarray}
 T = \frac{z_{h}^{3} e^{-3A(z_h)-B^2 z_{h}^{2}}}{4 \pi \int_0^{z_h} \, d\xi \ \xi^3 e^{-B^2 \xi^2 -3A(\xi) } } \,,  \qquad S_{BH} = \frac{V_3 e^{3 A(z_h)+B^2 z_{h}^{2}}}{4 G_{(5)} z_{h}^3 } \,,
\label{BHtemp}
\end{eqnarray}
%%%%%%%%%%%%%%%%%%%%%%%%%%%%%%%%%%%%%%%%%%%%%%%%%
where $V_3$ is the volume of the three-dimensional spatial volume. For the record, $B$ denotes the five-dimensional (bulk) magnetic field which carries dimension GeV, the boundary magnetic field in units GeV${}^2$ will be related to this $B$ via a suitable $L$-rescaling, see \cite{Jena2022Apr} for more about this, including extra references.

It is important to emphasize that the metric solution (\ref{gsol}) is in the Einstein frame. In the presence of a non-trivial dilaton field, the metric solution takes different forms in Einstein and String frames. Therefore, it is expected that the presence of a non-trivial dilaton field might modify the string dynamics in different ways in the two frames. In particular, the chaos observables related to string dynamics might exhibit different behavior in these two non-equivalent frames. As we will later see, the Lyapunov exponent indeed behaves differently in both frames when a magnetic field is present. This should be contrasted with \cite{Colangelo2022Apr}, in which calculations are based on the model of \cite{Li:2016gfn} where no dilaton field is present and the String and Einstein frame metrics thus coincide. To be more precise, the used metric solves the Einstein equations of motion up to the first non-trivial order in a $1/r$ expansion, as seminally presented in \cite{DHoker:2009ixq}.

The Nambu-Goto (NG) string action is usually evaluated in the string frame metric \cite{Kiritsis:2009hu}. According to the standard gauge/gravity duality prescription, the on-shell value of the Nambu-Goto (NG) string action in the string frame is related to the quark-antiquark potential. Therefore, it is important to write down the metric solution also in the string frame. The standard procedure to go from the Einstein  to string frame involves the dilaton transformation \cite{Bohra2020Feb}, i.e. $(g_s)_{MN}=e^{\sqrt{2/3}\phi} g_{MN}$. The metric solution (\ref{gsol}) in the string frame then reads
\begin{equation}\label{4s}
	ds^2=L^2 r^2 e^{2 A_s(r)} \left[-g(r) dt^2+\frac{dr^2}{r^4 g(r)}+dx_1^2+e^{\frac{B^2}{r^2}} \left(dx_2^2+dx_3^2\right)\right]
\end{equation}
where $A_s(r) = A(r) + \sqrt{\frac{1}{6}} \phi(r)$.
Let us write down the above string and Einstein frame metrics in the following general form
\begin{equation}\label{7}
	ds^2=g_{\text{tt}} dt^2+g_{11} dx_1^2+g_{22} dx_2^2+g_{33} dx_3^2+g_{\text{rr}} dr^2 \,,
\end{equation}
as this will allow us to write various expression related to the string dynamics in a unified manner. Note that in the string frame, we have:
\begin{eqnarray}\label{8}
g_{\text{tt}}=-r^{2}e^{2A_s(r)}g(r),~~g_{11}=r^{2}e^{2A_s(r)}h(r),~~g_{22}=g_{33}=r^{2}e^{2A_s(r)}q(r),~~g_{\text{rr}}=\frac{e^{2A_s(r)}}{r^2g(r)} \,,
\end{eqnarray}
with
\begin{equation}\label{10}
h(r) = 1,~~q(r) = e^{\frac{B^{2}}{r^{2}}} \,,
\end{equation}
with similar expressions existing in the Einstein frame.

Before we fully dwell into the computation of chaotic dynamics of string in different frames, let us mention that the parameter $a$ appearing in the metric solution is completely arbitrary and is the only free parameter in this model, i.e. Eqs.~(\ref{asol})-(\ref{Vsol}) form a self-consistent solution of the magnetised EMD action (\ref{2}) for any choice of $a$. In the context of our holographic QCD model, its value can be fixed by taking inputs from the dual boundary QCD theory. For instance, in \cite{Bohra2020Feb}, the magnitude of $a$ was fixed by demanding the confinement/deconfinement (or the dual Hawking/Page) transition temperature to be around $270~\text{MeV}$ in the pure glue sector. This fixes $a$ to $0.15~\text{GeV}^{2}$. Moreover, notice from Eq.~(\ref{phisol}) that the requirement of real-valuedness of the dilaton field also puts an upper bound on $B$. For example, for $a=0.15~\text{GeV}^2$, the largest attainable magnitude of $B$ is $B\backsimeq 0.6~\text{GeV}$ \cite{Bohra2021Apr}. However, it is important to mention that many of the QCD features in the presence of a magnetic field, such as inverse magnetic behaviour, remain qualitatively the same for different values of $a$.

Once the magnitude of $a$ is fixed, so does the form of $V(z)$ in Eq.~(\ref{Vsol}). Essentially, the form of $V(z)$ is fixed by demanding the self-consistency of the Einstein-Maxwell-Dilaton field equations as well as the requirement of desirable properties of the dual boundary field theory. Notice that at the asymptotic boundary $V(z)$ has a very simple expression, i.e., it reduces to the cosmological constant in five dimensions $V(z)\rvert_{z\rightarrow 0}=2\Lambda+\dots$, which makes sure that the asymptotic boundary is AdS. Similarly, the subleading dots contain the information about the mass of the dilaton field, which satisfies the Breitenlohner-Freedman bound for stability in AdS space \cite{Breitenlohner:1982jf}.  Further more, the potential also satisfy the Gubser criterion to have a well-defined dual boundary theory \cite{Gubser:2000nd}. In the Appendix of \cite{Bohra2020Feb},  it is also discussed, by reexpressing the potential in terms of field variable $\phi$ rather than the coordinate $z$, via the inversion of $\phi(z)$, that $V(\phi)$ has the desirable properties of a physical effective potential.

\subsection{Chaos of Perturbative String: Analysis in String frame}\label{sec:2.2}
In this section we study the string motion and its chaotic behaviour in the presence of a background magnetic field at finite temperature. We rely on \cite{Hashimoto2018Oct,Colangelo2022Apr}. The string motion is described by the Nambu-Goto (NG) action,
\begin{equation}\label{12}
S = -\frac{1}{2\pi\alpha'} \int dt d\ell \sqrt{-h} \,,
\end{equation}
where $\alpha'$ is the string tension, $h$ is the determinant of the induced metric $h_{ij} = (g_s)_{\text{MN}}\frac{\partial X^{M}}{\partial \xi^{i}}\frac{\partial X^{N}}{\partial \xi^{j}}$ on the string world sheet, $(g_s)$ is the metric tensor in the string frame (\ref{4s}), and $\xi^i$ are the worldsheet coordinates.

To set the stage, we first study the static string configuration. The location of the string in the static case is specified by $r(\ell)$ and $x_{i}(\ell)$, with endpoints of the string located at $x_{i}=\pm L/2$. Here $\ell$ corresponds to a proper length measured along the string. Notice that in the presence of magnetic field we have two interesting scenarios to align the string: (i) parallel to or (ii) perpendicular to the magnetic field\footnote{Evidently, all possible angles are possible, but we will focus on these two special orientations.}. These two scenarios correspond to a string parametrization along $i=1$ or $i=3$ directions, i.e.~for $i=1$, the string endpoints lie along the parallel direction whereas for $i=3$, they lie along the perpendicular direction relative to the magnetic field.

 With the above parametrization, the NG action in the static case reads
\begin{equation}\label{13}
S = -\frac{T}{2\pi\alpha'}\int dt d\ell \sqrt{|g_{tt}g_{ii}(x_{i}')^{2} + g_{tt}g_{rr}(r')^{2}|} \,,
\end{equation}
where the prime $'$ denotes the derivative with respect to $\ell$. Since $x_{i}$ is a cyclic coordinate, its conjugate momentum
\begin{equation}\label{14}
\frac{\partial L}{\partial x_{i}'} = -\frac{T}{2\pi\alpha'}\frac{|g_{tt}|g_{ii}x_{i}'}{\sqrt{|g_{tt}|g_{ii}(x_{i}')^{2} + |g_{tt}|g_{rr}(r')^{2}}} \,,
\end{equation}
is a constant of motion. The location of the tip of the string $r(\ell=0)=r_0$ can be determined by the equation $\frac{dr}{dx_{i}}|_{\ell=0}=0$. This gives us,
\begin{equation}\label{15}
	\frac{\sqrt{|g_{tt}|}g_{ii}x_{i}'}{\sqrt{g_{ii}(x_{i}')^{2}+g_{rr}(r')^{2}}} = \sqrt{|g_{tt}|g_{ii}}\Big|_{\ell=0} \,.
\end{equation}
Notice that this minimal value $r_0$ of coordinate $r$ is reached at $x_{i}=0$ (or $\ell=0$). The above equation along with following condition,
\begin{equation}\label{16}
	d\ell^{2} = g_{ii}dx_{i}^{2}+g_{rr}dr^{2} \,,
\end{equation}
allow us to determine the string profile via the following equations
\begin{equation}\label{17}
	x' = \pm\frac{\sqrt{-g_{tt}(r_{0})g_{ii}(r_{0})}}{\sqrt{-g_{tt}}g_{ii}} \,,
\end{equation}
\begin{equation}\label{18}
	r' = \pm\frac{\sqrt{-g_{tt}g_{ii}+g_{tt}(r_{0})g_{ii}(r_{0})}}{\sqrt{-g_{tt}g_{ii}g_{rr}}} \,.
\end{equation}
Now, using the boundary condition that the endpoints of the string lie on the asymptotic boundary at $x_{i}=\pm L/2$,  we can further relate the string length $L$ to $r_0$. This is given by,
\begin{equation}\label{19}
	L=2 \int_r^{\infty } \, dr \biggl( \frac{g_{ii}(r)}{g_{rr}(r)} \left(\frac{g_{tt}(r) g_{ii}(r)}{g_{tt}\left(r_0\right)
					g_{ii}\left(r_0\right)}-1\right) \biggr)^{-\frac{1}{2}} \,.
\end{equation}

%%%%%%%%%%%%%%%%%%%%%%%%%%%%%%
\begin{figure}[ht]
\begin{minipage}[b]{0.45\linewidth}
\centering
\includegraphics[width=2.8in,height=2.1in]{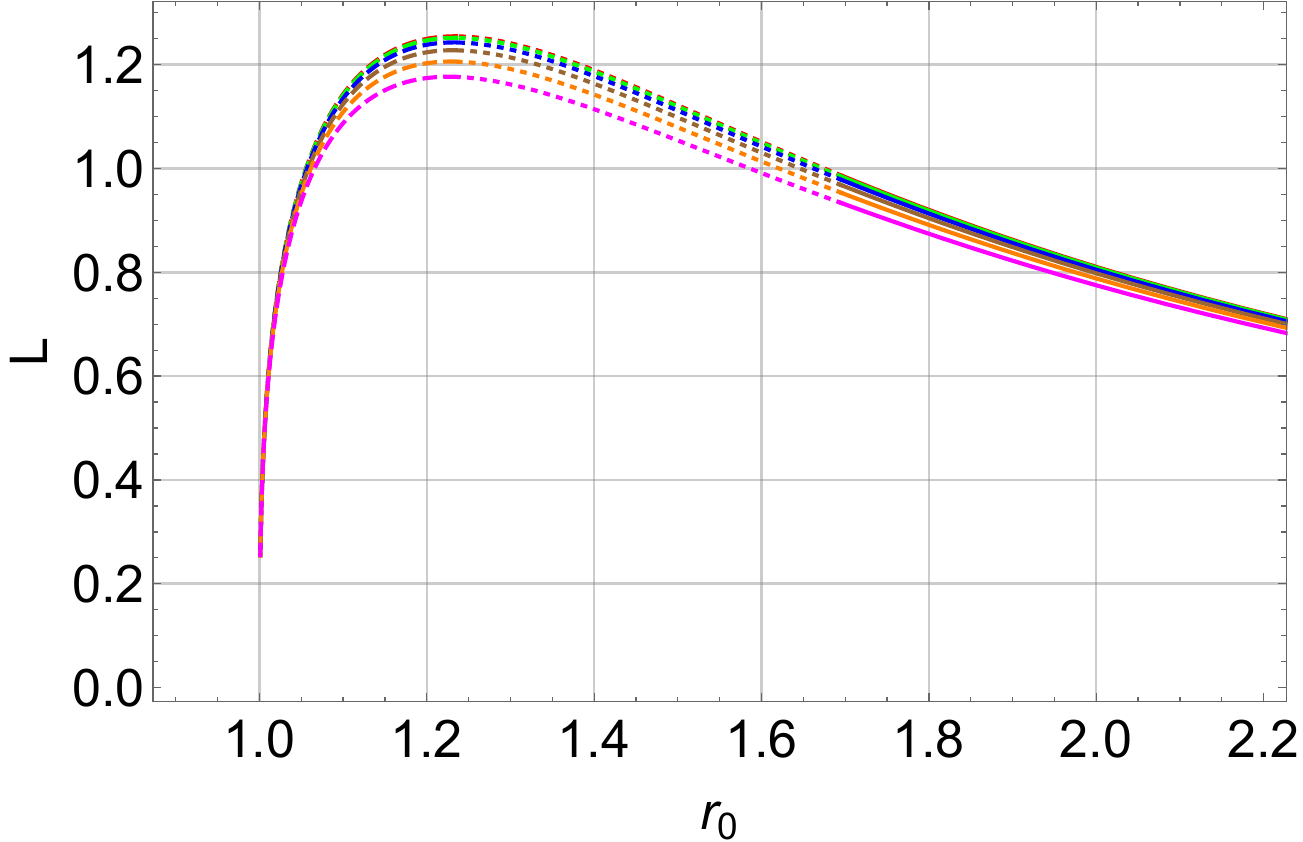}
\caption{$L$ as a function of $r_0$ for different values of $B$ in the parallel case. Here $r_h=1$ is used . The red, green, blue, brown, orange, and magenta
curves correspond to $B=0$, $0.1$, $0.2$, $0.3$, $0.4$, and $0.5$ respectively. In units of GeV.}
\label{r0vsLvsBrh1paraSF}
\end{minipage}
\hspace{0.4cm}
\begin{minipage}[b]{0.45\linewidth}
\centering
\includegraphics[width=2.8in,height=2.1in]{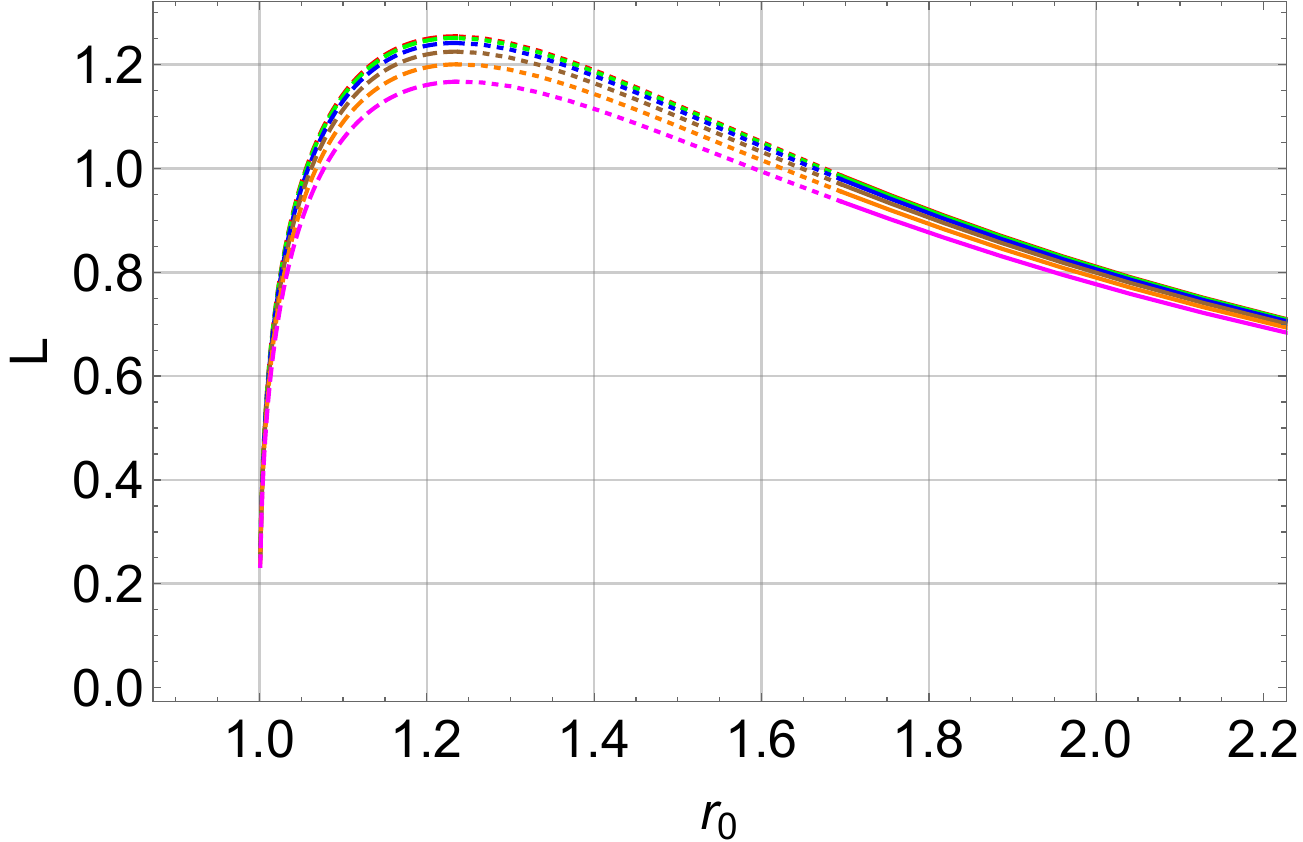}
\caption{$L$ as a function of $r_0$ for different values of $B$ in the perpendicular case. Here $r_h=1$ is used. The red, green, blue, brown, orange, and magenta
curves correspond to $B=0$, $0.1$, $0.2$, $0.3$, $0.4$, and $0.5$ respectively. In units of GeV.}
\label{r0vsLvsBrh1perpSF}
\end{minipage}
\end{figure}
%%%%%%%%%%%%%%%%%%%%%%%%%%%%%%
\begin{figure}[t]
	\centering
	\includegraphics[width=0.4\linewidth]{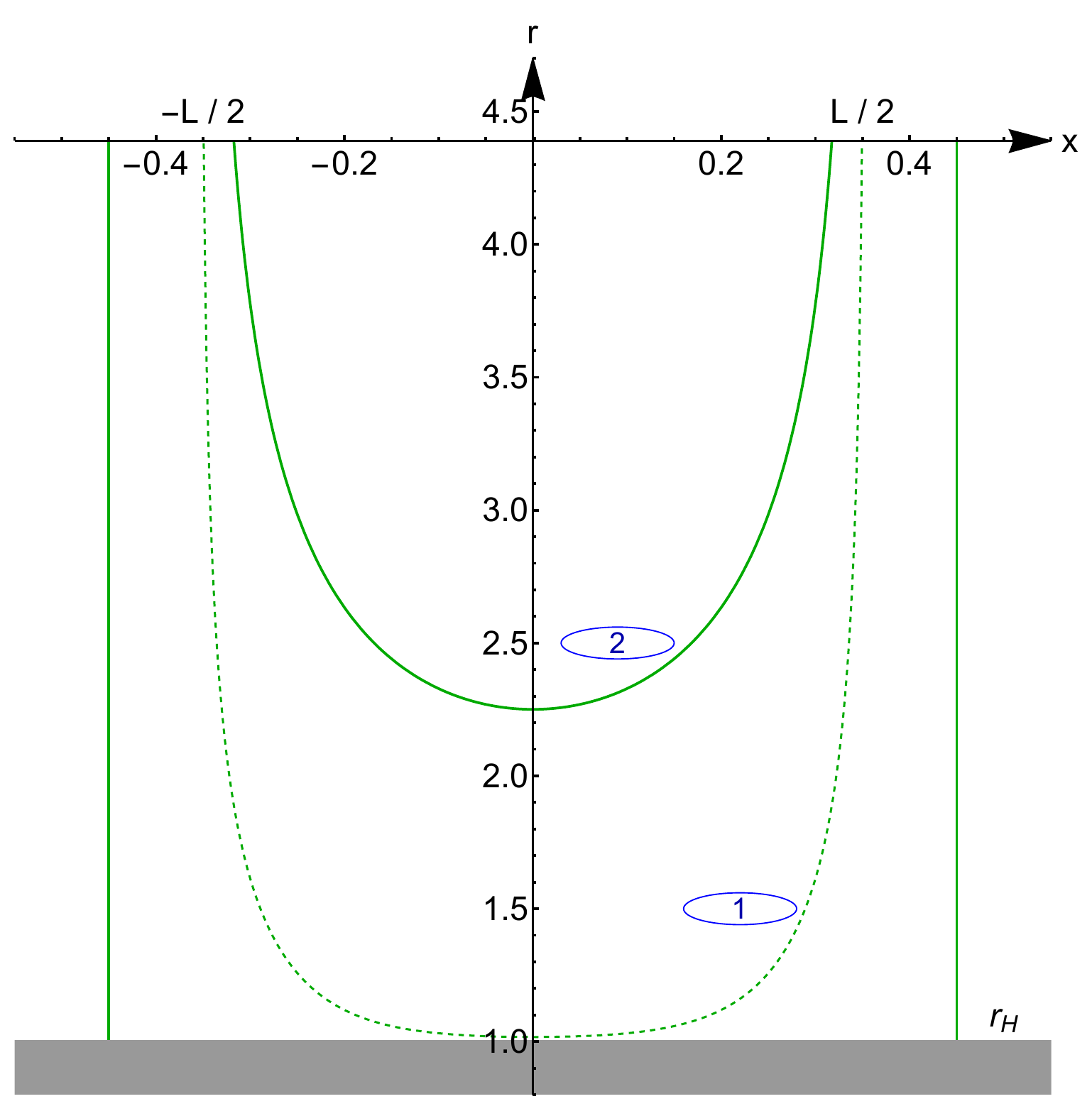}
	\caption{Three different shapes of the static suspended strings. The solid and dashed lines are connected strings corresponding to the local minima and maxima of the energy respectively. Two straight solid parallel lines correspond to two disconnected strings going from the asymptotic boundary to the horizon. Here we have used $B=0.1$, $r_{h}=1$, and $L=0.7$ for the parallel magnetic field. In units of GeV.}
	\label{stringprofilezh1BPt1}
\end{figure}

In Figs.~\ref{r0vsLvsBrh1paraSF} and \ref{r0vsLvsBrh1perpSF}, the behaviour of $L(r_0)$ for different values of $B$ in the parallel and perpendicular cases are shown. Here we have taken $r_h=1$ for illustrative purposes, but similar results appear for other horizon radius as well. These results suggests that, irrespective to the value of $B$, there is a maximum string length $L_{max}$ above which the string solution [(\ref{17}) and (\ref{18})] does not exist.  This $L_{max}$ is a magnetic field dependent quantity, which decreases for both parallel and perpendicular orientations. Moreover, below $L_{max}$, there are two solutions for each values of $L$. The one solution corresponding to large $r_0$ (indicated by solid lines) is closer to the asymptotic boundary whereas the second solution (indicated by dashed lines) is closer to the horizon, see Fig.~\ref{stringprofilezh1BPt1}. It turns out that the former solution actually corresponds to the local minimum of the energy whereas the latter solution corresponds to the local maximum of the energy. At this point, we note that the free energy spectrum of the string profile can be computed from the on-shell NG action. For the string solution in Eqs.~(\ref{17}) and (\ref{18}), the free energy is given by,
\begin{equation}\label{19}
F=-\frac{2}{2\pi\alpha'} \int_r^{\infty } \, dr \sqrt{-g_{tt}(r)g_{rr}(r)}\biggl( \frac{g_{tt}(r_0)g_{ii}(r_0)}{g_{tt}(r)g_{ii}(r)} \left(\frac{g_{tt}(r) g_{ii}(r)}{g_{tt}\left(r_0\right)
g_{ii}\left(r_0\right)}-1\right) \biggr)^{-\frac{1}{2}}\,.
\end{equation}
Notice that $F$ as usual, contains UV divergences in the limit $r \rightarrow \infty$. To cancel these divergences we subtract the free energy of two disconnected strings that are separated by a distance $L$ and extend them from the boundary to the horizon. One can think of these disconnected strings as the third string solution. The free energy expression of these two disconnected strings is
\begin{equation}\label{19}
F_{dis}=-\frac{2}{2\pi\alpha'} \int_r^{\infty } \, dr \sqrt{-g_{tt}(r)g_{rr}(r)}\,.
\end{equation}
Since the nature of the poles in the disconnected free energy is the same as in $F$, it thereby allows us to regularize the latter in a minimalistic way. Therefore, below $L_{max}$, there are three possible string configurations (two connected and one disconnected) whereas above $L_{max}$ only the disconnected configuration exists. This is more clearly illustrated in Fig.~\ref{stringprofilezh1BPt1}, where two connected solutions are indicated by solid and dashed lines whereas the disconnected solution is indicated by two parallel vertical lines.

%%%%%%%%%%%%%%%%%%%%%%%%%%%%%%
\begin{figure}[ht]
\begin{minipage}[b]{0.45\linewidth}
\centering
\includegraphics[width=2.8in,height=2.1in]{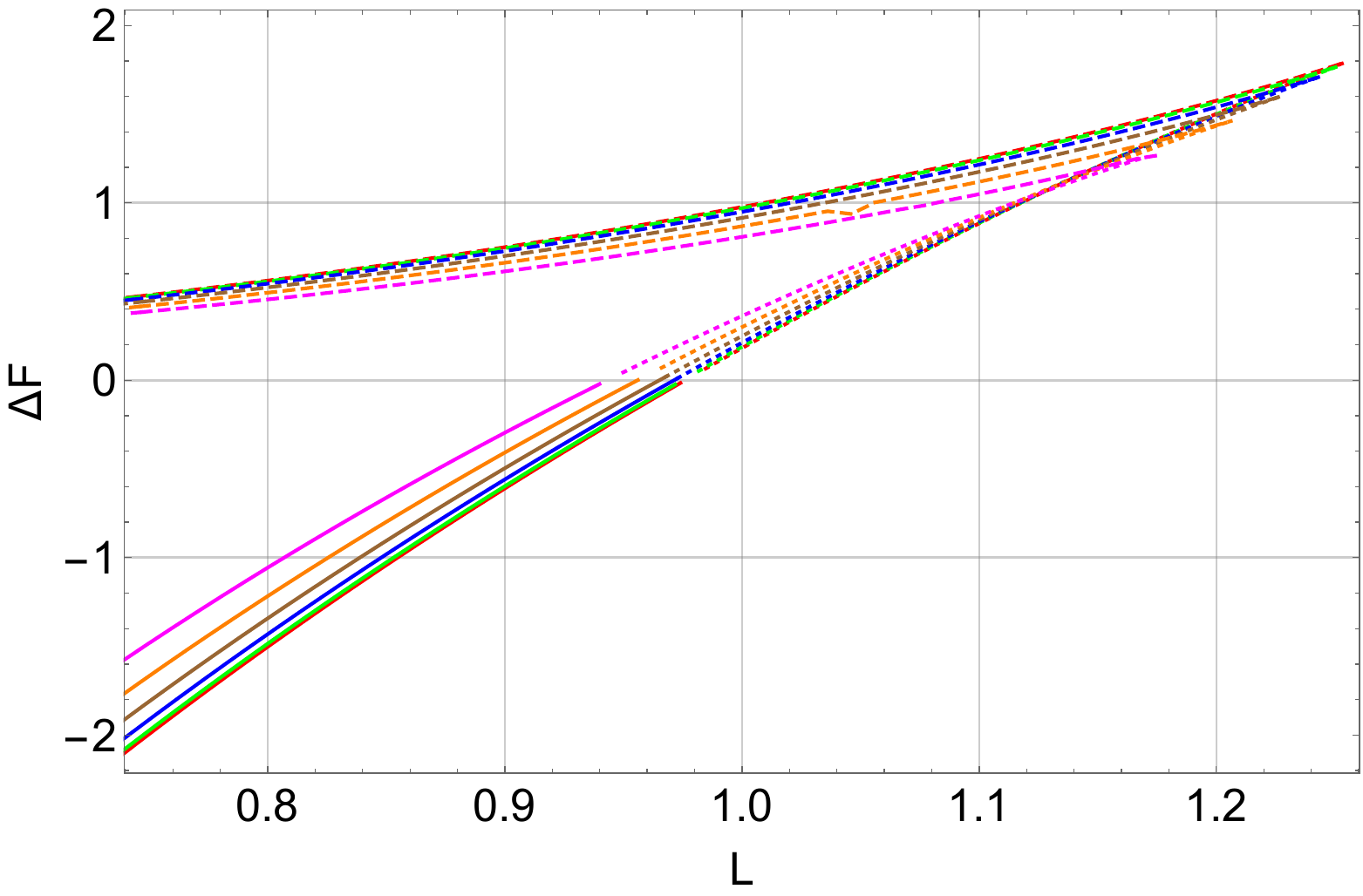}
\caption{$\Delta F$ as a function of $L$ for different values of $B$ in the parallel case. Here $r_h=1$ is used.  The red, green, blue, brown, orange, and magenta
curves correspond to $B=0$, $0.1$, $0.2$, $0.3$, $0.4$, and $0.5$ respectively. In units of GeV.}
\label{LvsDeltaFvsBrh1paraSF}
\end{minipage}
\hspace{0.4cm}
\begin{minipage}[b]{0.45\linewidth}
\centering
\includegraphics[width=2.8in,height=2.1in]{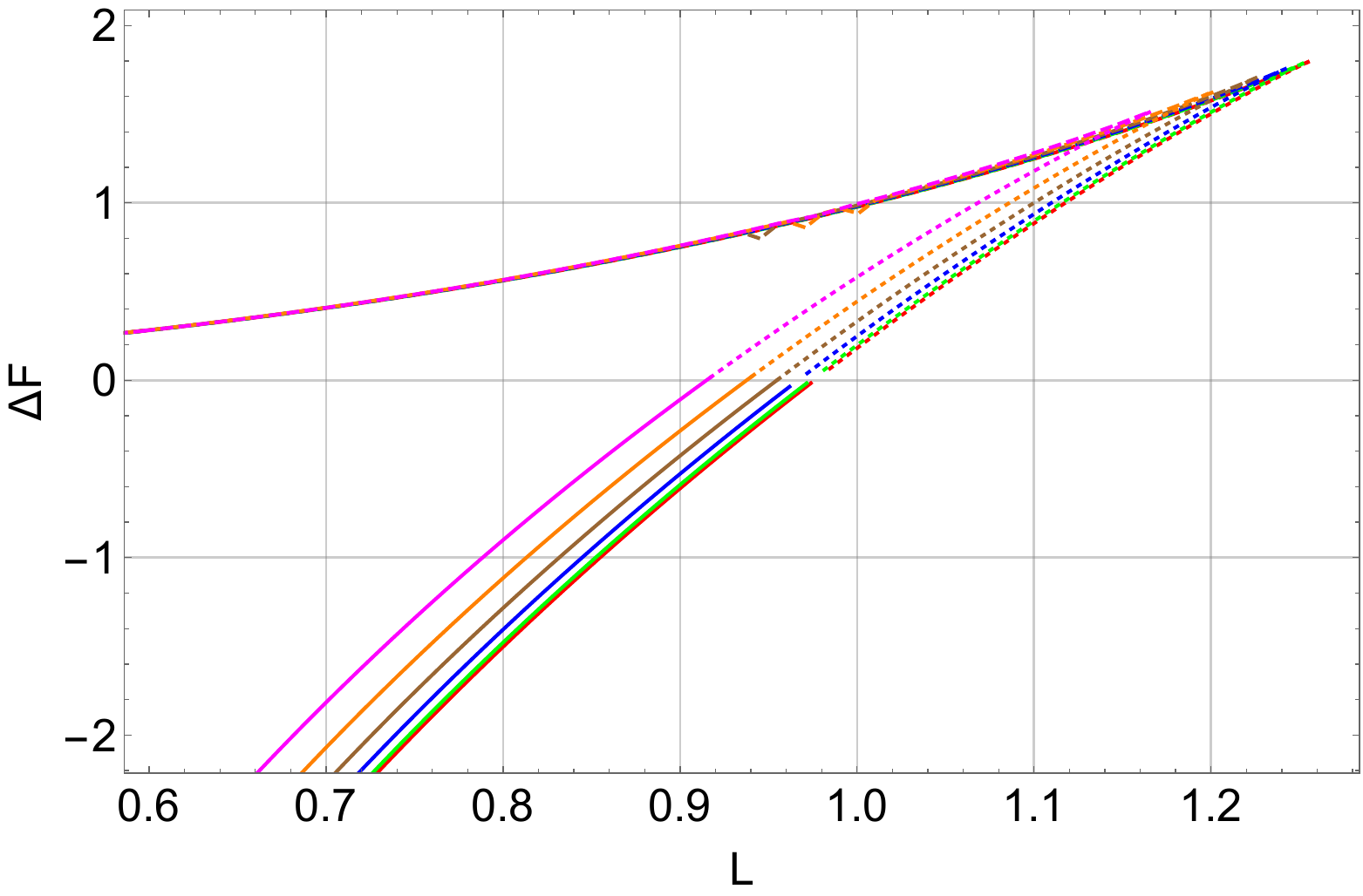}
\caption{$\Delta F$ as a function of $L$ for different values of $B$ in the perpendicular case. Here $r_h=1$ is used.  The red, green, blue, brown, orange, and magenta
curves correspond to $B=0$, $0.1$, $0.2$, $0.3$, $0.4$, and $0.5$ respectively. In units of GeV.}
\label{LvsDeltaFvsBrh1perpSF}
\end{minipage}
\end{figure}
%%%%%%%%%%%%%%%%%%%%%%%%%%%%%%

The free energy behaviour of these string solutions is shown in Figs.~\ref{LvsDeltaFvsBrh1paraSF} and \ref{LvsDeltaFvsBrh1perpSF}, where the free energy difference $\Delta F=F-F_{dis}$ is plotted. Here, dashed and solid lines again correspond to small and large $r_0$ solutions respectively. We observe that the small $r_0$ string solution always has a higher free energy than the large $r_0$ solution, suggesting that the latter actually corresponds to true minima of the solution. Moreover, depending on the length $L$, the free energy of the large $r_0$ solution can be greater, less, or equal to $F_{dis}$. In particular, below some critical length $L_{crit}$ the large $r_0$ solution minimizes the free energy whereas the above this critical length $F_{dis}$ minimizes the free energy. Accordingly, for lengths between $L_{crit}\leq L \leq L_{max}$, the large $r_0$ solution  actually corresponds to metastable phase. This metastable phase is indicated by dotted lines in Figs.~\ref{r0vsLvsBrh1paraSF} and \ref{LvsDeltaFvsBrh1paraSF}. However, the free energy of the small $r_0$ solution is always larger than $F_{dis}$.  This analysis suggests that the string profile near the horizon always corresponds to the local maximum of the energy. These results are true for both parallel and perpendicular orientations of the string.

To study chaos in the string motion and analyse the anisotropic effects of $B$ on it, we examine the background string solution corresponding to small $r_0$ (dashed lines), which is closer to the horizon, and perturb it. In particular, since black holes are thermal objects the closeness of the tip of the string to the black hole horizon might enhance its chaotic dynamics. Indeed, as we will see in the next section, the small $r_0$ unstable string solution does exhibit chaotic behaviour. For completeness, we also analyse the dynamics of large $r_0$ stable string solution and do not find any chaotic signature.

\subsection{Perturbing the static string}\label{sec:2.3}
\begin{figure}[t]
	\centering
	\includegraphics[width=0.4\linewidth]{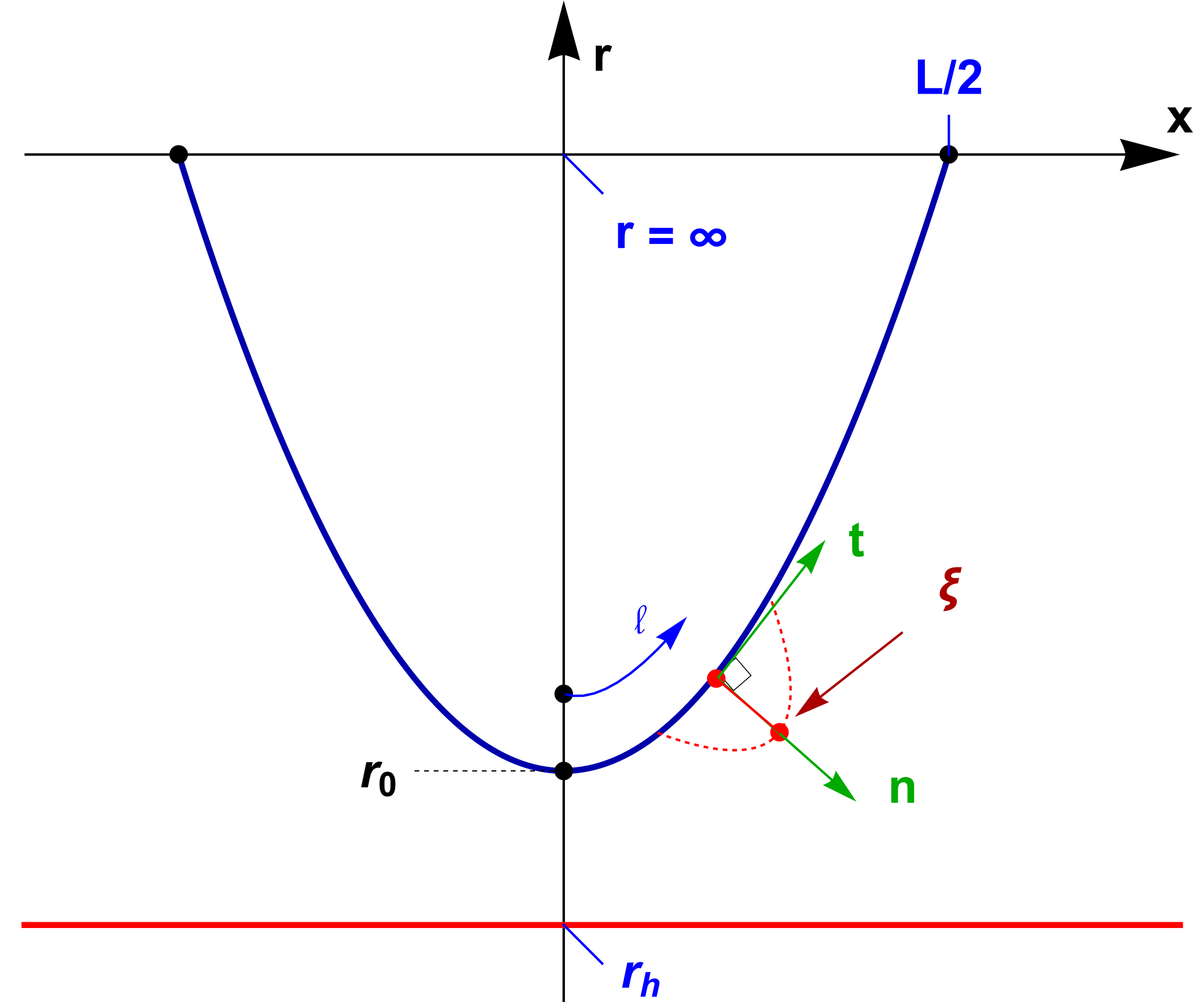}
	\caption{Illustration of the static string profile and normal perturbation along it.}
	\label{fig2}
\end{figure}

In order to analyse the chaotic dynamics of the string, we perturb it by a small time dependent effect. In particular, we consider perturbative motion around the unstable large $r_0$ string solution and construct the NG action up to next-to-leading order in perturbation. As we will see below, the leading order perturbation describes an oscillation whereas the next-to-leading order perturbation generates a trapping potential near the unstable point. Following \cite{Hashimoto2018Oct,Colangelo2020Oct}, we introduce a perturbation of the string along the normal direction by a proper distance $\xi(t,\ell)$ for both parallel ($i=1$) and perpendicular ($i=3$) magnetic field cases. The perturbation is shown in Fig.~\ref{fig2}.

The time-dependent perturbation $\xi(t,\ell)$ modifies the location $r$ and $x$ of the string as
\begin{eqnarray}\label{23}
	r(t,\ell)=r_b(\ell)+\xi(t,\ell)n^{r}(\ell)\,, \qquad	x(t,\ell)=x_b(\ell)+\xi(t,\ell)n^{x}(\ell),
\end{eqnarray}
where $r_b(\ell)$ and $x_b(\ell)$ correspond to the static solutions, which can be obtained by integrating Eqs.~(\ref{17}) and (\ref{18}). Here $x=x_1 (x_3)$ corresponds for the parallel (perpendicular) orientation of the string with respect to the magnetic field. $n^{M} = (0,n^{x},0,0,n^{r})$ is unit vector which is orthogonal to tangent vector $t^{M}$. Therefore, we have
\begin{eqnarray}
	g_{rr}(r)(n^{r})^{2}+g_{xx}(r)(n^{x})^{2} &=& 1  \label{20} \,,\nonumber\\
	r'(\ell)g_{rr}(r)n^{r}+x'(\ell)g_{xx}(r)n^{x} &=& 0\,. \label{21}
\end{eqnarray}
The above two equations can be used to find the normal components $n^{x}$ and $n^{r}$. For an outward perturbation, as in Fig.\ref{fig2}, these are given by
\begin{eqnarray}\label{22}
	n^{x}(\ell)=\sqrt{\frac{g_{rr}}{g_{xx}}}r'(\ell),~~~~ n^{r}(\ell)=-\sqrt{\frac{g_{xx}}{g_{rr}}}x'(\ell) \,.
\end{eqnarray}
To describe the dynamics of the small perturbation, we can expand the metric function, and hence the NG action, around the static solution $\{r_b(\ell), x_b(\ell)\}$ upto next-to-leading order in perturbation $\xi$. To this order, the NG action contains a quadratic term and a cubic term. The quadratic term has the expression
\begin{equation}\label{24}
	S^{(2)}=\frac{1}{2\pi\alpha'}\int dt \int_{-\infty}^{\infty} dl (C_{tt}^{x_{i}}\dot{\xi}^{2}+C_{\ell \ell}^{x_{i}} \xi'^{2}+C_{00}^{x_{i}}\xi^{2}) \,,
\end{equation}
where the dot $\dot{}$ denotes the derivative with respect to $t$. The coefficients $C_{tt}^{x_{i}}$, $C_{\ell \ell}^{x_{i}}$, $C_{00}^{x_{i}}$ depend on $\ell$ and their expressions are given by
\begin{eqnarray}\label{25}
	\begin{split}
		C_{tt}^{x_{i}}(\ell)=&\frac{e^{-A_s({r_b})}}{2 {r_b} \sqrt{g({r_b})}},\\
		C_{\ell \ell}^{x_{i}}(\ell)=&-\frac{1}{2} {r_b} e^{A_s({r_b})} \sqrt{g({r_b})},\\
		C_{00}^{x_1}(\ell)=&\frac{e^{-5 A_s({r_b})}}{8 {r_b}^3 g({r_b})^{3/2} h({r_b})^2}\\&
		\times \Biggl(r_0^4 e^{4 A_s(r_0)} g(r_0) h(r_0) \Bigl({r_b} g({r_b}) \Bigl(2 h({r_b}) \Bigl(\Bigl(3 {r_b} A_s'({r_b})+1\Bigr) g'({r_b})-{r_b} g''({r_b})\Bigr)\\&+{r_b} g'({r_b}) h'({r_b})\Bigr)+2 g({r_b})^2 \Bigl({r_b} \Bigl({r_b} A_s'({r_b})+1\Bigr) h'({r_b})+h({r_b}) \Bigl(8 {r_b} A_s'({r_b})\\&+{r_b}^2 \Bigl(6 A_s'({r_b})^2-2 A_s''({r_b})\Bigr)+4\Bigr)\Bigr)+2 {r_b}^2 h({r_b}) g'({r_b})^2\Bigr)\\&-{r_b}^4 e^{4 A_s({r_b})} g({r_b})^2 \Bigl(2 {r_b} h({r_b}) g'({r_b}) \Bigl(2 h({r_b}) \Bigl({r_b} A_s'({r_b})+1\Bigr)+{r_b} h'({r_b})\Bigr)\\&+g({r_b}) \Bigl(2 {r_b} h({r_b}) \Bigl(2 \Bigl({r_b} A_s'({r_b})+2\Bigr) h'({r_b})+{r_b} h''({r_b})\Bigr)+4 h({r_b})^2 \Bigl(4 {r_b} A_s'({r_b})\\&+{r_b}^2 \Bigl(A_s''({r_b})+A_s'({r_b})^2\Bigr)+2\Bigr)-{r_b}^2 h'({r_b})^2\Bigr)\Bigr)\Biggr) \,.
	\end{split}		
\end{eqnarray}
here, as mentioned earlier, $i=1$ and $i=3$, correspond to the parallel and perpendicular magnetic field orientations. The expressions of $C_{00}^{x_1}$  and $C_{00}^{x_3}$ are identical, except $h(r)$ is replaced by $q(r)$.

In order to analyze the chaotic dynamics, we first find the equation of motion from the action (\ref{24}),
\begin{equation}\label{26}
	C_{tt}^{x_{i}}\ddot{\xi}+\partial_{\ell}(C_{\ell \ell}^{x_{i}}\xi')-C_{00}^{x_{i}}\xi = 0 \,.
\end{equation}
Using the factorization $\xi(t,\ell)=\xi(\ell)e^{i\omega t}$, it can be recast into a Sturm-Liouville equation
\begin{equation}\label{27}
	\partial_{\ell}(C_{\ell \ell}^{x_{i}}\acute{\xi})-C_{00}^{x_{i}}\xi = \omega^{2}C_{tt}^{x_{i}}\xi,
\end{equation}
where $W(\ell) = -C_{tt}^{x_{i}}(\ell)$ is the weight function, with inner product
\begin{equation}\label{28}
	(\xi,\zeta) \equiv \int_{-\infty}^{\infty} W(\ell)\xi(\ell)\zeta(\ell) d\ell
\end{equation}
Next we numerically solve Eq.~(\ref{27}) for different values of magnetic field $B$. For this purpose, we impose the boundary condition $\xi(l) \xrightarrow{l\rightarrow \pm\infty}0$ and set $L=1.1$. Fixing $L$ to a particular value makes $r_0$ a $B$-dependent quantity\footnote{Note that this is a different strategy than adopted in \cite{Colangelo2022Apr}, where instead the value of $r_0$ was fixed, irrespective of $B$. This translates into a different interquark separation $L$ for different $B$. Since our motive is primarily to investigate the anisotropic effects of $B$ on chaos, we find it more reasonable to fix the value of $L$ and let the string bulk profiles take different $r_0$ value in the magnetised AdS background naturally for different $B$.}.

The values of $r_0$ for $L=1.1$ for different values of $B$ for the unstable string configuration have been collected in Table~\ref{tabler0vsB}. Notice that the $r_0$ values increase for both parallel and perpendicular magnetic field. This implies that the tip of the string is moving away from the horizon as the magnetic field increases. If the horizon were to be the source of chaos, as has been suggested in \cite{Hashimoto2018Oct}, then it would indicate less chaos with higher magnetic field for both parallel and perpendicular cases. As will be seen shortly, this expectation indeed turned out to be true and the string motion clearly becomes less chaotic with higher magnetic field values.

\begin{figure}[t]
	\centering
	\subfloat[Parallel configuration]{\label{fig:sub1}	\includegraphics[width=0.45\textwidth]{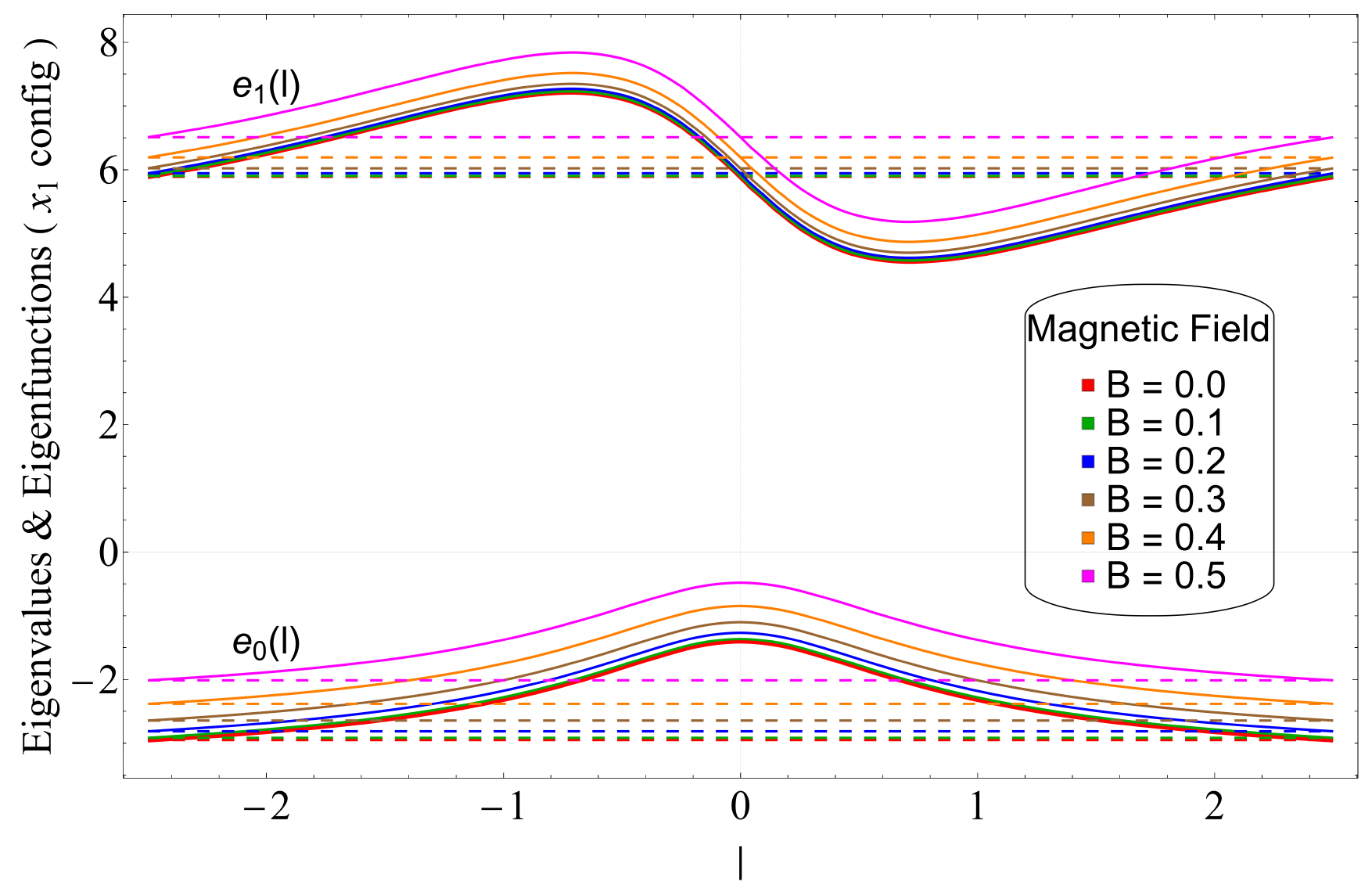}}
	\hfill
\subfloat[Perpendicular configuration]{\label{fig:sub2}
		\includegraphics[width=0.45\textwidth]{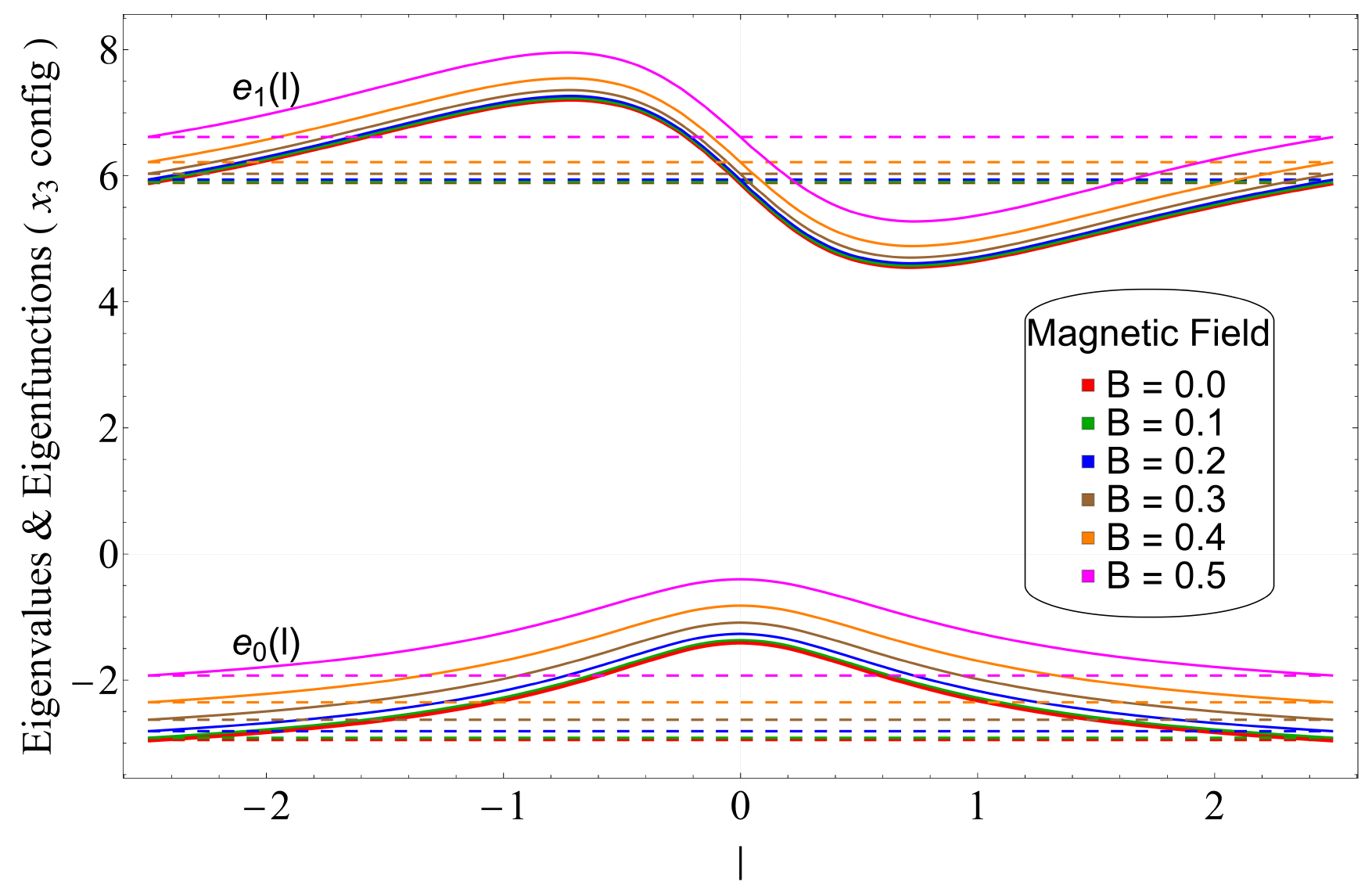}}
	\caption{\label{fig3}The magnetic field dependence of eigenfunctions $e_{0}(\ell)$ and $e_{1}(\ell)$ of Eq.~(\ref{27})  for parallel and perpendicular string configurations. Here $L=1.1$ is used. In units of GeV.}
\end{figure}

Solving Eq.~(\ref{27}), we can compute both eigenvalues and -functions of this system.  The two lowest eigenvalues $\omega_{0}^{2}$ and $\omega_{1}^{2}$ for different values of $B$ for the parallel and perpendicular string configurations are collected in Table \ref{table2}. The corresponding eigenfunctions $\xi(\ell)=e_{0}(\ell)$ and $\xi(\ell)=e_{1}(\ell)$ are shown in Fig.~\ref{fig3}. Note that $e_{0}(\ell)$ and $e_{1}(\ell)$ are even and odd functions of $\ell$, respectively. Generally one encounters negative eigenvalues in case of unstable systems. We find that the lowest eigenvalue is always negative for all values of $B$ in our system, indicating instability of the string configurations. We further find that $\omega_{0}^{2}$ increases with $B$, both in the ($x_1$) parallel configuration as well as in the ($x_3$) perpendicular configuration. This in turn implies that the magnetic field stabilizes the system for both orientations of the magnetic field. Moreover, the effect of $B$ is stronger for the string in the perpendicular direction compared to the parallel direction, suggesting that the magnetic field stabilizes the system in the former configuration more than in the latter. This should be contrasted with the results of \cite{Colangelo2022Apr}, where the magnetic field has produced more stability in the parallel configuration compared to the perpendicular configuration. Let us also mention for completeness that for the stable string configurations (solid line of Fig.~\ref{stringprofilezh1BPt1}), the eigenvalues are always positive, indicating the stability of these configurations against perturbations.

\begin{table}[t]
	\centering
		\begin{tabular}{|c| c ||c|}
 \hline
 $B$ & $r_0~(||)$ & $r_0~(\perp)$\\
 \hline
 		0 & 1.08318 &  1.08318\\
		0.1 & 1.08382 &  1.08429\\
		0.2 & 1.08581 &  1.08768\\
		0.3 & 1.08940 &  1.09417\\
		0.4 & 1.09598 &  1.10519\\
		0.5 & 1.10728 &  1.12473\\
		%\bottomrule
\hline
\end{tabular}
\caption{$r_0$ values for different values of $B$ for the parallel and perpendicular unstable string configurations. Here $L=1.1$ is used. In units of GeV.}
\label{tabler0vsB}
\end{table}

\begin{table}[t]
	\centering
	\begin{tabular}{|c| c |c||c|c|}
 \hline
 	$B$ & $\omega_{0}^{2}~(||)$ & $\omega_{1}^{2}~(||)$ & $\omega_{0}^{2}~(\perp)$ & $\omega_{1}^{2}~(\perp)$ \\
  \hline
		0 & $-2.9503$ & 5.8882 & $-2.9503$ & 5.8882\\
		0.1 & $-2.9166$ & 5.9010 &  $-2.9155$ & 5.9006\\
		0.2 & $-2.8152$ & 5.9426 &  $-2.8121$ & 5.9393\\
		0.3 & $-2.6448$ & 6.0229 &  $-2.6291$ & 6.0306\\
		0.4 & $-2.3834$ & 6.1918 &  $-2.3512$ & 6.2170\\
		0.5 & $-2.0130$ & 6.5096 &  $-1.9276$ & 6.6158\\
\hline
	\end{tabular}
	\caption{The magnetic field dependence of eigenvalues $\omega_{0}^{2}$ and $\omega_{1}^{2}$ of Eq.~(\ref{27}) for parallel and perpendicular string configurations. Here $L=1.1$ is used. In units of GeV.}
\label{table2}
\end{table}

Next we evaluate the action up to cubic order. Curtailing the perturbation up to the two lowest eigenfunctions
\begin{equation}\label{30}
	\xi(t,\ell) = c_{0}(t)e_{0}(\ell)+c_{1}(t)e_{1}(\ell),
\end{equation}
the cubic order action, up to the surface terms, reduces to
\begin{equation}\label{29}
	S^{(3)} = \frac{1}{2\pi\alpha^{'}}\int dt \int_{-\infty}^{\infty} d\ell \left[ D_{0}\xi^{3}+D_{1}^{x_{i}}\xi\acute{\xi}^{2}+D_{2}^{x_{i}}\xi\dot{\xi}^{2} \right],
\end{equation}
with $D_{0,1,2}^{x_{i}}$ functions of $\ell$, and the coefficients $c_{0}(t)$ and $c_{1}(t)$ encode the perturbation's time dependence. Using the above mentioned form of $\xi(t,\ell)$ we have,
\begin{equation}\label{31}
	\begin{split}
		S^{(3)} &= \frac{1}{2\pi\alpha^{'}}\int dt \int_{-\infty}^{\infty} d\ell \big[ \left(D_{0}^{x_{i}}e_{0}^{3}+D_{1}^{x_{i}}e_{0}\acute{e_{0}}^{2}\right)c_{0}^{3}(t)+\left(3 D_{0}^{x_{i}}e_{0}e_{1}^{2}+D_{1}^{x_{i}}\left(2\acute{e_{0}}e_{1}\acute{e{_{1}}}+e_{0}\acute{e_{1}}^{2}\right)\right)c_{0}c_{1}^{2}\\
		&+D_{2}^{x_{i}}\left(e_{0}e_{1}^{2}c_{0}\dot{c_{1}}^{2}+e_{0}^{3}c_{0}\dot{c_{0}}^{2}+2e_{0}e_{1}^{2}\dot{c_{0}}c_{1}\dot{c_{1}}\right)\big]\,.
	\end{split}
\end{equation}
Summing $S^{(2)}+S^{(3)}$, and integrating it for $\ell$, we can obtain the relevant action for $c_{0}(t)$ and $c_{1}(t)$,
\begin{equation}\label{32}
	\begin{split}
		S^{(2)}+S^{(3)} &= \frac{1}{2\pi\alpha^{'}}\int dt \bigg[\sum_{n=0,1}(\dot{c_{n}}^{2}-\omega_{n}^{2}c_{n}^{2})+K_{1}^{x_{i}}c_{0}^{3}+K_{2}^{x_{i}}c_{0}c_{1}^{2}+K_{3}^{x_{i}}c_{0}\dot{c_{0}}^{2}+K_{4}^{x_{i}}c_{0}\dot{c_{1}}^{2}+K_{5}^{x_{i}}\dot{c_{0}}c_{1}\dot{c_{1}}\bigg]\,.
	\end{split}
\end{equation}
The coefficients $K_{1,...,5}^{x_{i}}$ appearing in the above action can be computed numerically. These coefficients turn out to be $L$- and $B$- dependent. For $L=1.1$, these coefficients are given in Table \ref{table3} for various values of $B$ for the parallel and perpendicular string configurations.

\begin{table}[t]
	\centering
	\begin{tabular}{|c|c|c|c|c|c||c|c|c|c|c|}
		\hline
		$B$ & $K_{1}~(||)$ & $K_{2}~(||)$ & $K_{3}~(||)$ & $K_{4}~(||)$ & $K_{5}~(||)$     & $K_{1}~(\perp)$ & $K_{2}~(\perp)$ & $K_{3}~(\perp)$ & $K_{4}~(\perp)$ & $K_{5}~(\perp)$ \\
		\hline
		0 & 9.873 & 12.432 & 5.896 & 2.197 & 4.394& 9.873 & 12.432 & 5.896 & 2.197 & 4.394\\
		0.1 & 9.782 & 12.352 & 5.885 & 2.196 & 4.391& 9.771 & 12.340 & 5.872 & 2.197 & 4.395\\
		0.2 & 9.511 & 12.115 & 5.852 & 2.191 & 4.382& 9.476 & 12.069 & 5.804 & 2.198 & 4.397\\
		0.3 & 9.066 & 11.734 & 5.797 & 2.182 & 4.364& 8.982 & 11.611 & 5.684 & 2.199 & 4.397\\
		0.4 & 8.414 & 11.182 & 5.697 & 2.165 & 4.329& 8.293 & 10.961 & 5.510 & 2.197 & 4.393\\
		0.5 & 7.552 & 10.485 & 5.543 & 2.134 & 4.268& 7.378 & 10.074 & 5.261 & 2.187 & 4.373\\
\hline
\end{tabular}
\caption{The magnetic field dependence of  coefficients $K_{i}$ appearing in Eq.~(\ref{32}) for parallel and perpendicular string configurations. Here $L=1.1$ is used. In units of GeV.}
	\label{table3}
\end{table}

The action given by Eq.~(\ref{32}) provides us with the information of the string's motion. In particular, there is a trapping potential for the unstable string configuration. In that trap, we want to analyse the behaviour of $c_{0}$ and $c_{1}$. Unfortunately, the kinetic terms of $c_{0}$ and $c_{1}$ can become negative in some parts of the potential, hampering a numerical analysis. In order to make the kinetic terms positive definite, we follow the strategy adopted in \cite{Hashimoto2018Oct,Colangelo2020Oct} and do the following variable change: $c_{0,1}\rightarrow\tilde{c}_{0,1}$, where $c_{0}=\tilde{c_{0}}+\alpha_{1}\tilde{c_{0}}^{2}+\alpha_{2}\tilde{c_{1}}^{2}$ and $c_{1}=\tilde{c_{1}}+\alpha_{3}\tilde{c_{0}}\tilde{c_{1}}$. We neglect $\mathcal{O}(\tilde{c_{i}}^{4})$ terms, and choose appropriate values for $\alpha_{i}$, to make sure that the kinetic terms are now positive definite. One example of such a choice is $\alpha_{1}=-1.45$, $\alpha_{2}=-0.5$ and $\alpha_{3}=-1$, which gives us the modified action:
\begin{equation}\label{33}
	\begin{split}
		S^{(modified)} &= \frac{1}{2\pi\alpha^{'}}\int dt \bigg[\sum_{n=0,1}(\dot{\tilde{c_n}}{}^2-\omega_{n}^{2}\tilde{c_n}^{2}) +\tilde{K_{1}}^{x_{i}}\tilde{c_{0}}^{3}+\tilde{K_{2}}^{x_{i}}\tilde{c_{0}}\tilde{c_{1}}^{2}+\tilde{K_{3}}^{x_{i}}\tilde{c_{0}}\dot{\tilde{c_0}}{}^2
+\tilde{K_{4}}^{x_{i}}\tilde{c_{0}}\dot{\tilde{c_1}}{}^2+\tilde{K_{5}}^{x_{i}}\dot{\tilde{c_0}}\tilde{c_{1}}\dot{\tilde{c_1}}\bigg] \,.
	\end{split}
\end{equation}
This variable change makes the time evolution of the system well-posed without affecting the dynamics. In addition to that the chaotic behaviour also shows up in the modified action.

The above action can be used to find the equations of motion of $\tilde{c_0}$ and $\tilde{c_1}$. Let us explicitly write down these equations as they will be useful in the analysis of the Lyapunov exponent at the unstable fixed point,
\begin{eqnarray}
		&&\ddot{\tilde{c_0}}=\frac{1}{4 \Bigl(\tilde{c_0} \Bigl(4 \alpha _1+\tilde{K_3}\Bigr)+1\Bigr) \Bigl(\tilde{c_0} \Bigl(2 \alpha _3+\tilde{K_4}\Bigr)+1\Bigr)-\tilde{c_1}{}^2 \Bigl(4 \alpha _2+2 \alpha _3+\tilde{K_5}\Bigr){}^2}\times\nonumber\\
		&& \Biggl(-\Bigl(\Bigl(2 \tilde{c_0} \Bigl(2 \alpha _3+\tilde{K_4}\Bigr)+2\Bigr) \Bigl(4 \alpha _1 \dot{\tilde{c_0}}{}^2+4 \alpha _2 \dot{\tilde{c_1}}{}^2+\tilde{c_1}{}^2 \Bigl(2 \alpha _2 \omega _0^2+2 \alpha _3 \omega _1^2-\tilde{K_2}\Bigr)\nonumber\\
		&&-3 \tilde{c_0}{}^2 \Bigl(\tilde{K_1}-2 \alpha _1 \omega _0^2\Bigr)+\tilde{K_3} \dot{\tilde{c_0}}{}^2-\tilde{K_4} \dot{\tilde{c_1}}{}^2+\tilde{K_5} \dot{\tilde{c_1}}{}^2+2 \omega _0^2 \tilde{c_0}\Bigr)-2 \tilde{c_1} \Bigl(4 \alpha _2+2 \alpha _3+\tilde{K_5}\Bigr)\nonumber\\
		&& \Bigl(\tilde{c_0} \tilde{c_1} \Bigl(2 \alpha _2 \omega _0^2+2 \alpha _3 \omega _1^2-\tilde{K_2}\Bigr)+\dot{\tilde{c_0}} \dot{\tilde{c_1}} \Bigl(2 \alpha _3+\tilde{K_4}\Bigr)+\omega _1^2 \tilde{c_1}\Bigr)\Bigr)\Biggr) \,,
\label{c0eom}
	\end{eqnarray}
next to
\begin{eqnarray}
		&&\ddot{\tilde{c_1}}=\frac{1}{4 \tilde{c_0}{}^2 \Bigl(4 \alpha _1+\tilde{K_3}\Bigr) \Bigl(2 \alpha _3+\tilde{K_4}\Bigr)+4 \tilde{c_0} \Bigl(4 \alpha _1+2 \alpha _3+\tilde{K_3}+\tilde{K_4}\Bigr)-\tilde{c_1}{}^2 \Bigl(4 \alpha _2+2 \alpha _3+\tilde{K_5}\Bigr){}^2+4}\times\nonumber\\
		&&\Biggl(\tilde{c_1}{}^3 \Bigl(4 \alpha _2+2 \alpha _3+\tilde{K_5}\Bigr) \Bigl(2 \alpha _2 \omega _0^2+2 \alpha _3 \omega _1^2-\tilde{K_2}\Bigr)+\tilde{c_1} \Bigl(\dot{\tilde{c_0}}{}^2 \Bigl(4 \alpha _1+\tilde{K_3}\Bigr) \Bigl(4 \alpha _2+2 \alpha _3+\tilde{K_5}\Bigr)\nonumber\\
		&&-\dot{\tilde{c_1}}{}^2 \Bigl(-4 \alpha _2+\tilde{K_4}-\tilde{K_5}\Bigr) \Bigl(4 \alpha _2+2 \alpha _3+\tilde{K_5}\Bigr)-4 \omega _1^2\Bigr)-\tilde{c_0}{}^2 \tilde{c_1} \Bigl(8 \alpha _1 \alpha _2 \omega _0^2-12 \alpha _1 \alpha _3 \omega _0^2\nonumber\\
		&&+32 \alpha _1 \alpha _3 \omega _1^2-6 \alpha _1 \tilde{K_5} \omega _0^2+8 \alpha _2 \tilde{K_3} \omega _0^2+8 \alpha _3 \tilde{K_3} \omega _1^2-4 \tilde{K_2} \Bigl(4 \alpha _1+\tilde{K_3}\Bigr)+3 \tilde{K_1} \Bigl(4 \alpha _2+2 \alpha _3+\tilde{K_5}\Bigr)\Bigr)\nonumber\\
		&&+2 \tilde{c_0} \Bigl(\tilde{c_1} \Bigl(2 \alpha _3 \omega _0^2-8 \alpha _1 \omega _1^2-4 \alpha _3 \omega _1^2+\tilde{K_5} \omega _0^2-2 \tilde{K_3} \omega _1^2+2 \tilde{K_2}\Bigr)-2 \dot{\tilde{c_0}} \dot{\tilde{c_1}} \Bigl(4 \alpha _1+\tilde{K_3}\Bigr) \Bigl(2 \alpha _3+\tilde{K_4}\Bigr)\Bigr)\nonumber\\
		&&-4 \dot{\tilde{c_0}} \dot{\tilde{c_1}} \Bigl(2 \alpha _3+\tilde{K_4}\Bigr)\Biggr)\,.
\label{c1eom}
	\end{eqnarray}

\subsection{Poincar\'{e} sections}\label{sec:2.4}
\begin{figure}[p]
		\centering
	\begin{tabular}{c c}
		\textbf{Parallel Configuration} & \textbf{Perpendicular Configuration} \\
		\includegraphics[width=0.48\linewidth]{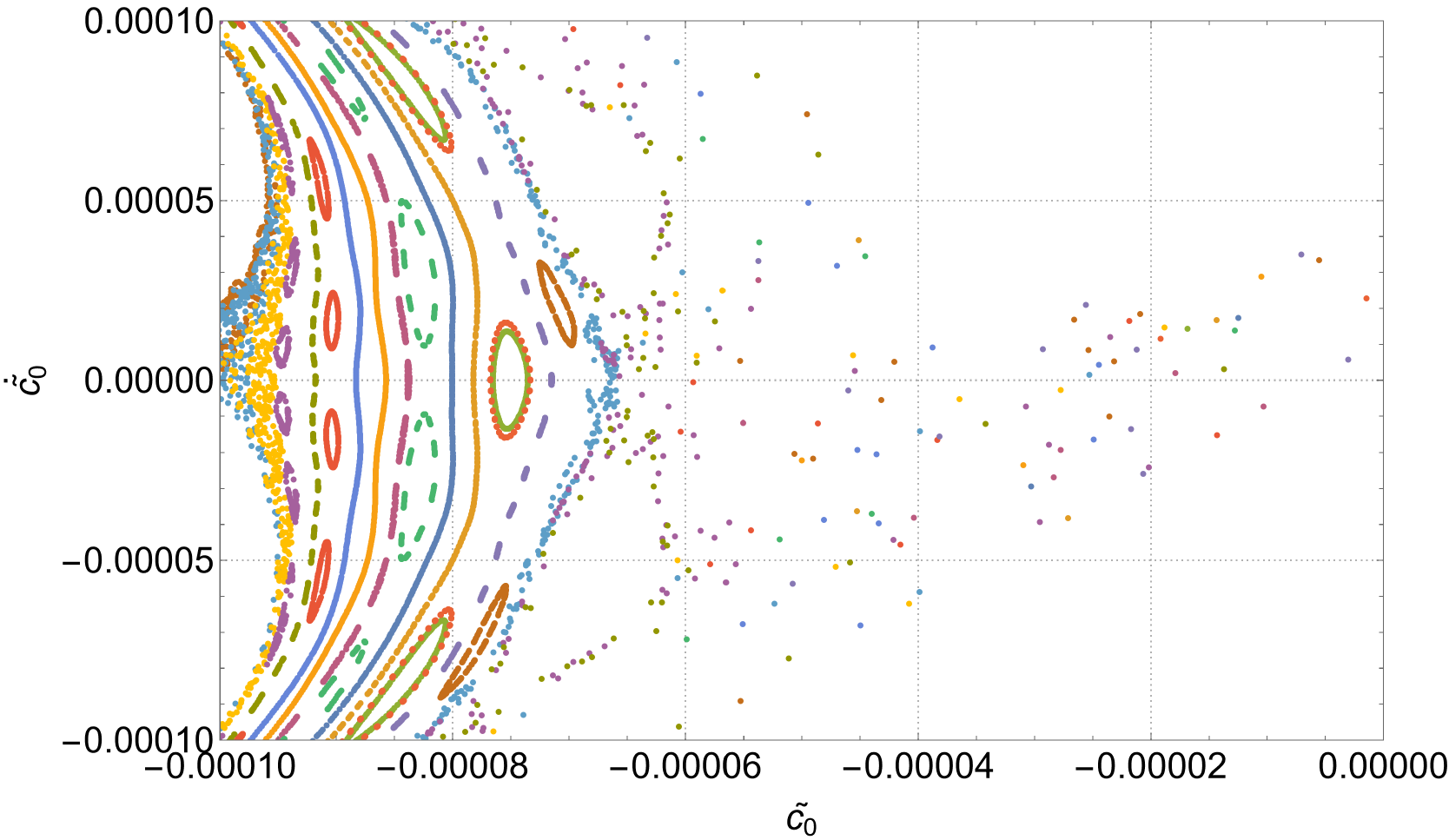} & \includegraphics[width=0.48\linewidth]{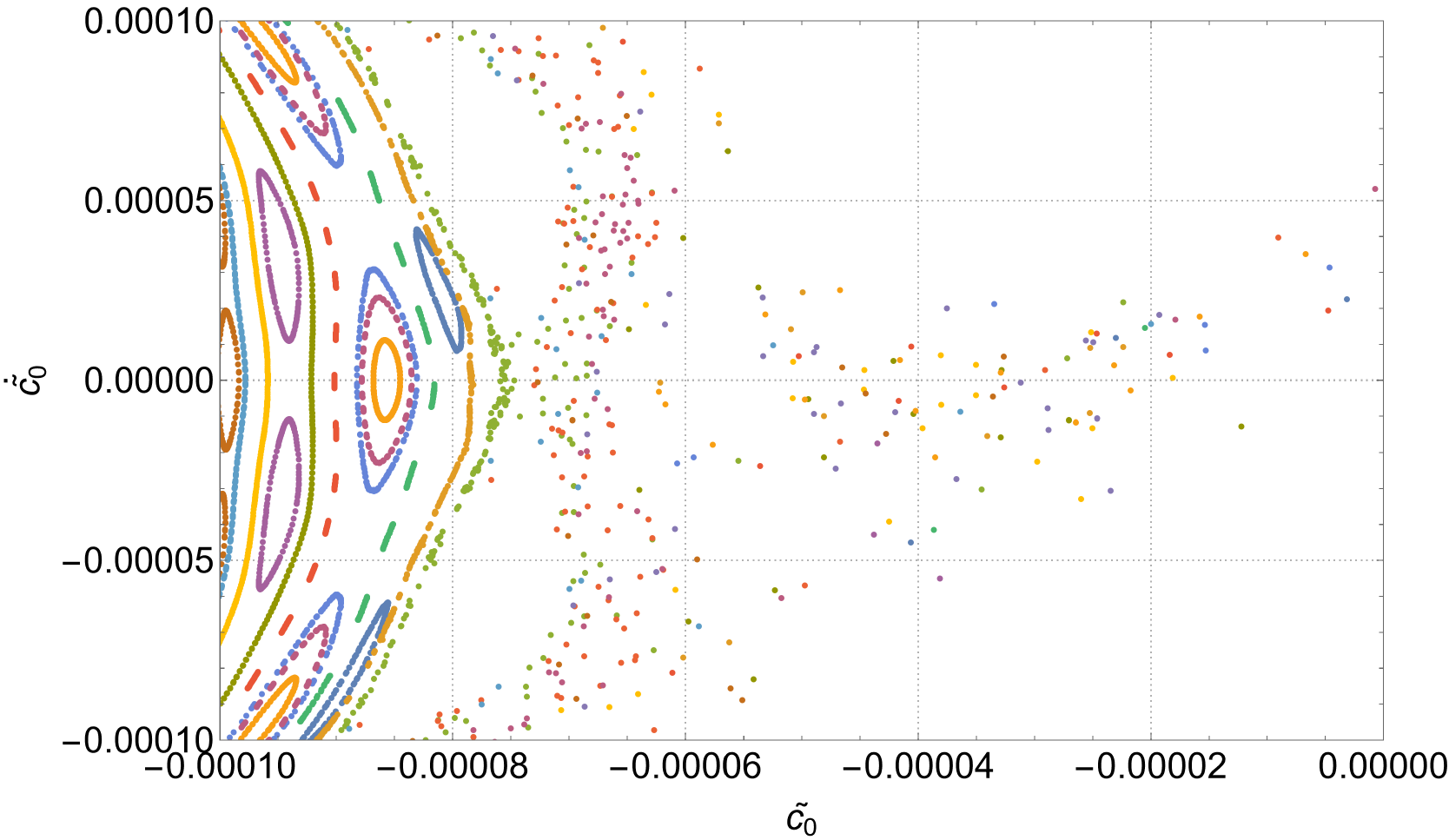} \\
		\includegraphics[width=0.48\linewidth]{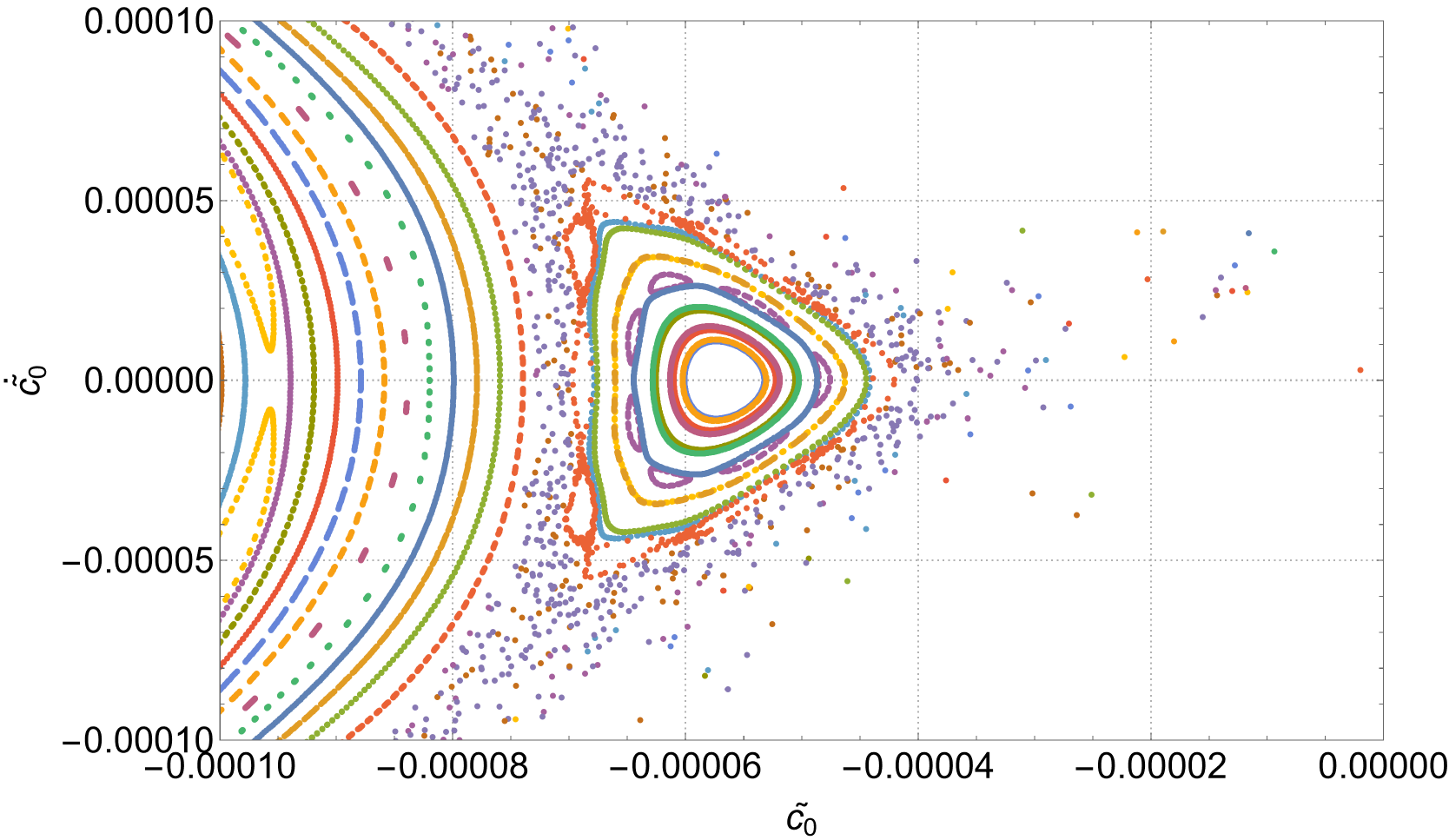} & \includegraphics[width=0.48\linewidth]{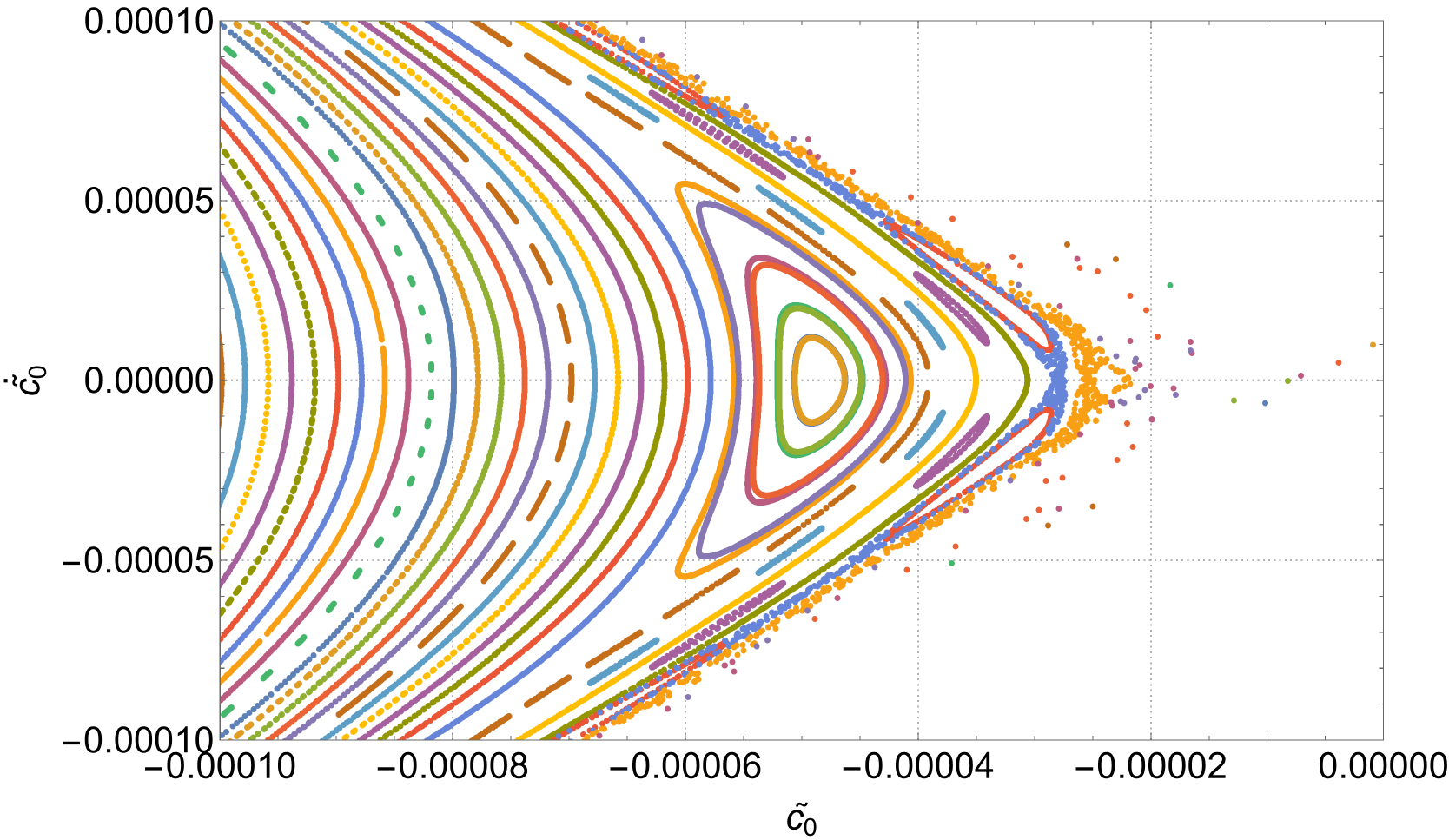} \\
		\includegraphics[width=0.48\linewidth]{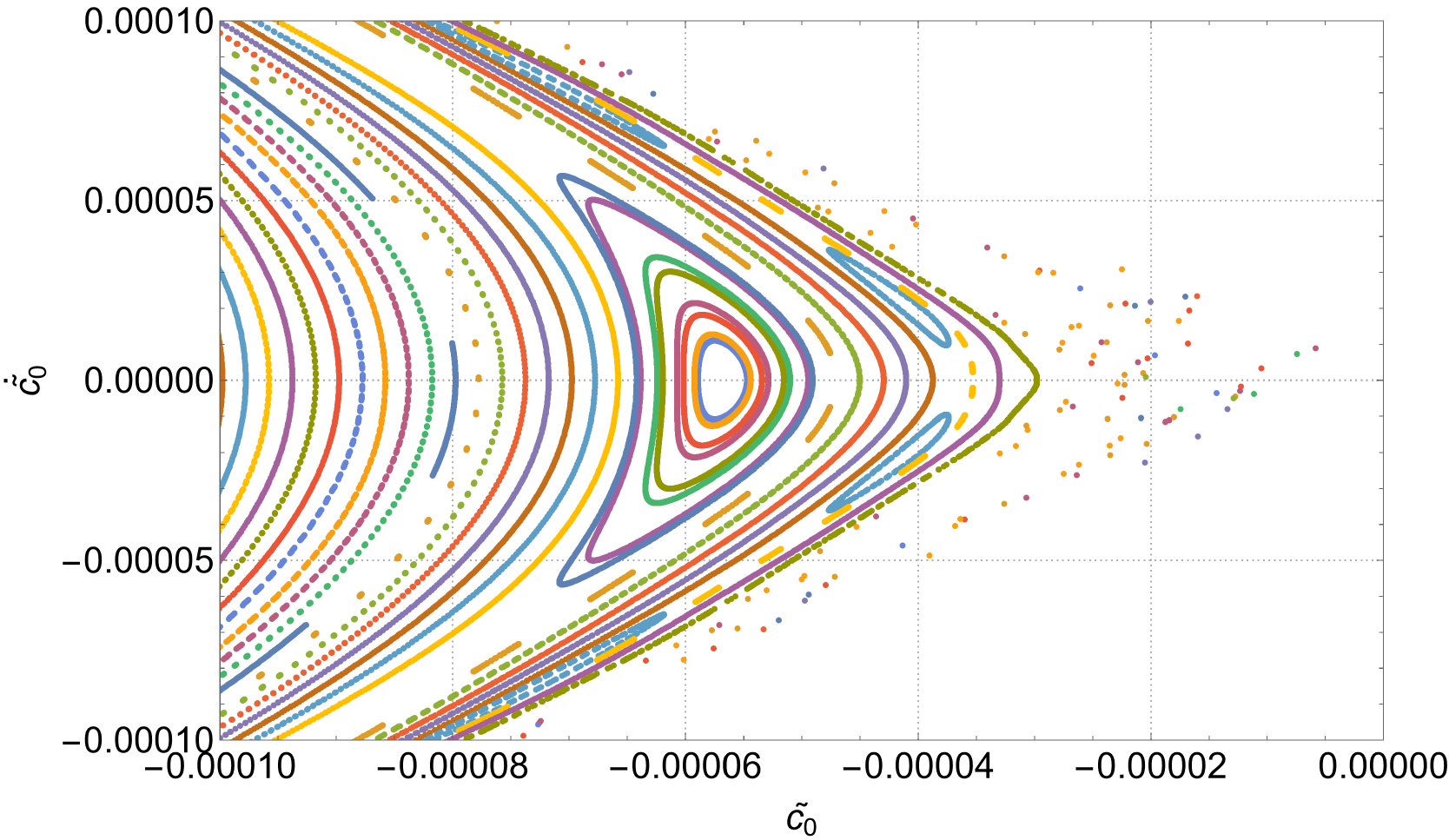} & \includegraphics[width=0.48\linewidth]{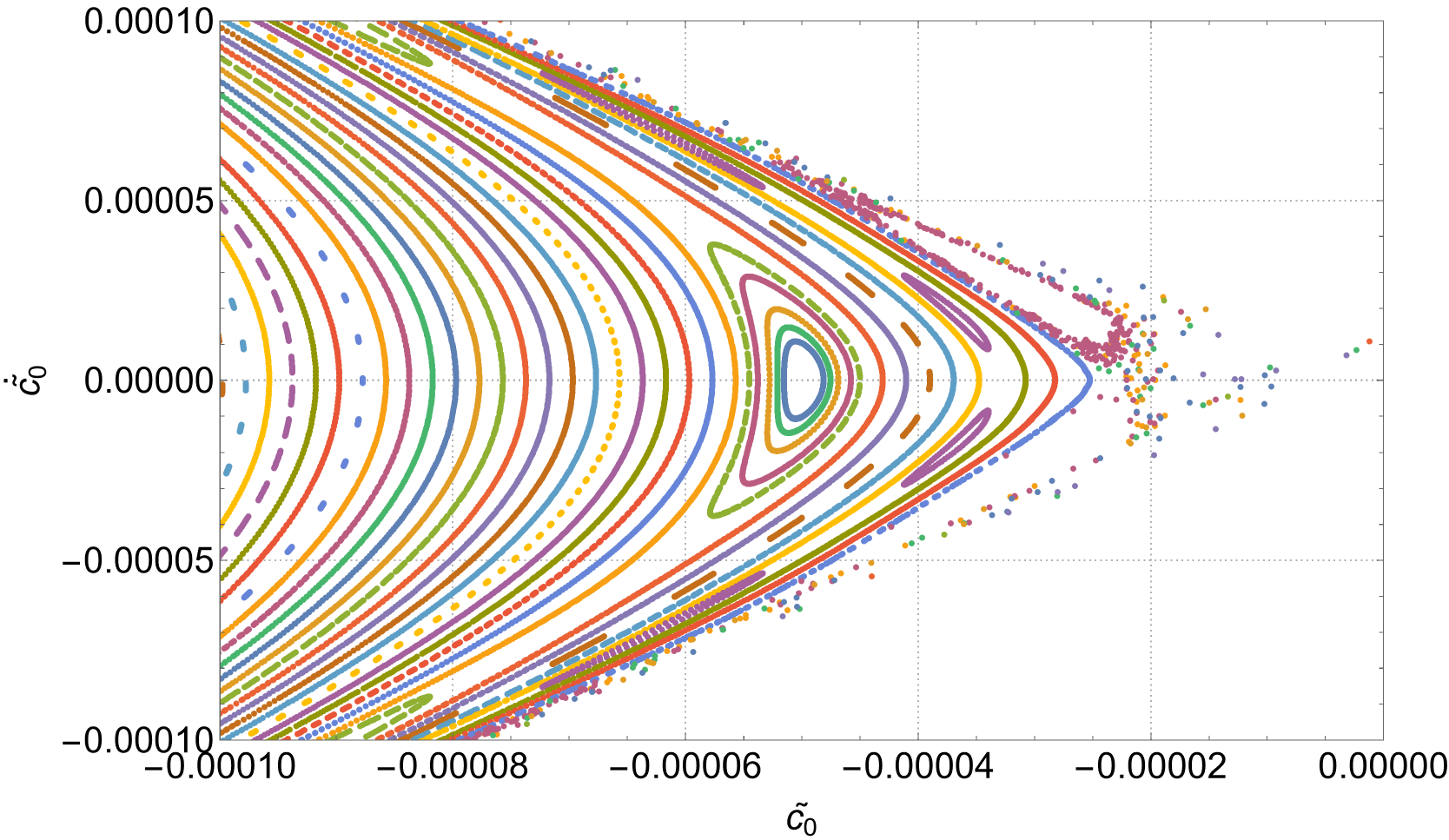} \\
	\end{tabular}
	\caption{Poincar\'{e} sections corresponding to $\tilde{c_{1}}=0$ and $\dot{\tilde{c_{1}}}\geq 0$ for parallel (left column) and perpendicular (right column) string configurations for the orbits with fixed energy $E=10^{-5}$ and fixed $L=1.1$. $B$ is varied from $B=0.1$ (top row) to $B=0.3$ (middle row) and $B=0.5$ (bottom row). In units of GeV.}
	\label{fig4}
\end{figure}
We can qualitatively observe whether the system is chaotic or not by solving the time evolution of the action (\ref{33}) and constructing a Poincar\'{e} section for the bound orbits within the trapping potential defined by $\tilde{c_{1}}(t)=0$ and $\dot{\tilde{c_{1}}}(t)\geq 0$. Our numerical results for the Poincar\'{e} sections for different values of magnetic fields for parallel and perpendicular string orientations are shown in Fig.~\ref{fig4}. This is a close-up view of the entire Poincar\'{e} section near the origin, where most of the fascinating stuff happens. Here we have used $L=1.1$ but similar results occur for other values of $L$ as well. Here the initial conditions are changed with fixed energy $E=10^{-5}$ and time interval  $0<t<15000$, and the points coming in different colours belong to the numerical data of orbits for the different starting conditions. This specific value of $E$ is chosen for illustrative purposes.

For $\tilde{c_{0}}$ near zero, we find that there are scattered points which show strong dependence on initial conditions. As we increase the magnetic field $B$, the scattered points transform to regular paths, which shows that the effect of the magnetic field is to reduce the chaotic behaviour in both parallel and perpendicular configurations. Also, the Poincar\'{e} sections get more and more structured and regular, with less scattered points, as we increase the magnetic field from $B=0.1$ to $B=0.5$. This implies that the effect of turning on the magnetic field is to lessen up the chaotic behaviour. This is true for both orientations of the string. Moreover, this lessen of chaotic behaviour is slightly greater in case of the perpendicular configuration than the parallel one. These results correlates well with our earlier observation from the lowest negative eigenvalues, which were not only found to be decreasing with $B$ but also were found to be decreasing more for the perpendicular case.

\begin{figure}[b]
	\centering
	\subfloat[Parallel configuration]{\label{saddlepoint_x1_stringframe}	\includegraphics[width=0.45\textwidth]{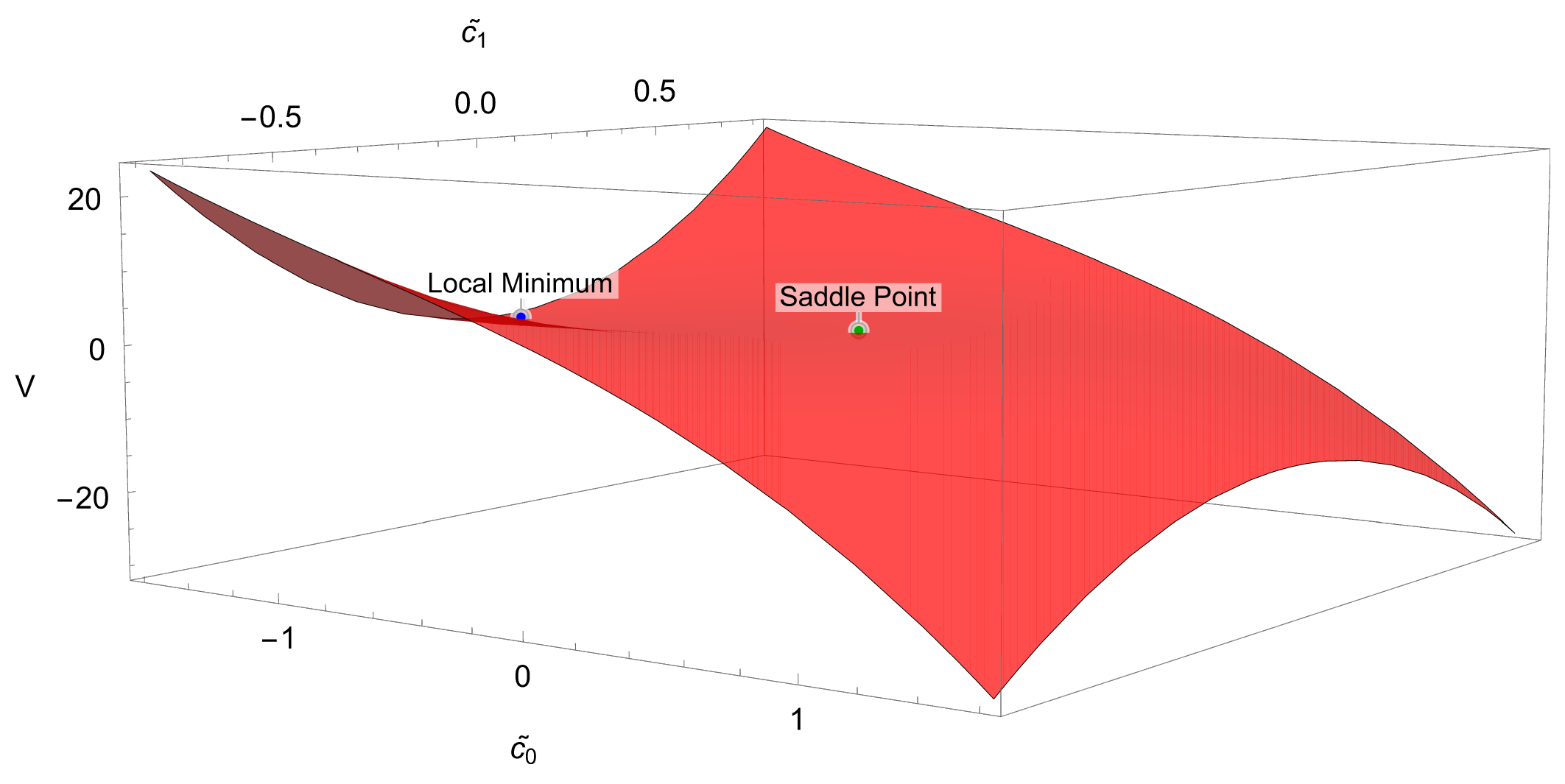}}
	\hfill
	\subfloat[Perpendicular configuration]{\label{saddlepoint_x3_stringframe}
		\includegraphics[width=0.45\textwidth]{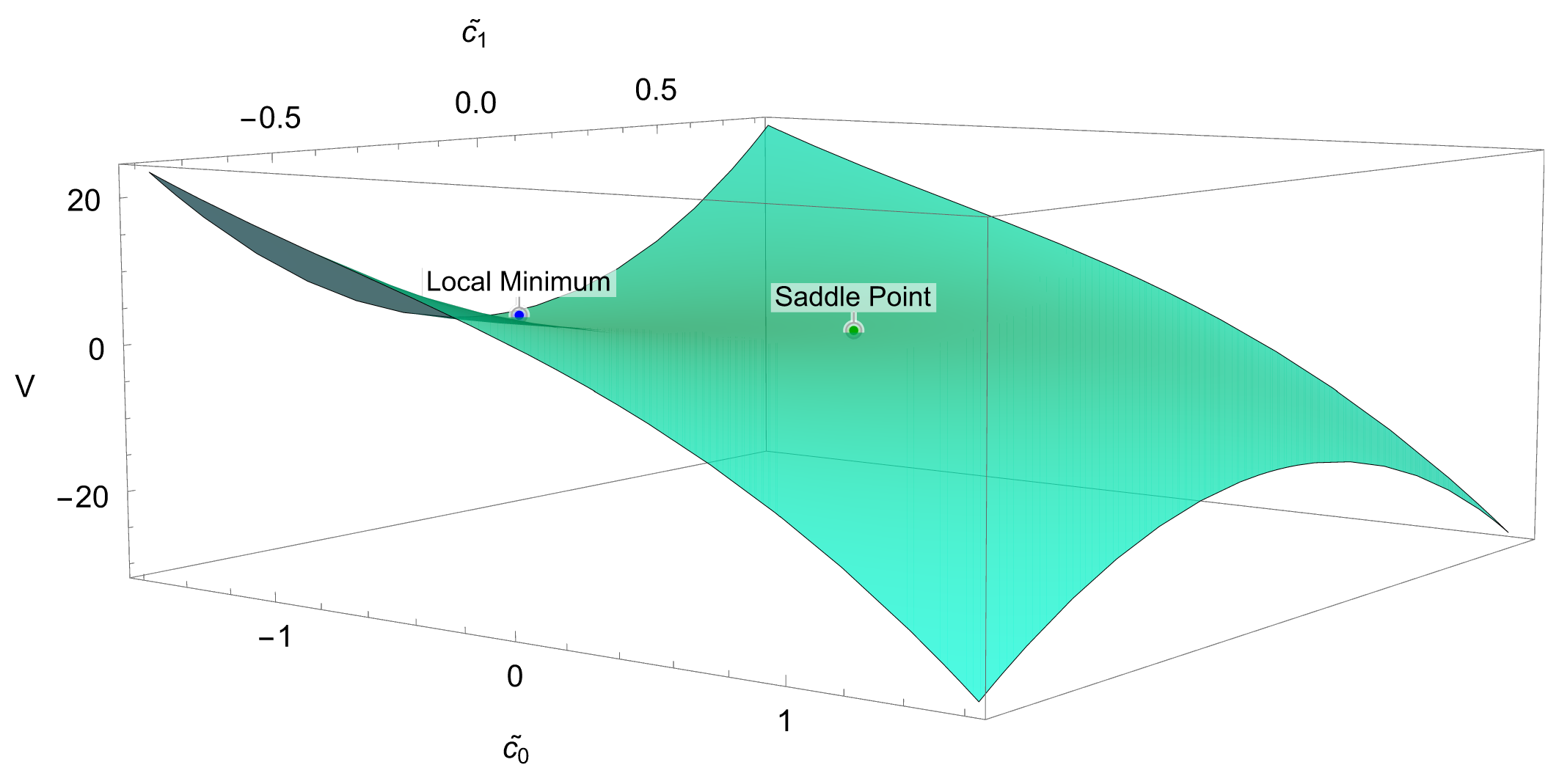}}
	\caption{\label{saddlepoint_stringframe}Potentials obtained from Eq.~(\ref{33}) for $B=0.1$. In units of GeV. }
\end{figure}

In the Poincar\'{e} plots we have set $\tilde{c_{1}}=0$. This generates a trap in the potential for the case $\tilde{c_{0}}<0$. The nature of the potential is illustrated in Fig.~\ref{saddlepoint_stringframe}. The perturbative conditions $(\tilde{c_{1}}=0, \tilde{c_{0}}<0)$ imply $(c_{1}=0, c_{0}<0)$, and correspond to a string moving away from the black hole horizon. Accordingly, for $(\tilde{c_{1}}=0, \tilde{c_{0}}=0)$, the tip of suspended string is the point closet to the horizon. This suggests that the source of chaos is the black hole horizon, and the dynamics of the string is less chaotic if we increase the magnetic field, in both orientations of the magnetic field. Moreover, we have further calculated the Poincar\'{e} section of the stable string configuration (not shown here for brevity), which appears to be away from the horizon, and we find only stable orbits without scattered points. This further lends support to the notion that black hole horizon is the source of chaos of the unstable string dynamics.

\subsection{Lyapunov exponents}\label{sec:2.5}
\begin{figure}[p]
	\centering
	\begin{tabular}{c c}
		\textbf{Parallel Configuration} & \textbf{Perpendicular Configuration} \\
		\includegraphics[width=0.48\linewidth]{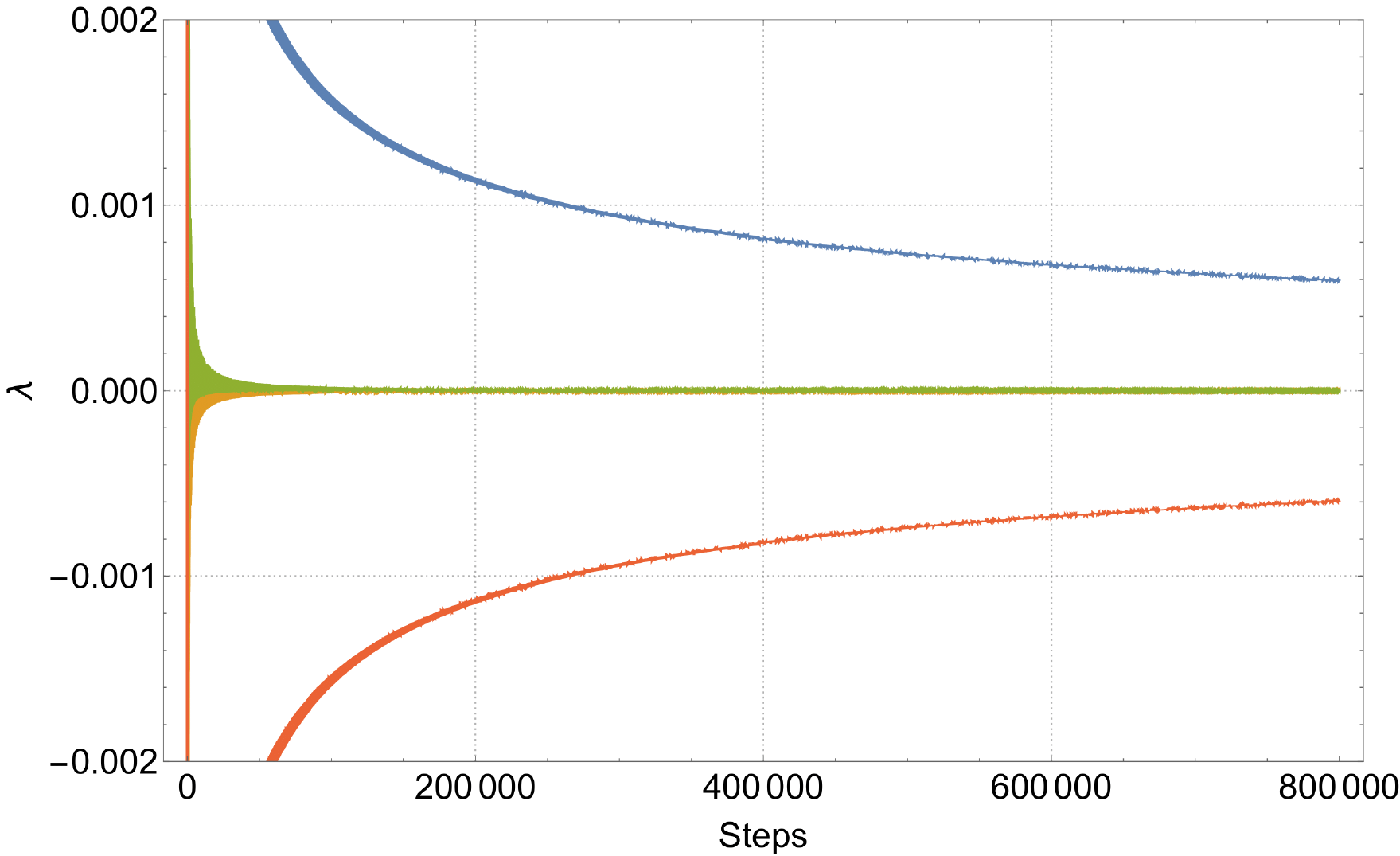} & \includegraphics[width=0.48\linewidth]{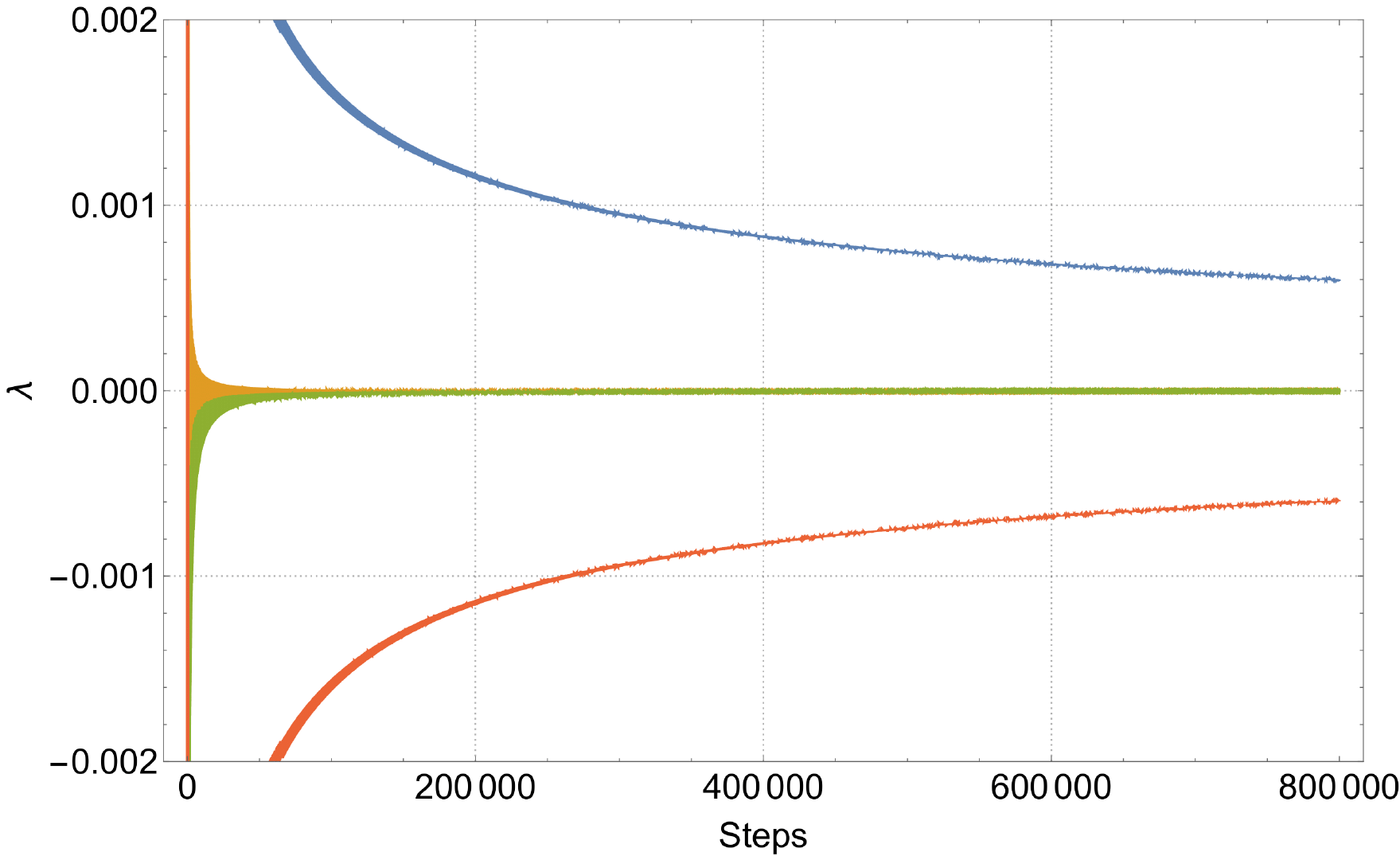} \\
		\includegraphics[width=0.48\linewidth]{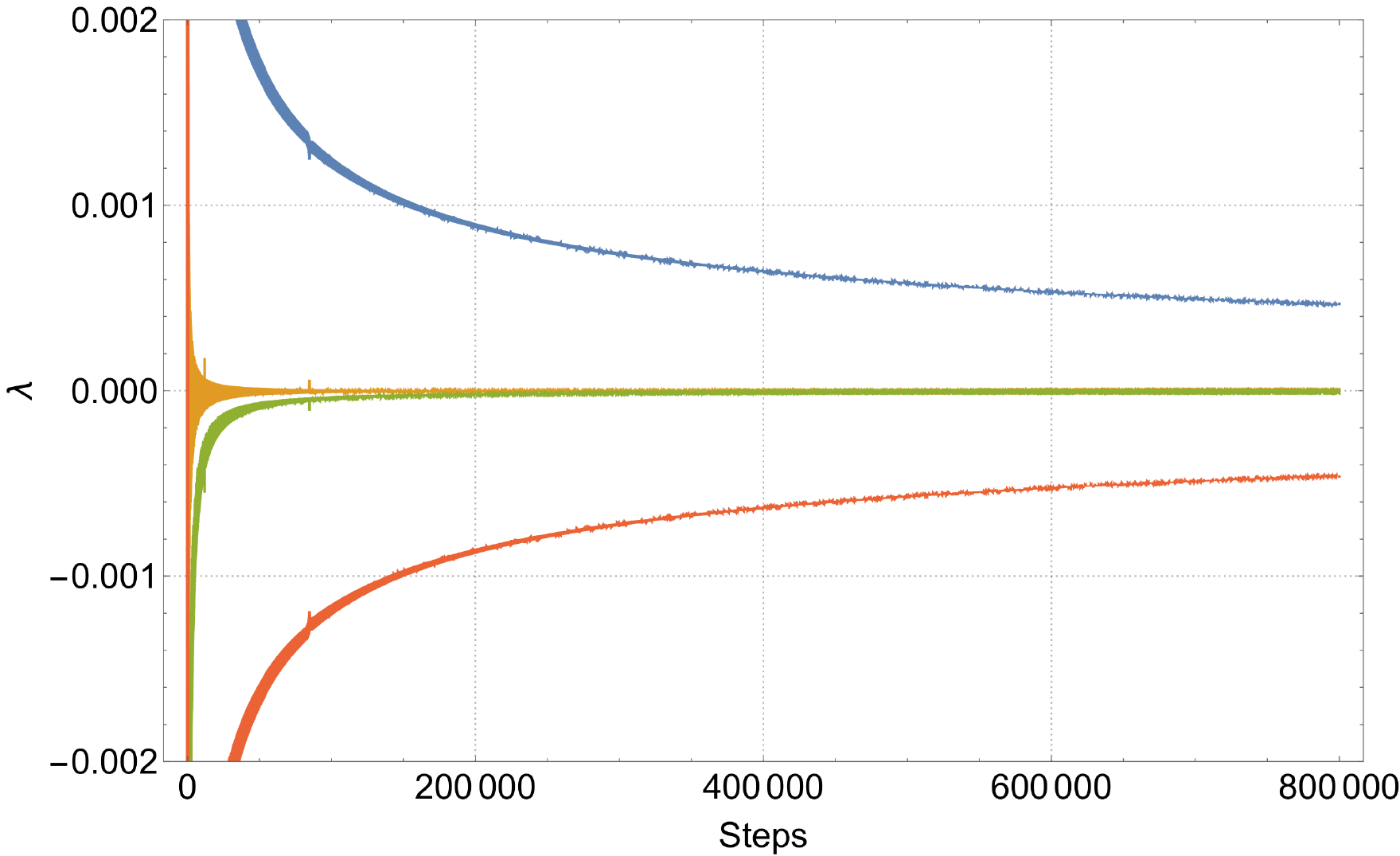} & \includegraphics[width=0.48\linewidth]{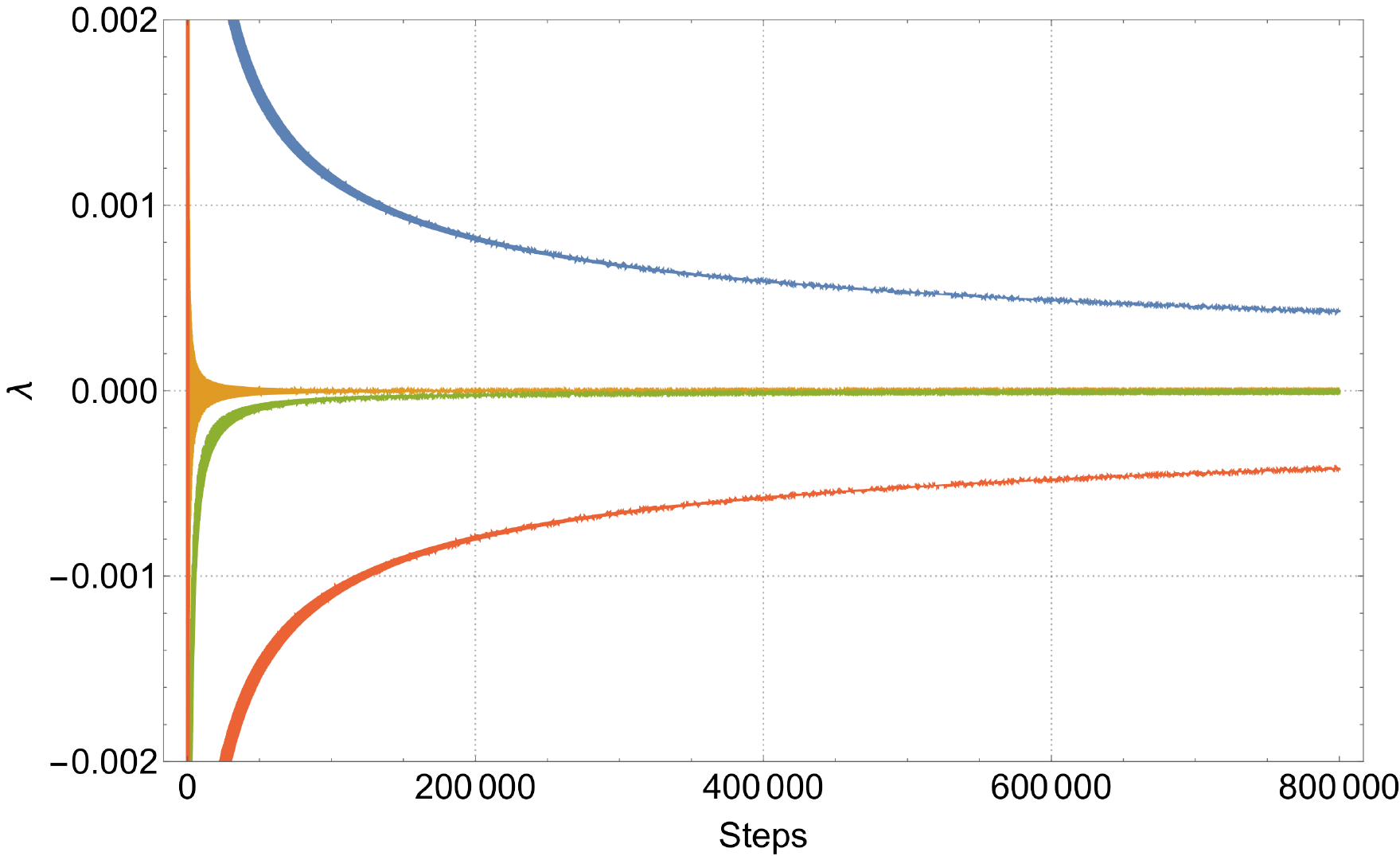} \\
		\includegraphics[width=0.48\linewidth]{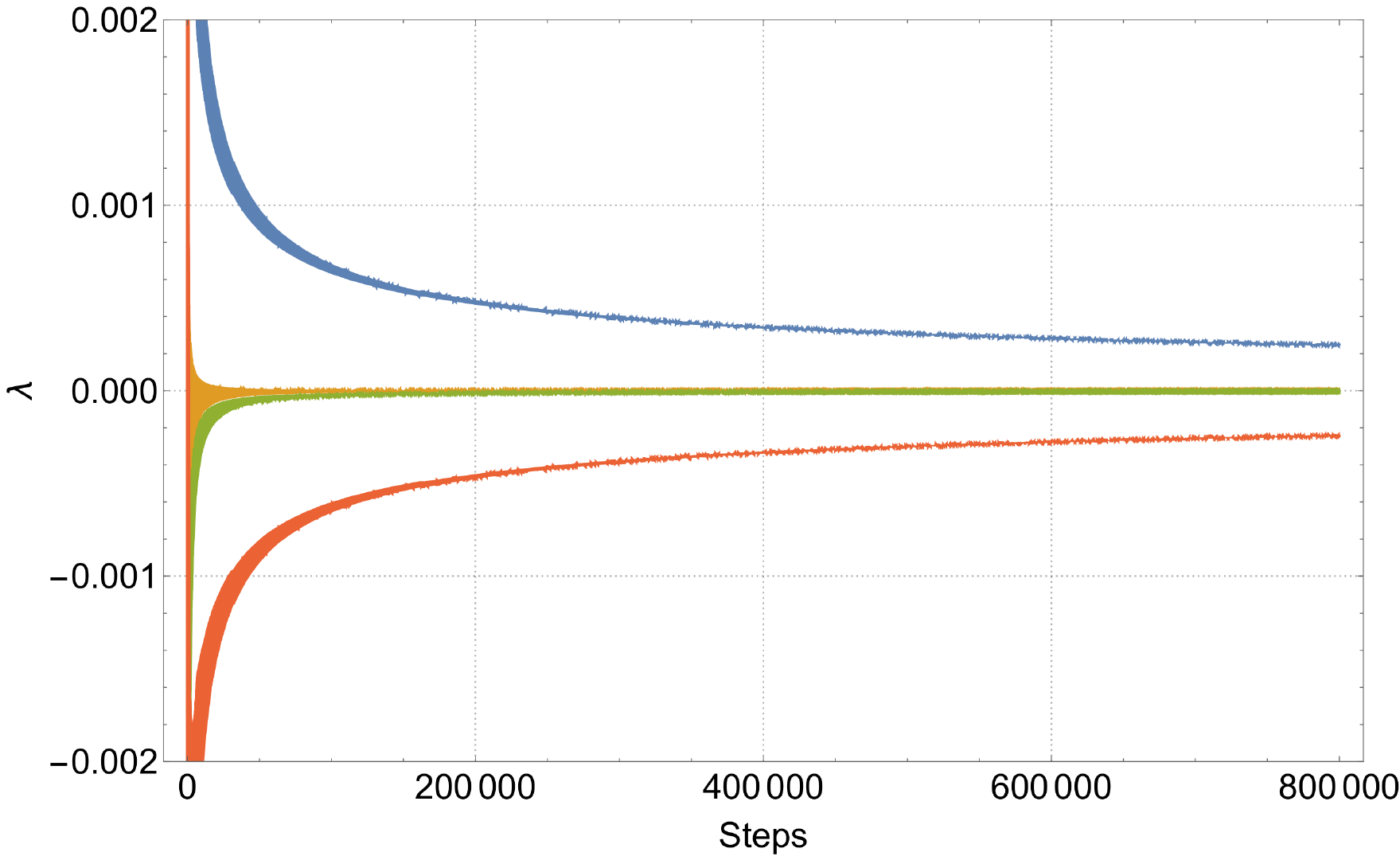} & \includegraphics[width=0.48\linewidth]{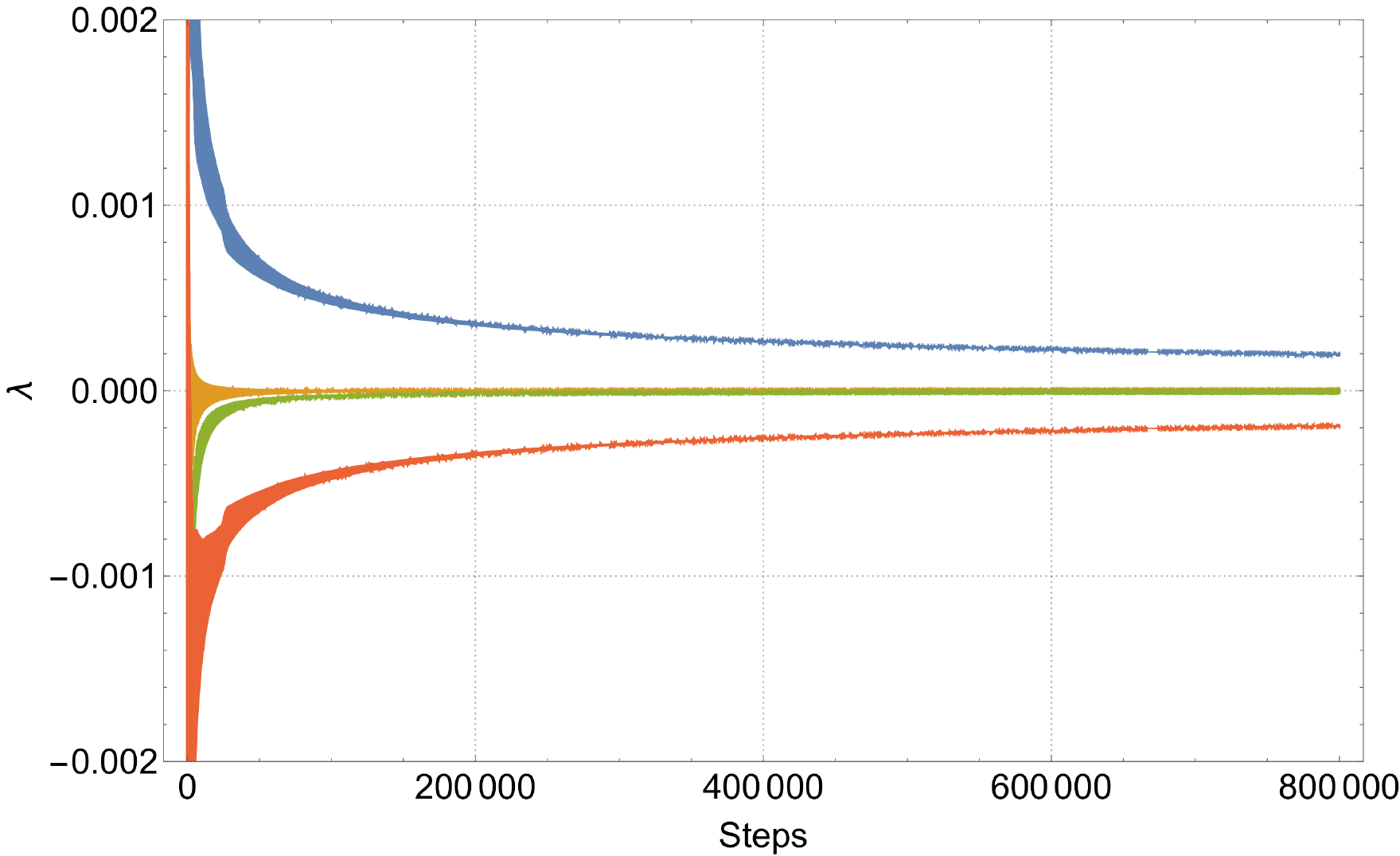} \\
		\includegraphics[width=0.48\linewidth]{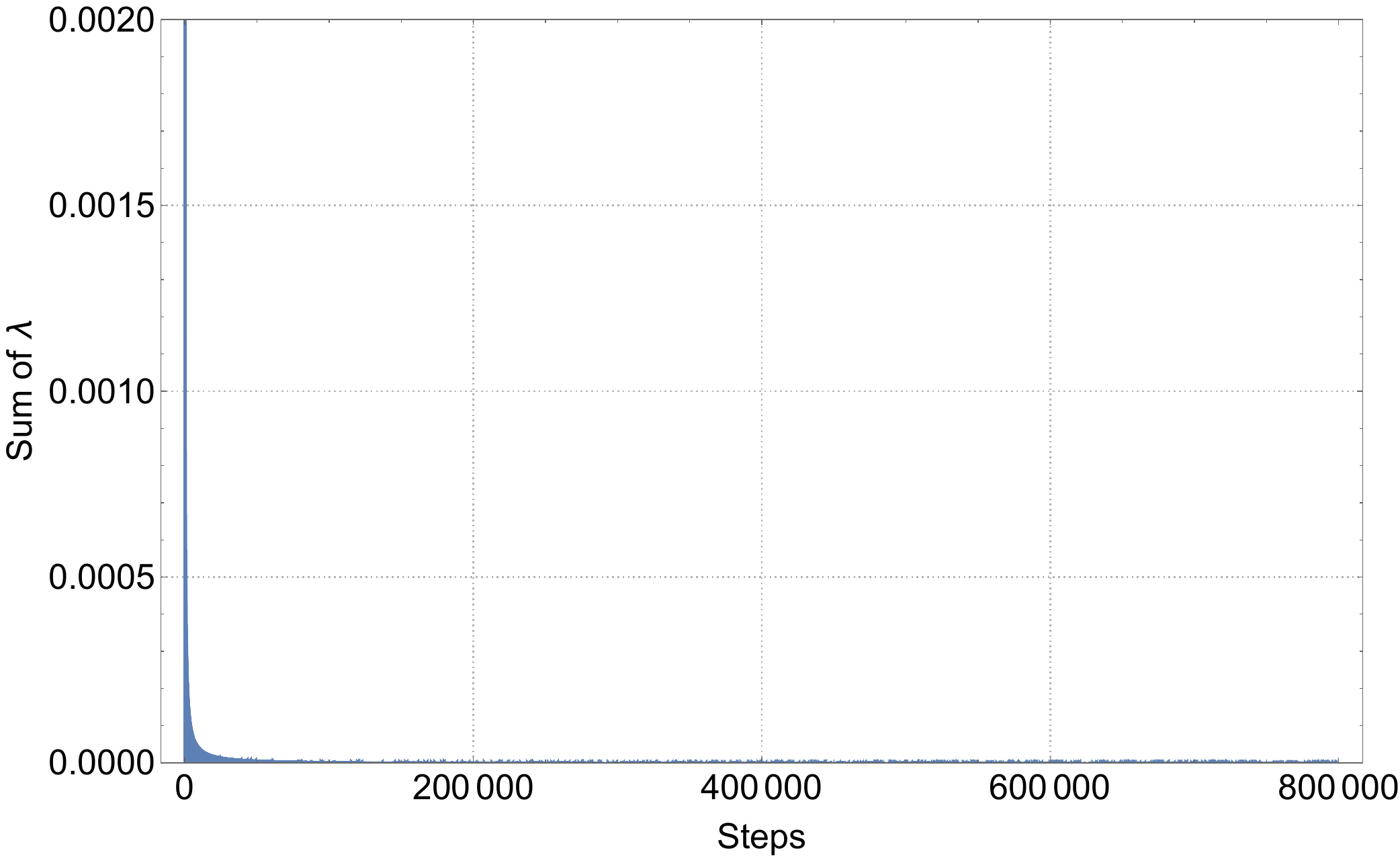} & \includegraphics[width=0.48\linewidth]{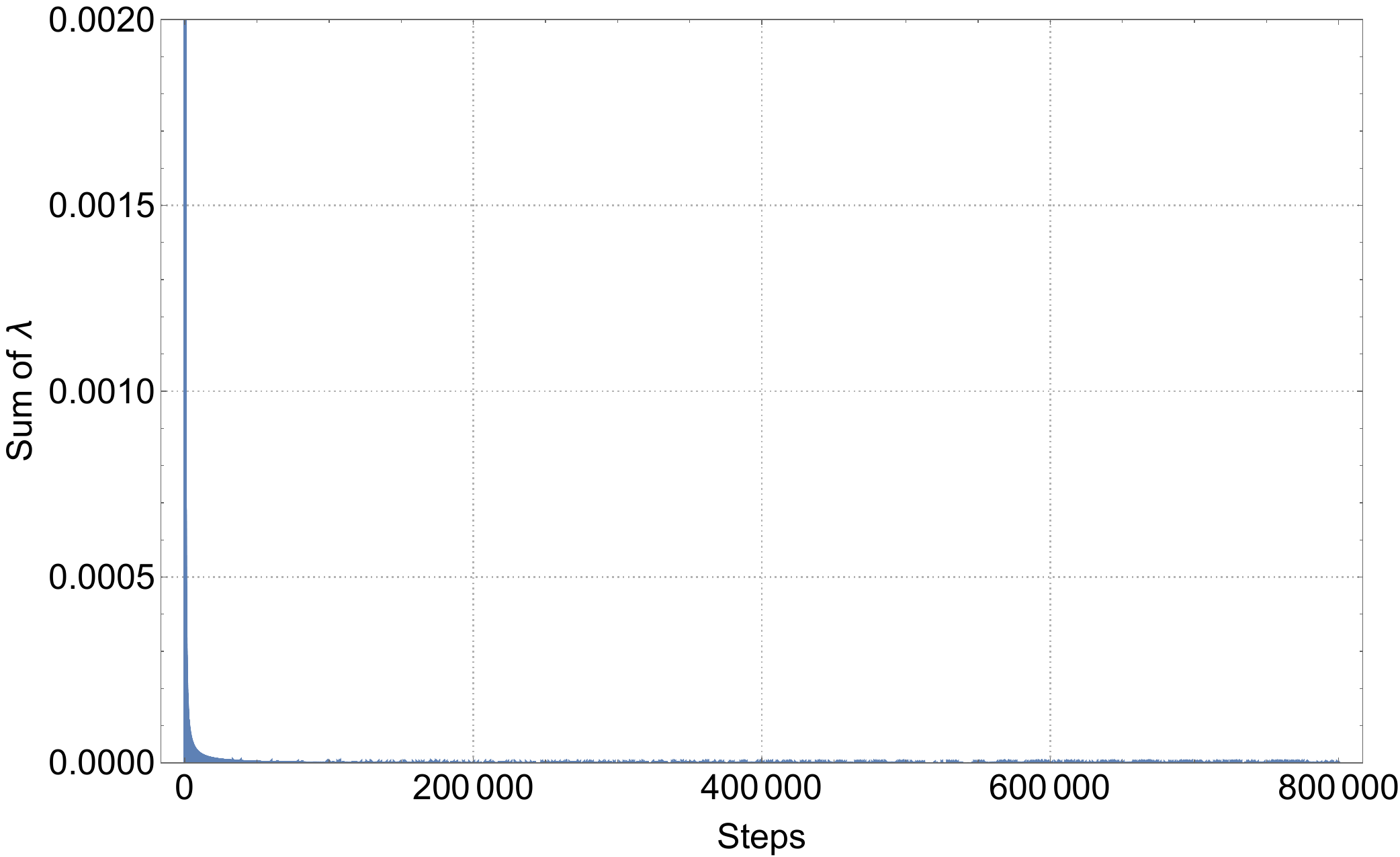} \\
	\end{tabular}
	\caption{Convergence plots of the four Lyapunov exponents for a string orientated in the parallel (left column) and perpendicular (right column) directions with respect to the magnetic field. Here we have taken $8 \times 10^{5}$ time steps. For initial conditions the energy is set to $E=10^{-5}$, along with $\tilde{c_{0}}=-0.0001$, $\dot{\tilde{c_{0}}}=0.00001$, and $\tilde{c_{1}}=0.0013$. Here $L=1.1$ is fixed and $B$ is varied from $B=0.1$ (first row) to $B=0.3$ (second row) and $B=0.5$ (third row). The Sum of Lyapunov exponents are shown in the last row for $B=0.3$. In units of GeV.}
	\label{fig5}
\end{figure}
\begin{figure}[t]
	\centering
	\includegraphics[width=0.5\linewidth]{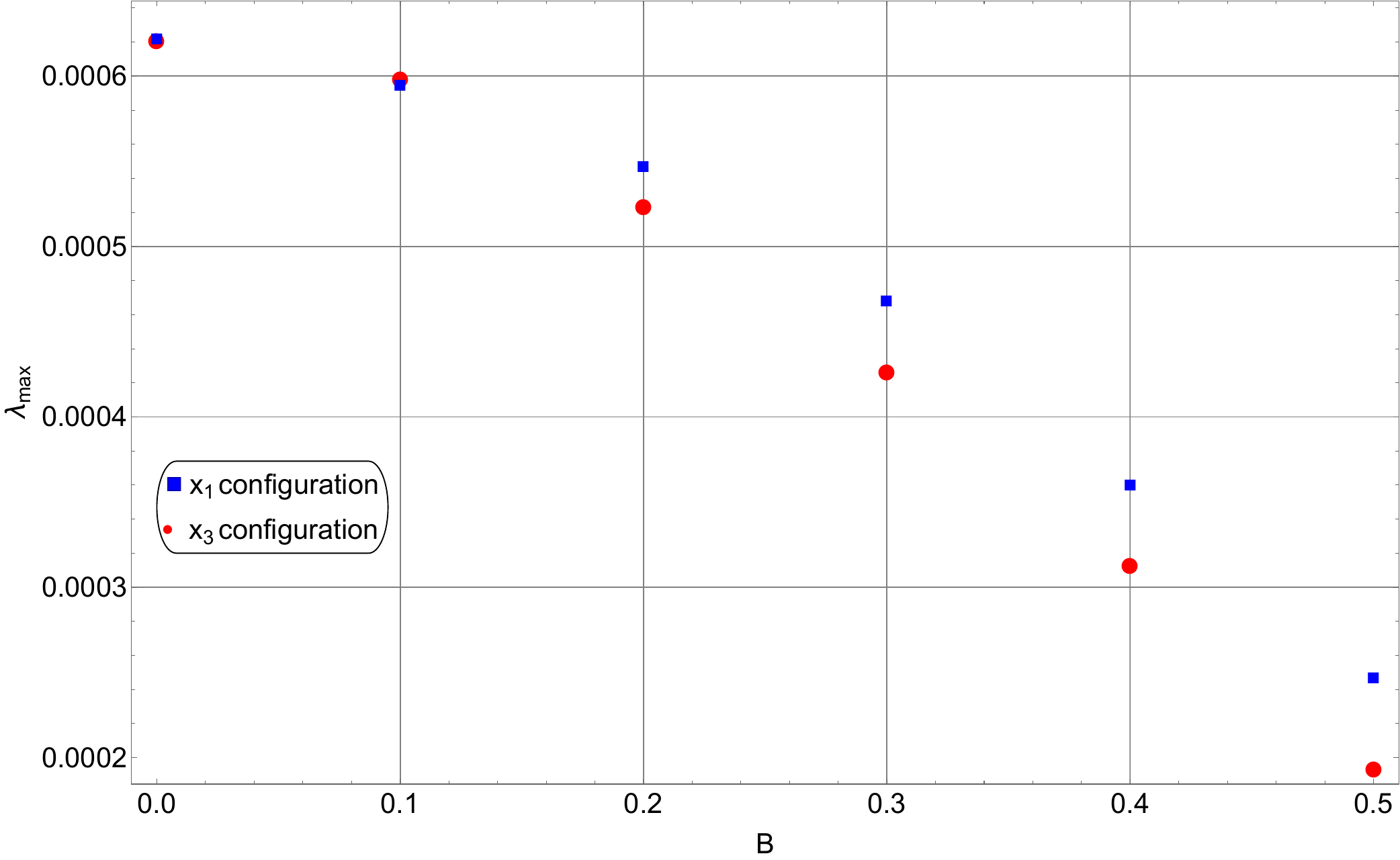}
	\caption{The largest Lyapunov exponent $\lambda_{max}$ for different values of $B$ at $L=1.1$ for two different orientations of the string. In units of GeV.}
	\label{lypmaxvsB_stringframe}
\end{figure}

Lyapunov exponents can be used to give a quantitative analysis of the chaotic dynamics. They can be computed in the four dimensional ($\tilde{c_{0}}$, $\tilde{c_{1}}$) phase-space with the help of numerical methods described in \cite{sandri1996numerical,Wolf1985Jul}. We calculate the Lyapunov exponents for both orientations of the strings for different values of the magnetic field and our numerical results for their convergency are shown in Fig.~\ref{fig5}. Here, we have focused on the system with $L=1.1$ and $E=10^{-5}$, and have computed the convergency plots of the four Lyapunov coefficients, one for each direction of the phase space,  together with their sum. Although not visible on these plots due to our high resolution/accuracy, the convergence rate is a damped oscillating function. The system is found to be conservative since the sum of Lyapunov exponents converges to zero with evolution as can be seen in the last row of the Fig.~\ref{fig5}.  This is true for all values of the magnetic field and its relative orientation with respect to the string. The value of the largest Lyapunov exponent $\lambda_{max}$ can be extrapolated from the plot by taking a large number of time steps and fitting the maximum in each oscillation. The $\lambda_{max}$ values obtained from this fit are shown in Fig.~\ref{lypmaxvsB_stringframe}. For the same string length $L$, $\lambda_{max}$ is found to be decreasing for both parallel and perpendicular orientations of the string. Moreover, for the same value of the magnetic field, $\lambda_{max}$ is found to be smaller for the perpendicular compared to the parallel case. This result confirms in a more qualitative way our previous analysis using the Poincar\'{e} sections, which suggested that the dynamics of the string becomes less chaotic when the magnetic field increases for both parallel and perpendicular cases.

Let us now compare the obtained Lyapunov exponents with the MSS bound. The MSS bound, from Eq.~(\ref{BHtemp}) can be rewritten as
\begin{equation}\label{35}
	\lambda_{MSS} = \frac{{r_h}^2 g'({r_h})}{2} = \frac{\left(B^2-3 a\right)^2}{r_h^3 e^{\frac{B^2-3a}{r_h^2}}-r_h \left(-3 a+B^2+r_h^2\right)}   \,.
\end{equation}
We find that the largest Lyapunov exponent at the stable fixed point is around three orders of magnitude smaller than $\lambda_{MSS}$ for all $B$, suggesting that it always satisfies the MSS bound for all values of magnetic field irrespective of its orientation with the respect to the string.

We also like to mention that we have presented the coefficients $K_{1,\ldots,5}^{x_{i}}$ in Table~\ref{table3} up to a $10^{-3}$ level of accuracy. We did not observe any apparent effect when enforcing greater accuracy. The algorithm in \cite{sandri1996numerical} was used to calculate the Lyapunov exponents with a step size of $0.001$ and a total number of steps $8 \times 10^{5}$, making our analysis numerically sufficiently accurate.

\subsection{Analysis of the saddle point and testing the MSS bound}\label{sec:2.6}
From our analysis in the previous sections, it can be firmly stated that the dynamics of the string under consideration exhibits chaos since the value of the largest Lyapunov exponent is positive. To complete our discussion on the Lyapunov exponent and its relative comparison with the MSS bound, here we discuss the Lyapunov exponent at the unstable fixed points. Notice that, as mentioned earlier, the potentials obtained from the action (\ref{33}) contain two fixed points: an unstable and a stable one. The stable fixed point corresponds to the local minimum of the potential whereas the unstable fixed point corresponds to the saddle point. These two fixed points are similar in terms of their position and appearance for the parallel as well as for perpendicular configurations, see Fig.~\ref{saddlepoint_stringframe}.

\begin{figure}[!htbp]
	\centering
	\includegraphics[width=0.5\linewidth]{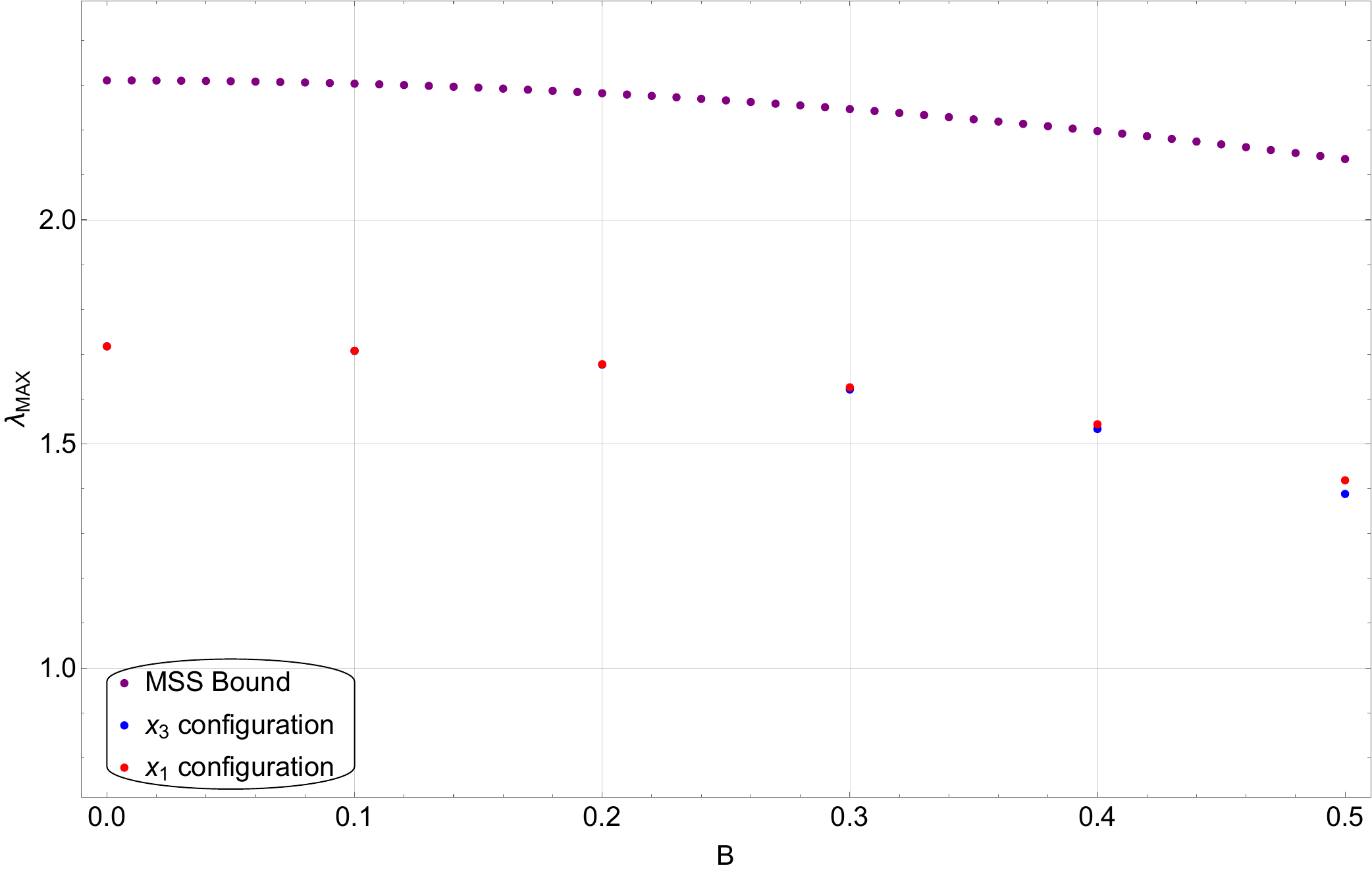}
	\caption{Comparison between MSS bound and largest Lyapunov exponent at the saddle point as a function of $B$ for $L = 1.1$. In units of GeV.}
	\label{mssbound_stringframe}
\end{figure}

Now, the dynamics of the action (\ref{33}) is governed by the equation $\dot{\vec{y}}=\vec{F}$, with $\vec{y}=(\tilde{c_0},\dot{\tilde{c_0}},\tilde{c_1},\dot{\tilde{c_1}})$. The two fixed points correspond to the case where $\vec{F}=0$. For the unstable string with energy $E=0$, the unstable fixed point can be located at $\vec{y}=(0,0,0,0)$. We compute the Lyapunov exponents at the unstable fixed point $\vec{y}=(0,0,0,0)$ numerically in a similar way as in the previous section and find that they asymptotically converges to $(\sqrt{-\omega_{0}^{2}},-\sqrt{-\omega_{0}^{2}},0,0)$. Using the values of $\omega_{0,1}^2$ from Table~\ref{table2}, we find that these Lyapunov exponents again satisfy the MSS bound. The comparison between the largest Lyapunov exponents at the unstable fixed point and the MSS bound for different magnetic field values is shown in Fig.~\ref{mssbound_stringframe}.

Interestingly, the Lyapunov exponents can be obtained analytically at the fixed point as the real part of the eigenvalues of the Jacobian matrix of $\vec{F}$ \cite{sandri1996numerical}. Here we can use Eqs.~(\ref{c0eom}) and (\ref{c1eom}) to find the required Jacobian. For the unstable fixed point $\vec{y}=(0,0,0,0)$, the Jacobian matrix reads
\begin{equation}\label{jacobian}
J = \begin{pmatrix}
0 & - \omega_{0}^2 & 0 & 0 \\
1 & 0 & 0 & 0 \\
0 & 0 & 0 & - \omega_{1}^2 \\
0 & 0 & 1 & 0 \\
\end{pmatrix}
\end{equation}
Interestingly, the same form of the Jacobian was found in \cite{Colangelo2022Apr}. The eigenvalues of this Jacobian metric are $(-i\sqrt{\omega_{0}^2}$, $i\sqrt{\omega_{0}^2}$, $-i\sqrt{\omega_{1}^2}$, $i\sqrt{\omega_{1}^2})$. This implies that the Lyapunov exponents vanish for $\omega_{0,1}^2>0$. Since in our case $\omega_{0}^2<0,\omega_{1}^2>0$, we have two non-vanishing Lyapunov exponents $(\sqrt{-\omega_{0}^2},-\sqrt{-\omega_{0}^2},0,0)$. The largest Lyapunov exponent $\lambda_{max}=\sqrt{-\omega_{0}^2}$ at the unstable fixed point again satisfies the MSS bound for all values of magnetic field. In particular, although the Lyapunov exponent at the unstable point is large compared to the stable fixed point, however, it remains below the MSS bound. \footnote{Recently the violation of the analog MSS bound for pointlike particles has been found in \cite{Giataganas:2021ghs}.} This result further suggests that even at the unstable fixed point the magnetic field again tries to soften the chaotic behaviour. Notice that our numerical results for the Lyapunov exponent at the unstable point agrees well with the analytic results. This provides further support to the accuracy of the numerical procedure, and hence the corresponding numerical results, considered in this work.

Let us also mention that, using similar arguments as above, one finds all the Lyapunov exponents vanish for the stable string configuration (as both $\omega_{0}^2$ and $\omega_{1}^2$ are now positive), indicating no chaos in the string which is away from the horizon.  This gives further credit to the notion that the black hole event horizon is indeed acting as a source for chaos in the string dynamics.

To summarize, our analysis in the string frame suggests that the chaos is produced in the proximity of the event horizon, visible from the Poincar\'{e} plots, and  that the effect of the magnetic field is to weaken the dependence on the initial conditions, making the string dynamics less chaotic. Furthermore, the string chaotic dynamics gets softened more along the perpendicular than the parallel direction relative to the magnetic field.

\section{Einstein frame}\label{sec:3}
Having thoroughly investigated the chaotic string dynamics in the String frame, we now move on to discuss them in the Einstein frame. The corresponding black hole metric is given in Eq.~(\ref{gsol}). Since most of the numerical procedures and computations are analogous to the string frame case, here we will be brief and mainly focus on the results. With the exception of the form of the scale factor $A_s(r)$, all the expressions in the Einstein frame metric take the same form as in the string frame metric. In particular, by replacing $A_s(r)\rightarrow A(r)$, and making appropriate changes wherever necessary, we can get the relevant expressions for the Einstein frame case.

%%%%%%%%%%%%%%%%%%%%%%%%%%%%%%
\begin{figure}[ht]
\begin{minipage}[b]{0.45\linewidth}
\centering
\includegraphics[width=2.8in,height=2.1in]{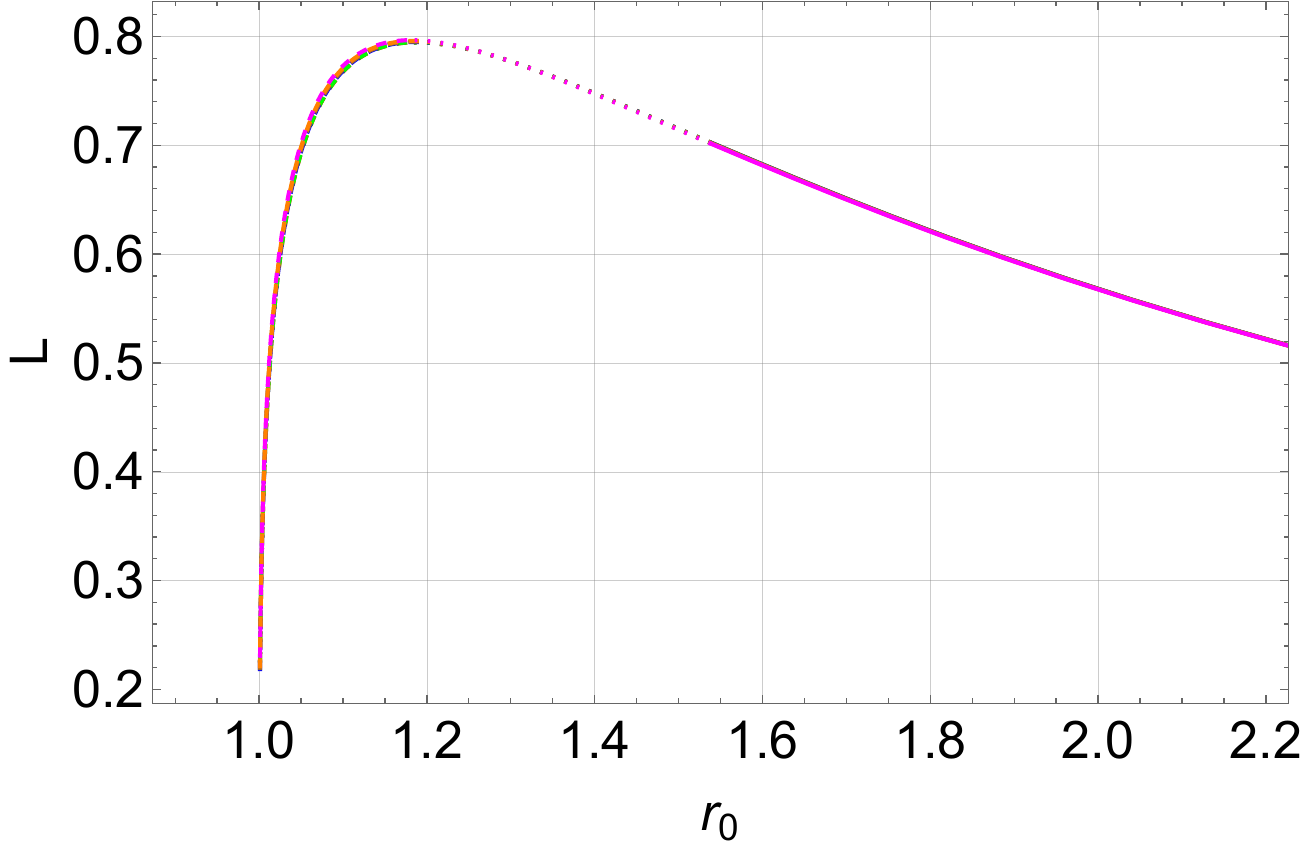}
\caption{$L$ as a function of $r_0$ for different values of $B$ in the parallel case. Here $r_h=1$ is used. The red, green, blue, brown, orange, and magenta
curves correspond to $B=0$, $0.1$, $0.2$, $0.3$, $0.4$, and $0.5$ respectively. In units of GeV.}
\label{r0vsLvsBrh1paraEF}
\end{minipage}
\hspace{0.4cm}
\begin{minipage}[b]{0.45\linewidth}
\centering
\includegraphics[width=2.8in,height=2.1in]{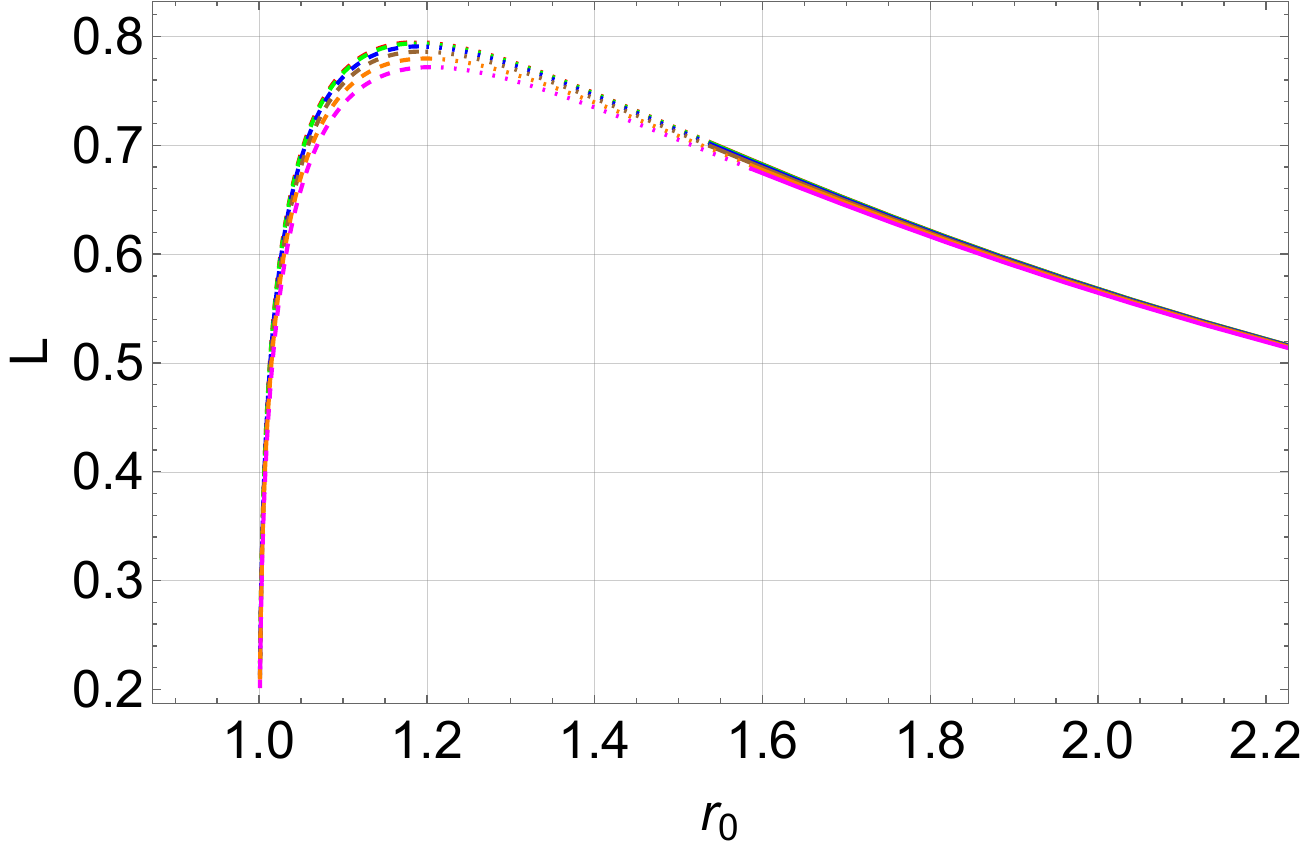}
\caption{$L$ as a function of $r_0$ for different values of $B$ in the perpendicular case. Here $r_h=1$ is used. The red, green, blue, brown, orange, and magenta
curves correspond to $B=0$, $0.1$, $0.2$, $0.3$, $0.4$, and $0.5$ respectively. In units of GeV.}
\label{r0vsLvsBrh1perpEF}
\end{minipage}
\end{figure}
%%%%%%%%%%%%%%%%%%%%%%%%%%%%%%

The static string profile can be computed from Eqs.~(\ref{17})-(\ref{18}). The variation of the string length $L(r_{0})$ for different values of $B$ for parallel and perpendicular configurations is shown in Fig.~\ref{r0vsLvsBrh1paraEF} and \ref{r0vsLvsBrh1perpEF}. Most of the results here are similar to the string frame case. In particular, there again appears a maximum string length $L_{max}$ above which no connected string solution exists, and below which $L_{max}$ there are two connected string solutions for each values of $L$.  The magnitude of $L_{max}$ again decreases with $B$ for both orientations of $B$, however, its magnitude is slightly smaller compared to the string frame case. The unstable string solution (indicated by dashed lines) corresponds to the local maximum of the energy and appears for small $r_0$ whereas the stable string solution (indicated by solid lines) corresponds to local minimum of the energy and appears for large $r_0$. The dotted lines indicate the metastable strings. The free energies of these stable and unstable strings have similar features as discussed in Figs.~\ref{LvsDeltaFvsBrh1paraSF} and \ref{LvsDeltaFvsBrh1perpSF}.

\begin{table}[t]
	\centering
	\begin{tabular}{|c|c||c|}
\hline
		$B$ & $r_0~(||)$ &  $r_0~(\perp)$\\
		\hline
		0 & 1.08235 &  1.08235\\
		0.1 & 1.08214 & 1.08351\\
		0.2 & 1.08149 & 1.08693\\
		0.3 & 1.08037 & 1.09303\\
		0.4 & 1.07896 & 1.10255\\
		0.5 & 1.07710 & 1.11701\\
\hline
	\end{tabular}
	\caption{$r_0$ values for different values of $B$ for the parallel and perpendicular unstable string configurations in the Einstein frame case. Here $L=0.75$ is used. In units of GeV.}
\label{table4}
\end{table}

Compared to the string frame case, there are a few interesting differences as well. Notice that, for a fixed $L$, the $r_0$ value decreases slightly with $B$ for the parallel case whereas it increases with $B$ for the perpendicular case. For $L=0.75$, the $r_0$ values are shown in Table~\ref{table4}. This result is different from the string frame case where $r_0$ value increases with $B$ for the parallel case as well. Therefore, in the Einstein frame, the tip of the string is moving closer and closer to the horizon as the magnetic field increases for the parallel case, whereas it is moving away from the horizon for higher magnetic field values for the perpendicular case. This already indicates that substantial changes might appear in the chaotic dynamics of the string in the parallel case compared to the perpendicular case in the Einstein frame. In particular, the chaotic behaviour might increase/decrease with the magnetic field in the parallel/perpendicular directions.

\subsection{Perturbing the static string}\label{sec:3.3}
We now repeat the perturbative analysis in the Einstein frame. We again mainly concentrate on the unstable strings (dashed lines of Figs. \ref{r0vsLvsBrh1paraEF} and \ref{r0vsLvsBrh1perpEF}), where tips are more closer to the horizon, to analyse the chaotic behaviour. For this purpose, we fix the length of the string $L=0.75$ and horizon radius $r_h=1$. It again makes the tip of the string $r_0$ a $B$ dependent quantity, shown in Table~\ref{table4}.

\begin{figure}[t]
	\centering
	\subfloat[Parallel configuration]{\label{fig:sub5}\includegraphics[width=0.45\textwidth]{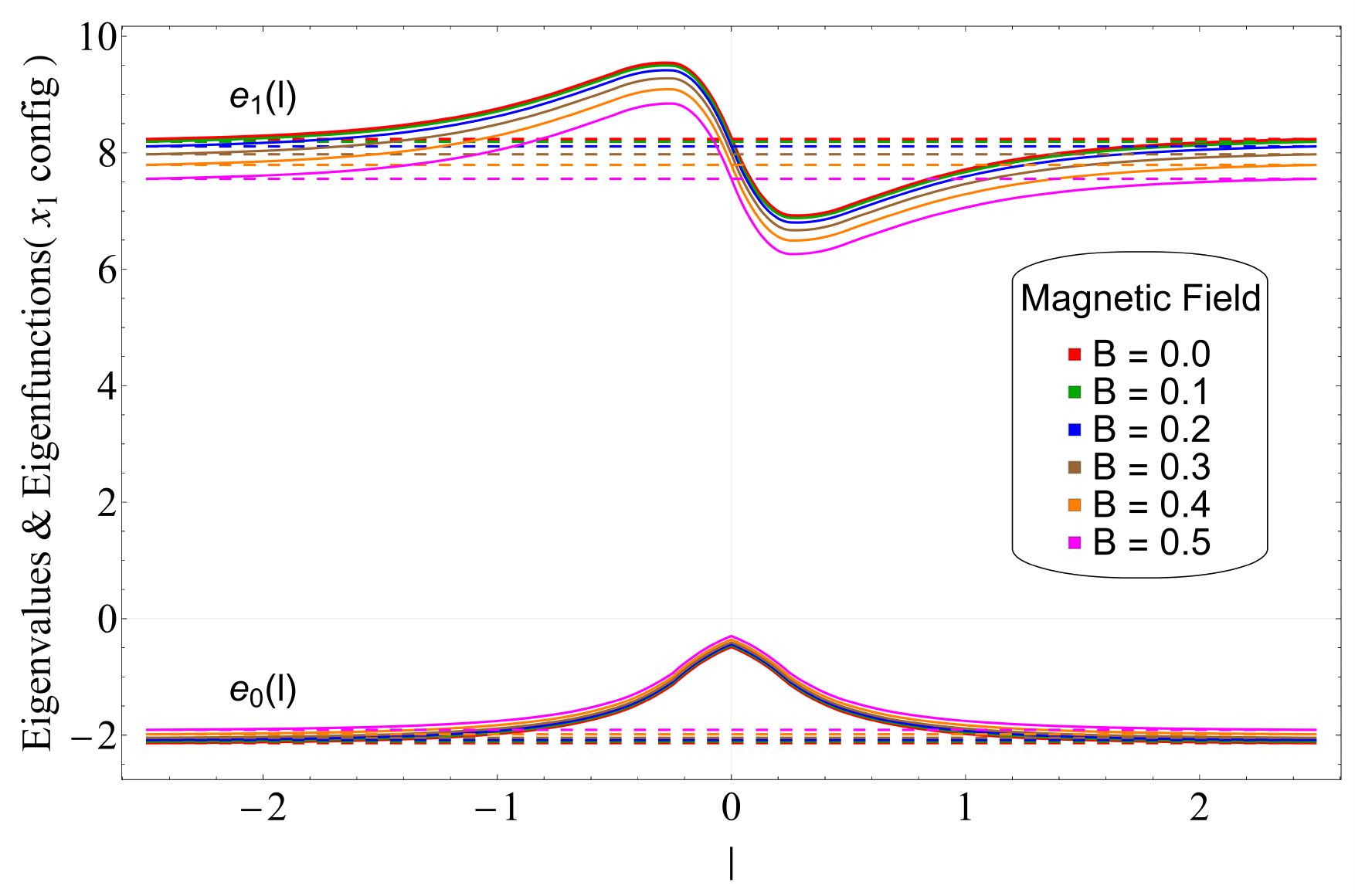}}
	\hfill
	\subfloat[Perpendicular configuration]{\label{fig:sub6}\includegraphics[width=0.48\textwidth]{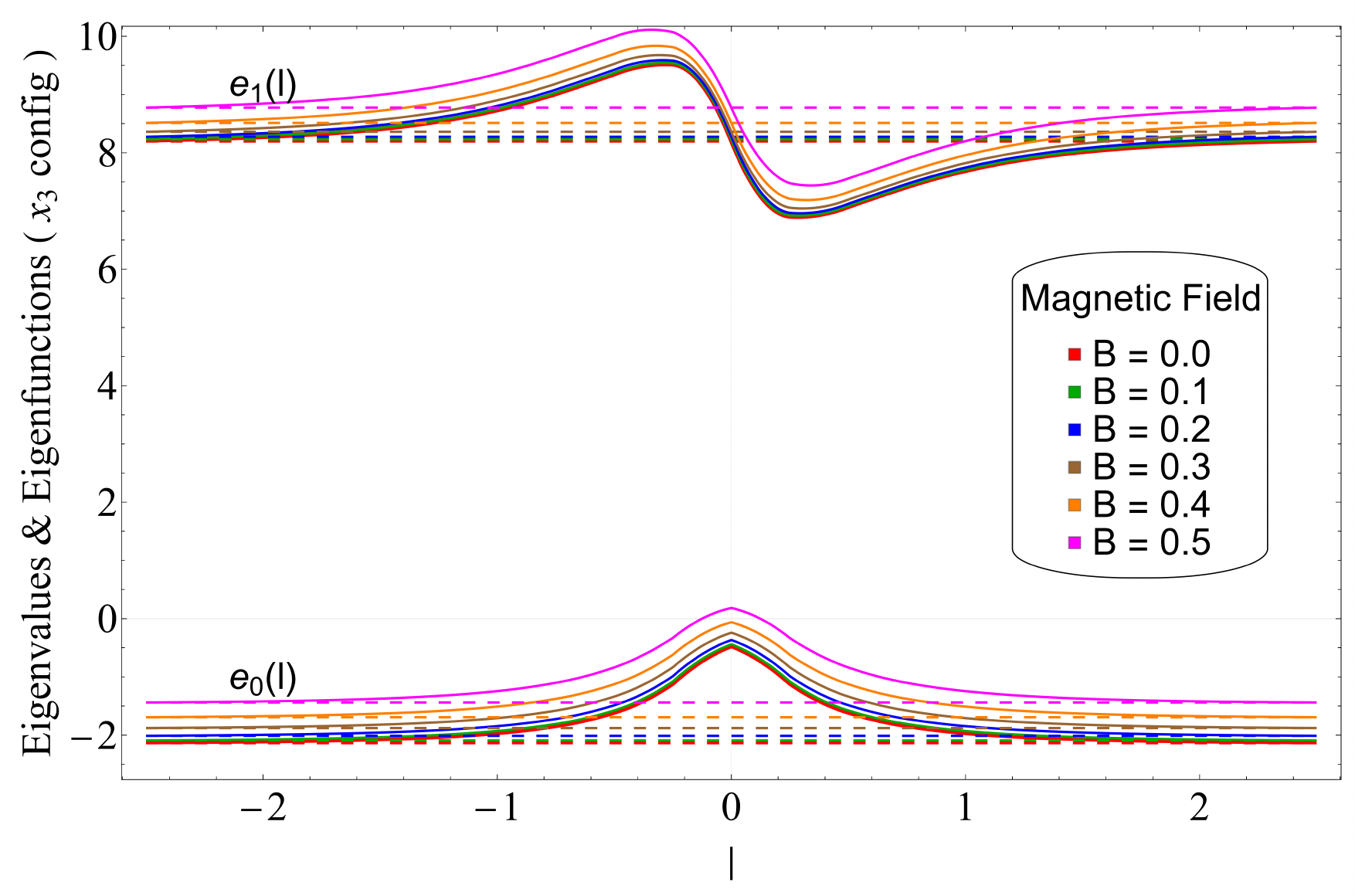}}
	\caption{Eigenfunctions $e_{0}(\ell)$ and $e_{1}(\ell)$ of Eq.~(\ref{27}) in the Einstein frame for $L=0.75$. In units of GeV.}
	\label{fig10}
\end{figure}

We can similarly compute the action up to second and third order in perturbations. The second order action again  translates into a Sturm-Liouville problem for the perturbations (see Eq.~(\ref{27})), with coefficients $C_{tt}^{x_{i}}$, $C_{\ell \ell}^{x_{i}}$, $C_{00}^{x_{i}}$ taking similar form as in Eq.~(\ref{25}), albeit with $A_s(r)$ replaced by $A(r)$. The eigenvalues are presented in Table~\ref{table5}. In the Einstein frame as well, the lowest eigenvalue $\omega_{0}^{2}$ again turns out to be negative for all values of $B$ for both parallel and perpendicular orientations of the string, suggesting chaos in the string dynamics in the Einstein frame as well. Moreover, the behaviour of these lowest eigenvalues is again quite similar to the string frame case. In particular, it not only decreases with magnetic field but also stronger in the perpendicular direction. The structure of the corresponding eigenfunctions  $\xi(l)=e_{0}(\ell)$ and $\xi(l)=e_{1}(\ell)$ are again similar to the string frame case and is shown in Fig.~\ref{fig10} for different values of magnetic field.

\begin{table}[t]
	\centering
	\begin{tabular}{|c|c|c||c|c|}
\hline
		$B$ & $\omega_{0}^{2}~(||)$ & $\omega_{1}^{2}~(||)$ & $\omega_{0}^{2}~(\perp)$ & $\omega_{1}^{2}~(\perp)$ \\
		\hline
		0 & $-2.11832$ & 8.21425 &  $-2.11832$ & 8.21425\\
		0.1 & $-2.11005$ & 8.18818 &  $-2.09053$ & 8.23044\\
		0.2 & $-2.08547$ & 8.10913 &  $-2.01133$ & 8.27463\\
		0.3 & $-2.04554$ & 7.97425 &  $-1.87938$ & 8.36054\\
		0.4 & $-1.98570$ & 7.79146 &  $-1.69241$ & 8.51136\\
		0.5 & $-1.90991$ & 7.55148 &  $-1.44083$ & 8.77507\\
\hline
	\end{tabular}
	\caption{The magnetic field dependence of eigenvalues $\omega_{0}^{2}$ and $\omega_{1}^{2}$ of Eq.~(\ref{27}) for parallel and perpendicular string configurations in the Einstein frame case. Here $L=0.75$ is used. In units of GeV.}
\label{table5}
\end{table}

\begin{table}[t]
	\centering
	\begin{tabular}{|c|c|c|c|c|c||c|c|c|c|c| }
\hline
		$B$ & $K_{1}~(||)$ & $K_{2}~(||)$ & $K_{3}~(||)$ & $K_{4}~(||)$ & $K_{5}~(||)$& $K_{1}~(\perp)$ & $K_{2}~(\perp)$ & $K_{3}~(\perp)$ & $K_{4}~(\perp)$ & $K_{5}~(\perp)$\\
		\hline
	0 & 19.200 & 33.873 & 13.721 & 3.783 & 7.567&19.200 & 33.873 & 13.721 & 3.783 & 7.567\\
	0.1 & 19.093 & 33.762 & 13.725 & 3.790 & 7.581& 18.904 & 33.616 & 13.603 & 3.768 & 7.535\\
	0.2 & 18.784 & 33.423 & 13.740 & 3.812 & 7.624& 18.075 & 32.875 & 13.271 & 3.725 & 7.449\\
	0.3 & 18.281 & 32.861 & 13.767 & 3.849 & 7.697& 16.769 & 31.704 & 12.739 & 3.652 & 7.305\\
	0.4 & 17.565 & 32.050 & 13.786 & 3.896 & 7.793& 15.092 & 30.193 & 12.032 & 3.551 & 7.102\\
	0.5 & 16.666 & 30.990 & 13.808 & 3.959 & 7.917& 13.161 & 28.491 & 11.170 & 3.419 & 6.838\\
		\hline
\end{tabular}
	\caption{The magnetic field dependence of  coefficients $K_{i}$ appearing in Eq.~(\ref{32}) for parallel and perpendicular string configurations in the Einstein frame. Here $L=0.75$ is used. In units of GeV.}
\label{table6}
\end{table}

The action up to cubic terms (\ref{32}) leads to trapping potential for the string in the Einstein frame as well. The action (\ref{32}) can further be used to study the dynamics of $c_0$ and $c_1$ within the trap. The values of the coefficients $K_{i}$ appearing in Eq.~(\ref{32}) are given in Table~\ref{table6}. These values are given up to third decimal place but in actual numerics these values have been computed with higher accuracy. We notice that, just like in the string frame case, the kinetic term of the Lagrangian is not positive definite everywhere in the parameter space. Here we again make the change of variables: $c_{0,1}\rightarrow\tilde{c}_{0,1}$, with $c_{0}=\tilde{c_{0}}+\alpha_{1}\tilde{c_{0}}^{2}+\alpha_{2}\tilde{c_{1}}^{2}$ and $c_{1}=\tilde{c_{1}}+\alpha_{3}\tilde{c_{0}}\tilde{c_{1}}$ to bypass this problem \cite{Hashimoto2018Oct,Colangelo2020Oct}. We neglect $\mathcal{O}(\tilde{c_{i}}^{4})$ terms, and choose appropriate values for $\alpha_{i}$, to make sure that the kinetic term is now positive definite. One example of such a choice is $\alpha_{1}=-3$, $\alpha_{2}=-1$ and $\alpha_{3}=-1.5$. This variable change makes the time evolution of the system well-posed without affecting the dynamics. The modified action takes the same form as in (\ref{33}), albeit with different magnitudes of coefficients $\tilde{K_{j}}^{x_{i}}$.

%We can see that there are striking differences in the $C_{jj}^{x_{i}}$ coefficients, $r_0$ values, the Eigenvalues, the eigenfunctions and the $K_i$ coefficients between the String frame and the Einstein frame which signifies different chaotic dynamics in both frames.

\subsection{Poincar\'{e} sections}\label{sec:3.4}
\begin{figure}[!]	
	\centering
	\begin{tabular}{c c}
		\textbf{Parallel Configuration} & \textbf{Perpendicular Configuration} \\
		\includegraphics[width=0.48\linewidth]{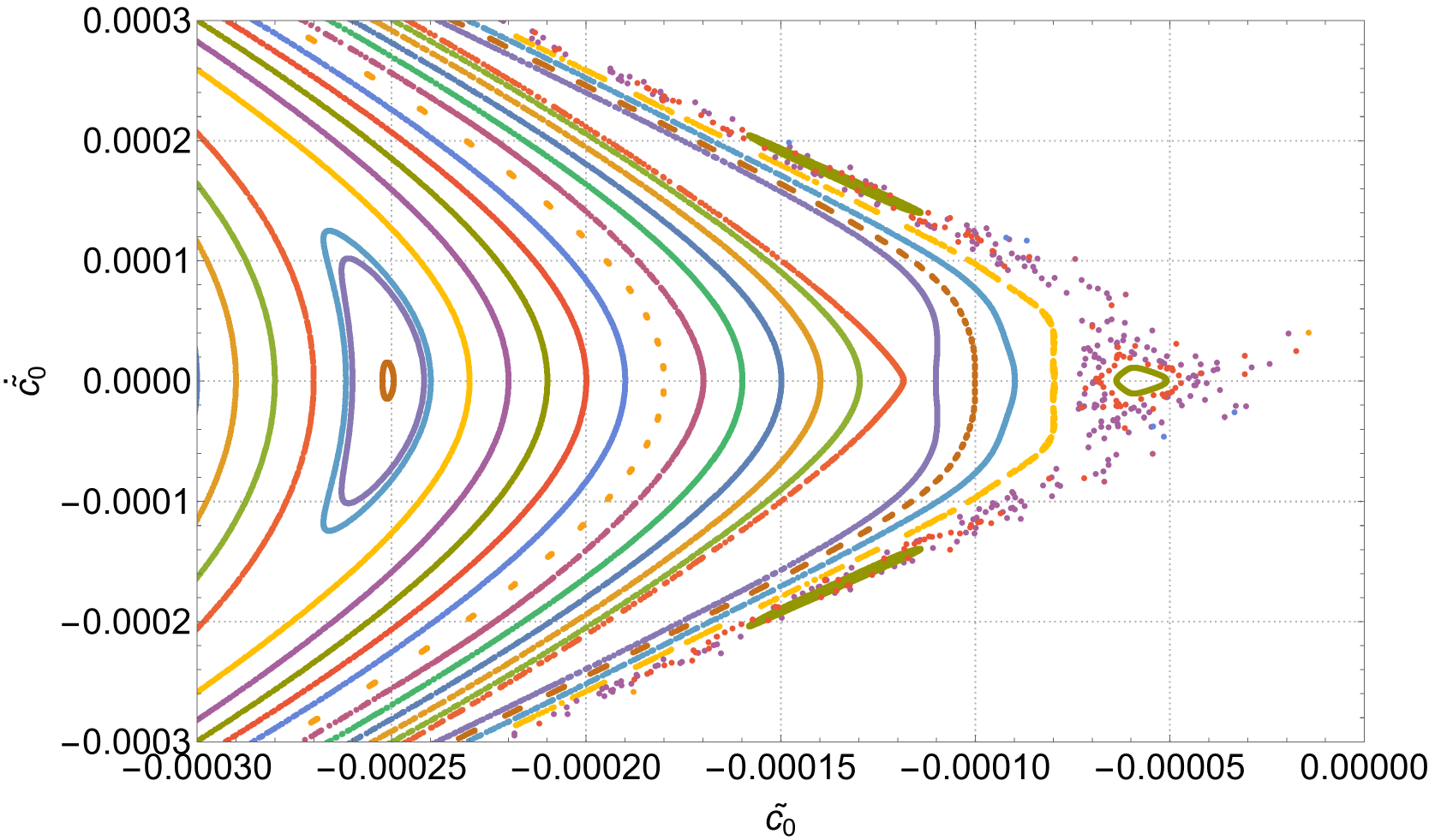} & \includegraphics[width=0.48\linewidth]{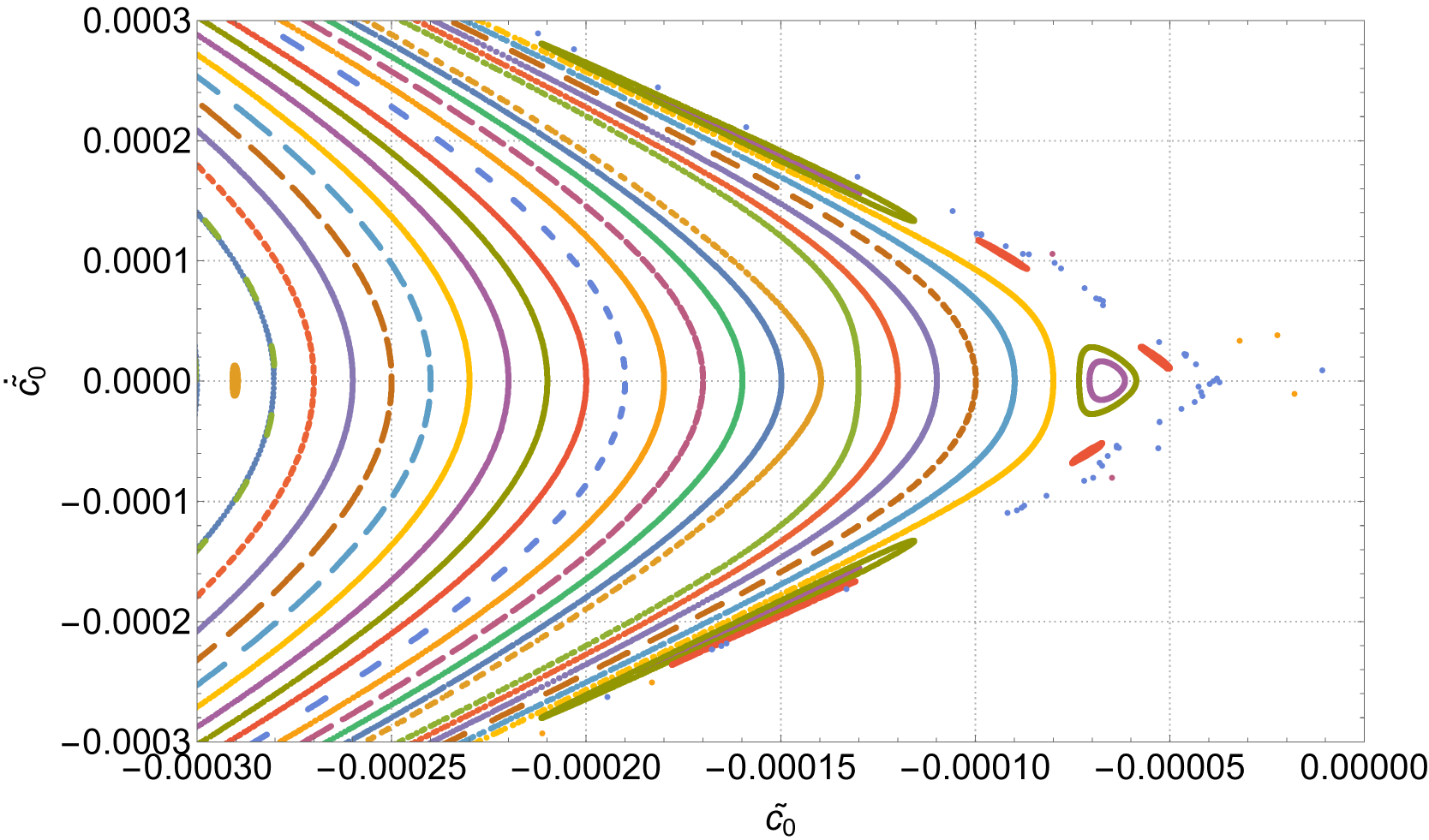} \\
		\includegraphics[width=0.48\linewidth]{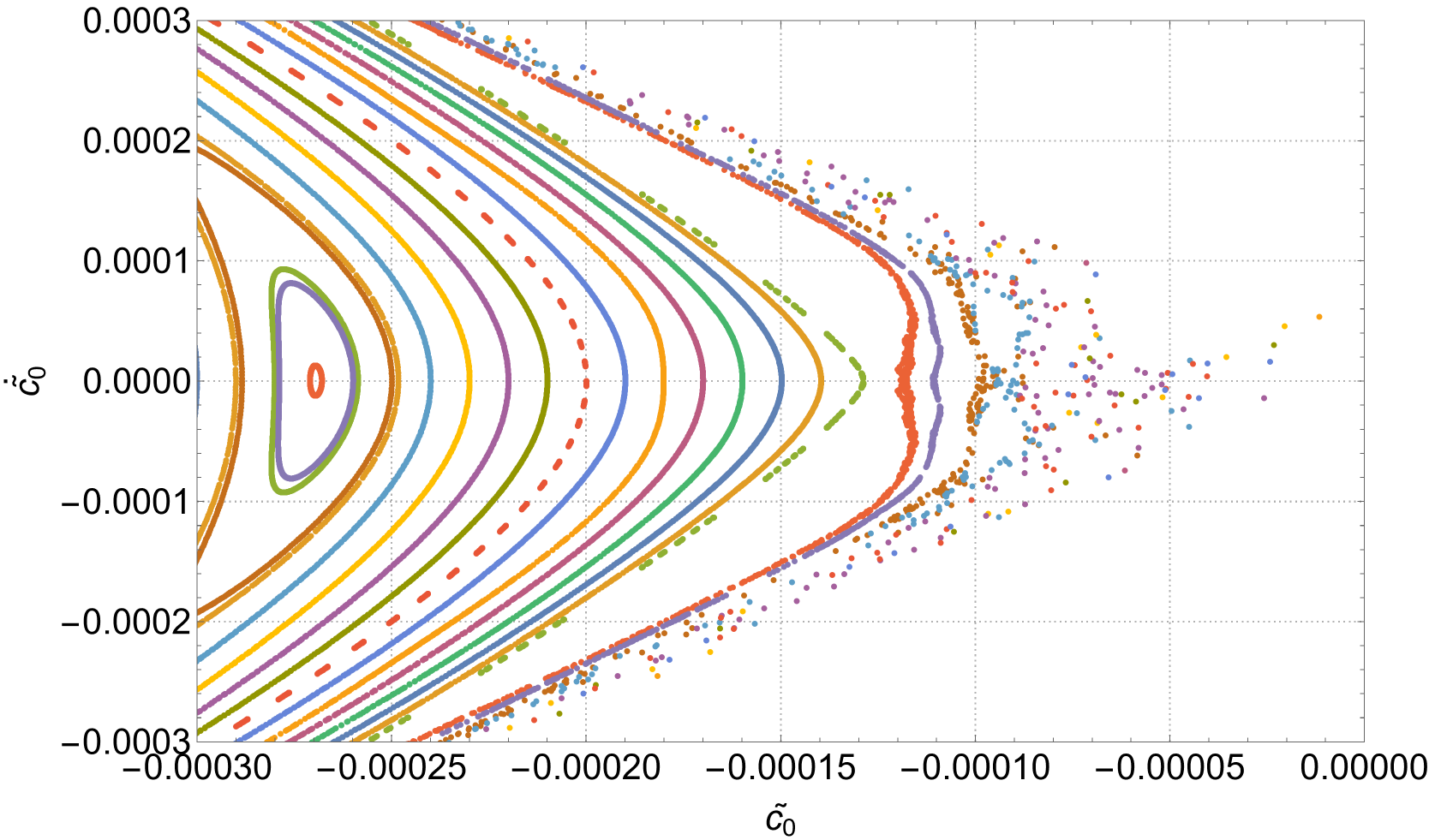} & \includegraphics[width=0.48\linewidth]{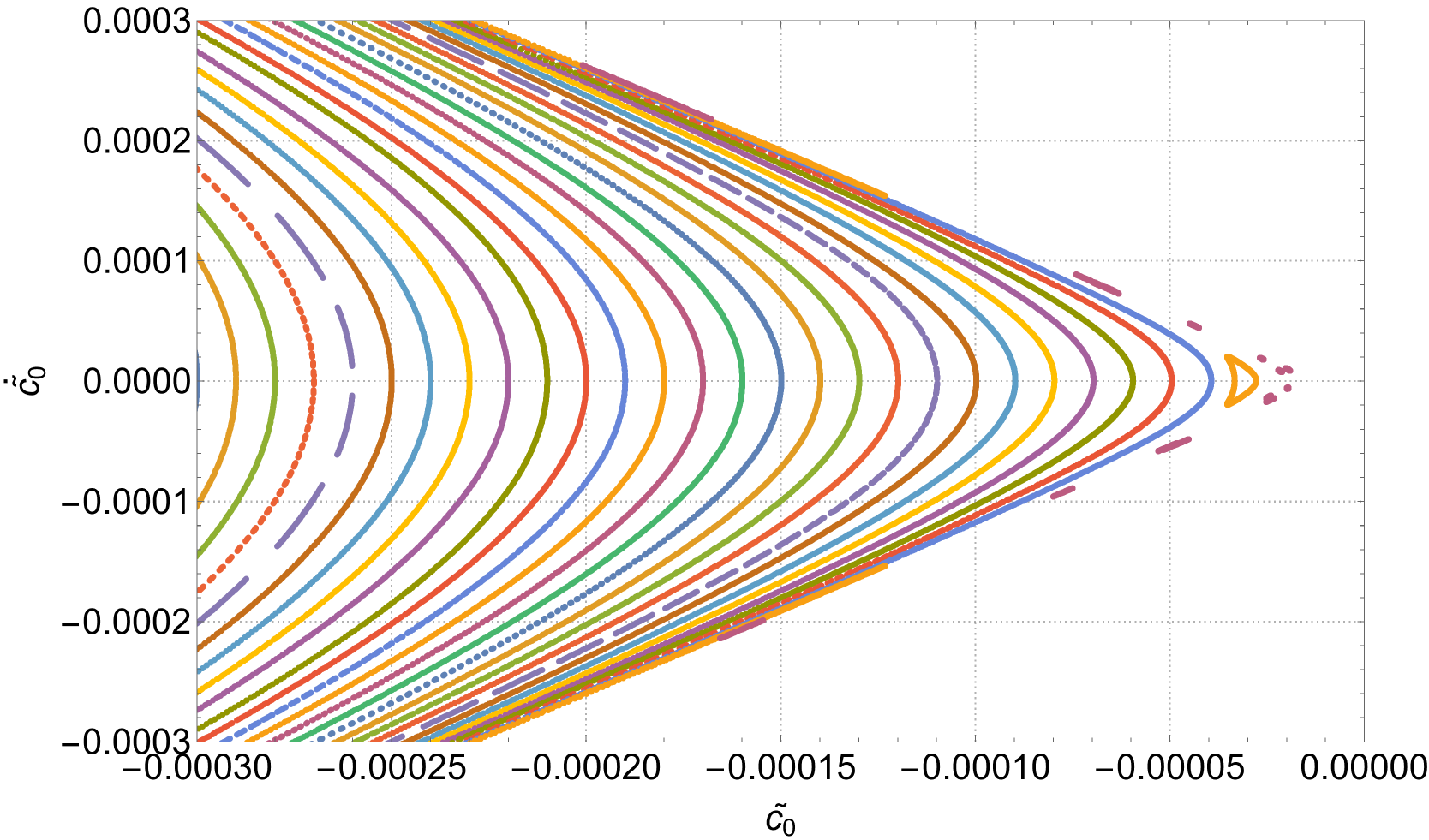} \\
		\includegraphics[width=0.48\linewidth]{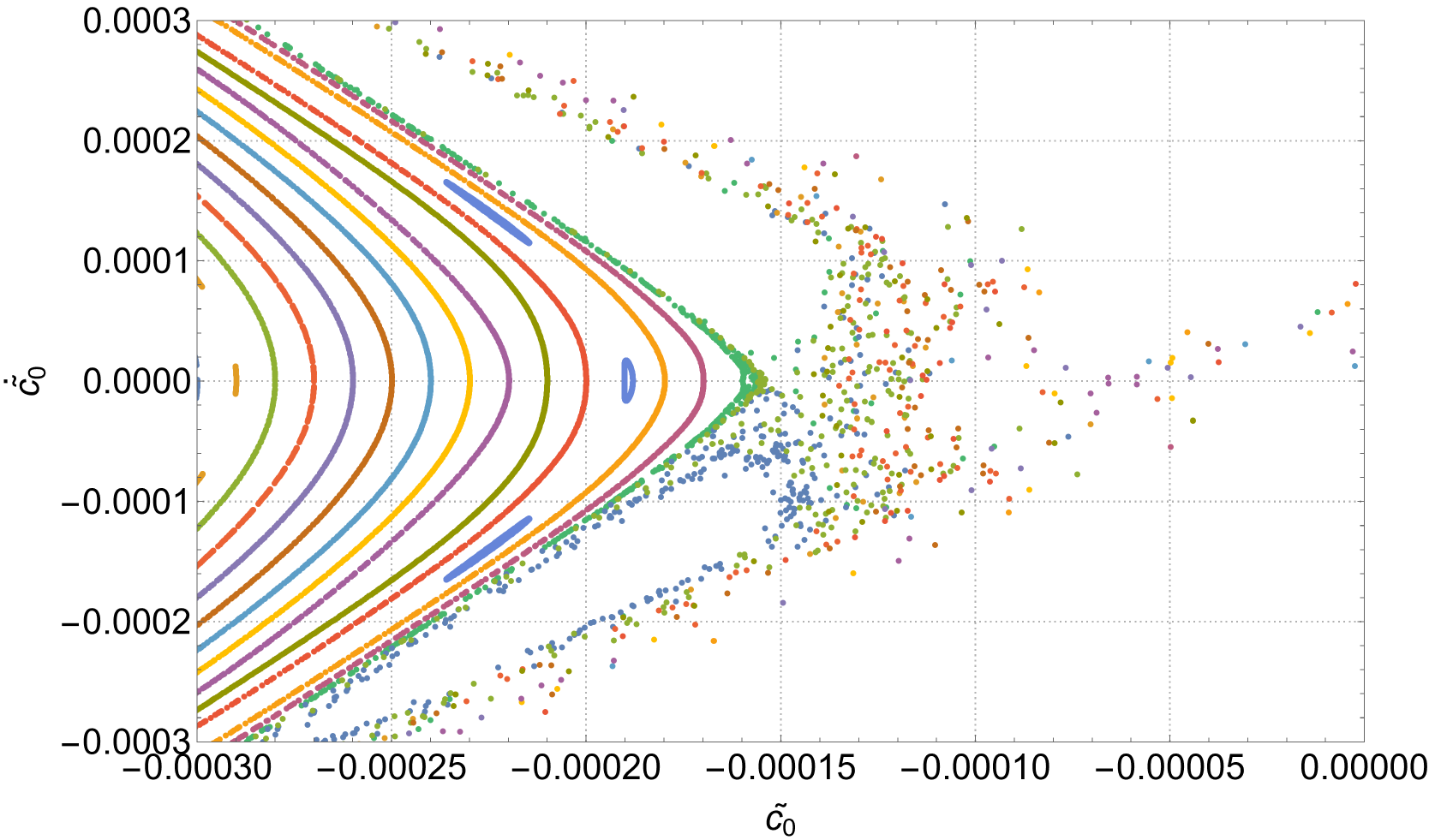} & \includegraphics[width=0.48\linewidth]{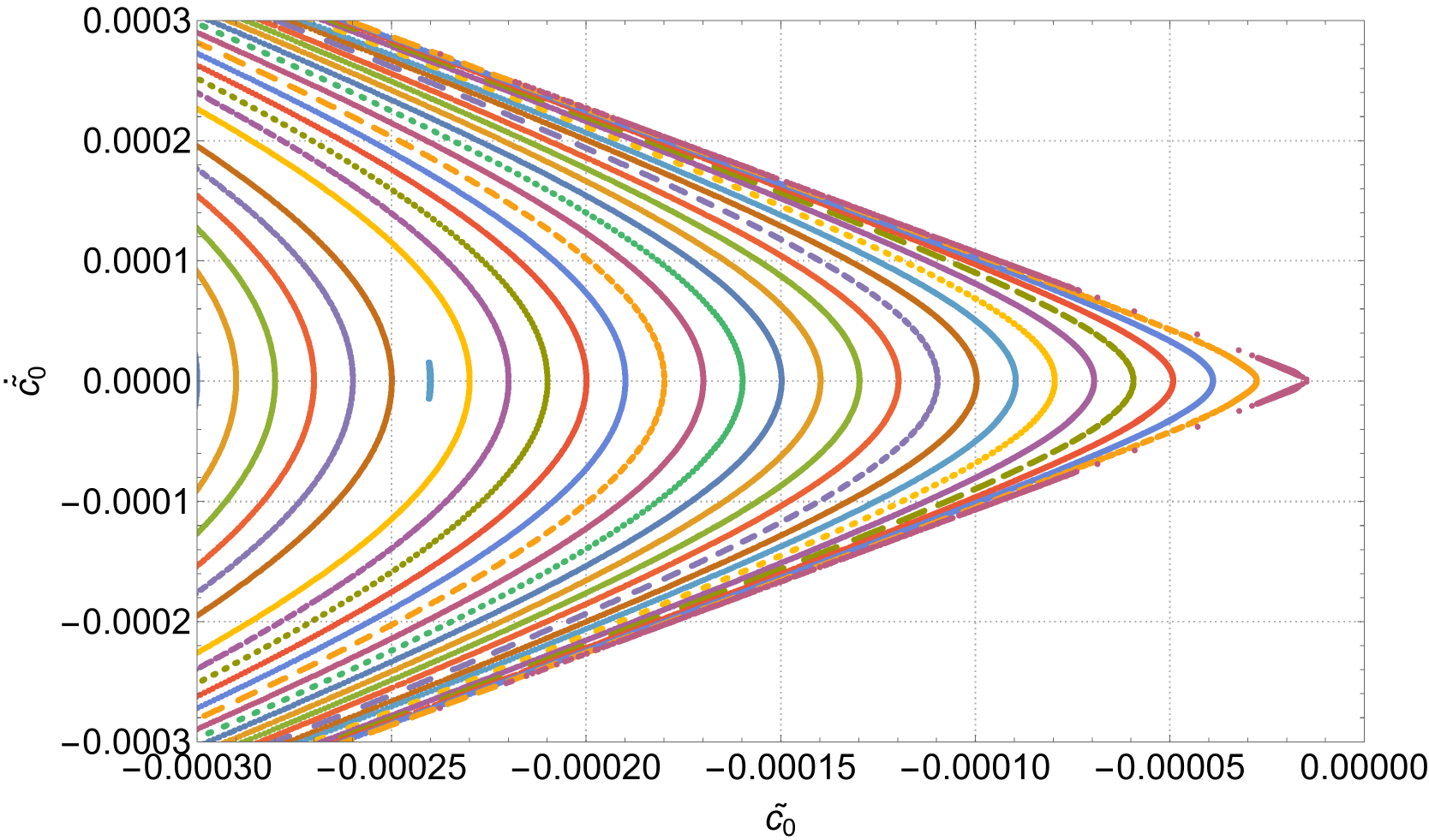} \\
	\end{tabular}
	\caption{Poincar\'{e} sections corresponding to $\tilde{c_{1}}=0$ and $\dot{\tilde{c_{1}}}\geq 0$ for parallel (left column) and perpendicular (right column) string
configurations for the orbits with fixed energy $E=10^{-5}$ and fixed $L=0.75$. $B$ is varied from $B=0.1$ (top row) to $B=0.3$ (middle row) and $B=0.5$ (bottom row). In units of GeV.}
	\label{fig11}
\end{figure}
We now discuss the Poincar\'{e} section defined by $\tilde{c_{1}}(t)=0$ and $\dot{\tilde{c_{1}}}(t)\geq 0$  for the bound orbits within the trapping potentia. The magnetic field dependence of the Poincar\'{e} section is shown in Fig.~\ref{fig11} for both parallel and perpendicular orientations. Here we have used fixed $L=0.75$, fixed energy $E=10^{-5}$, and $0<t<15000$.  The points in different colours belong to the numerical data of orbits for different starting conditions.

For the parallel configuration, when we increase $B$ there are scattered points near zero $\tilde{c_{0}}$, which show a strong dependence on initial conditions. For higher magnetic field $B$, the scattered points amplify in the system, which shows that the effect of turning on the magnetic field is to aggravate the chaotic behaviour in the parallel configuration. On the other hand, for the perpendicular configuration, when we increase $B$, the scattered points transform to regular paths, which shows that the effect of turning on the magnetic field is to reduce the chaotic behaviour. The perturbative conditions $(\tilde{c_{0}}<0,\tilde{c_{1}}=0)$ imply $(c_{0}<0,c_{1}=0)$, and corresponds to a string moving away from the black hole horizon. Therefore, for $(\tilde{c_{0}}=0,\tilde{c_{1}}=0)$, the tip of suspended string is the point closest to the horizon. This suggests that the source of chaos is again the black hole horizon. Moreover, the Poincar\'{e} section of the stable string configuration (not shown here for brevity) again contains only stable orbits without scattered points, lending further support to the notion that black hole horizon is the source of chaos of the unstable string dynamics in the Einstein frame as well.

Our overall analysis of the Poincar\'{e} section in the Einstein frame suggests that the dynamics of the string is more chaotic if we increase the magnetic field in the parallel direction, whereas it is less chaotic in the perpendicular direction.

\subsection{Lyapunov exponents}\label{sec:3.5}
\begin{figure}[!]
	\centering
	\begin{tabular}{c c}
		\textbf{Parallel Configuration} & \textbf{Perpendicular Configuration} \\
		\includegraphics[width=0.48\linewidth]{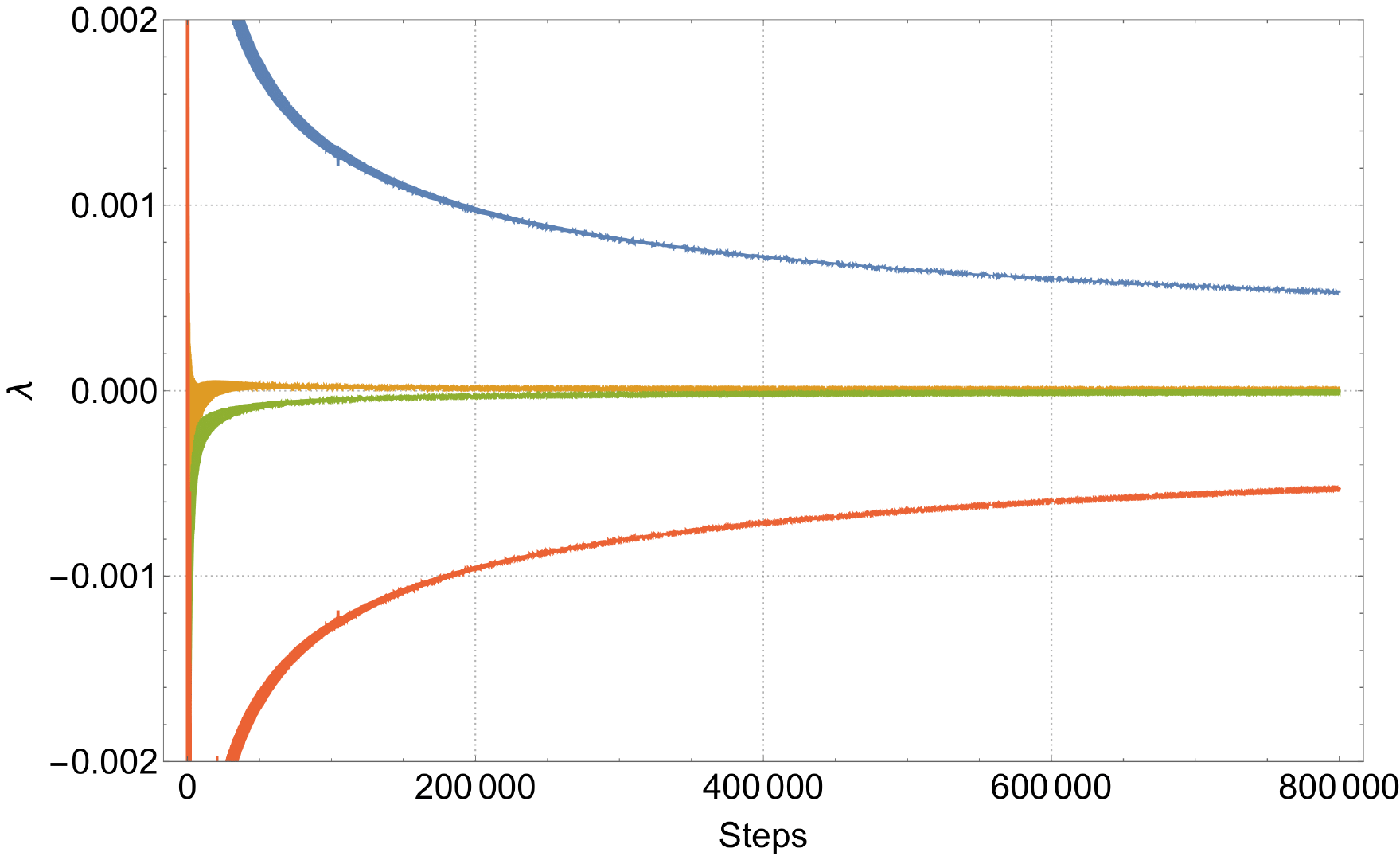} & \includegraphics[width=0.48\linewidth]{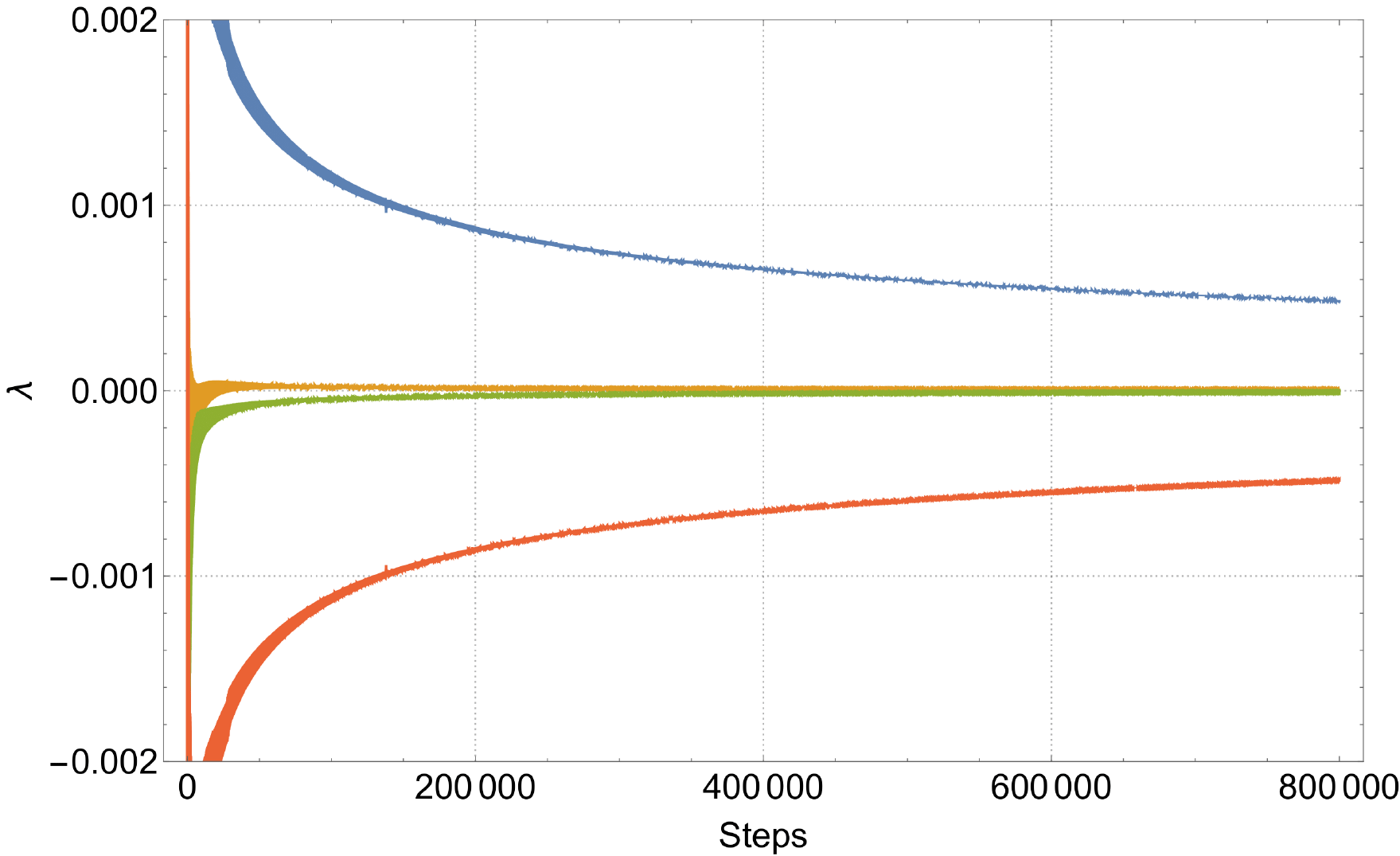} \\
		\includegraphics[width=0.48\linewidth]{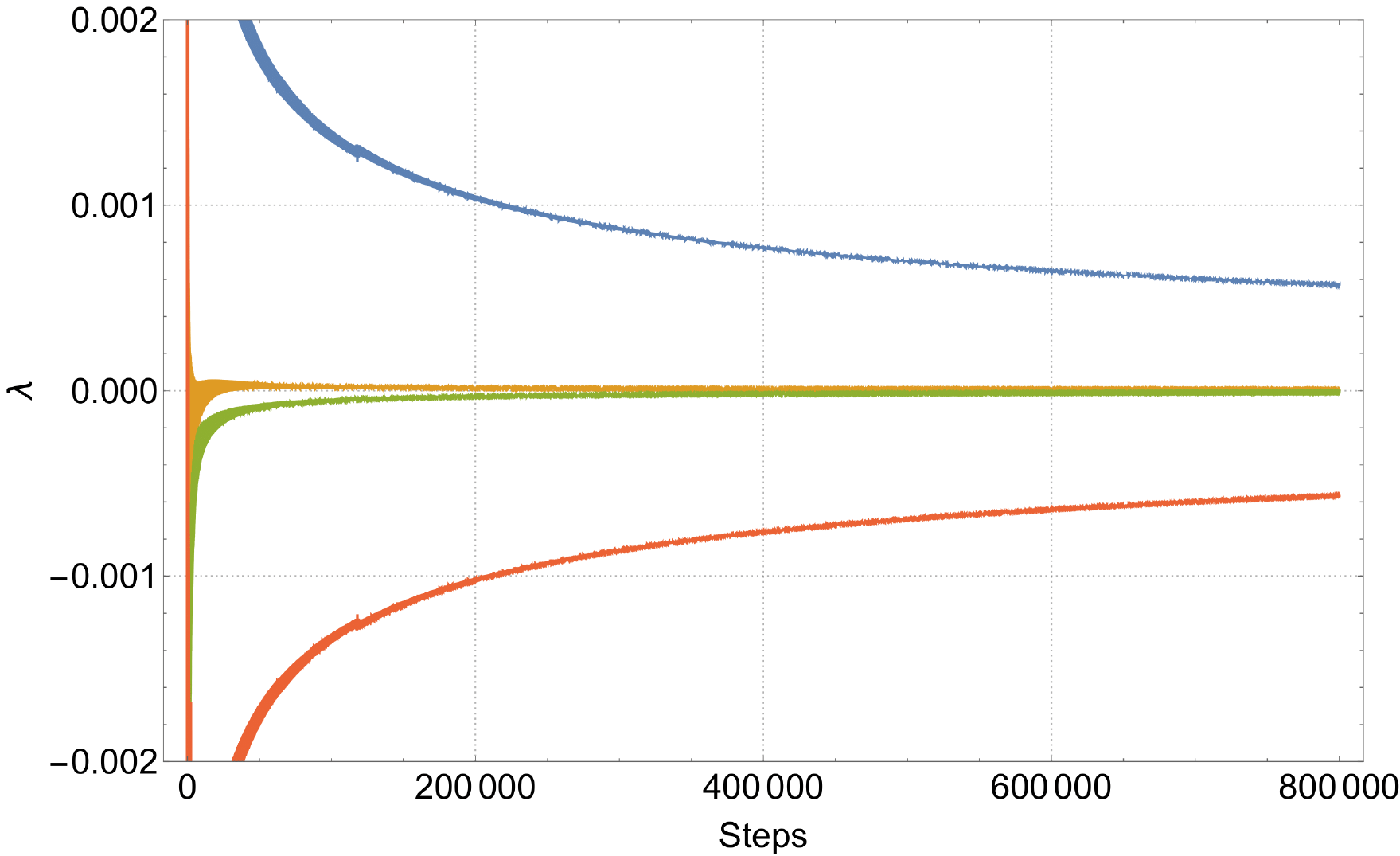} & \includegraphics[width=0.48\linewidth]{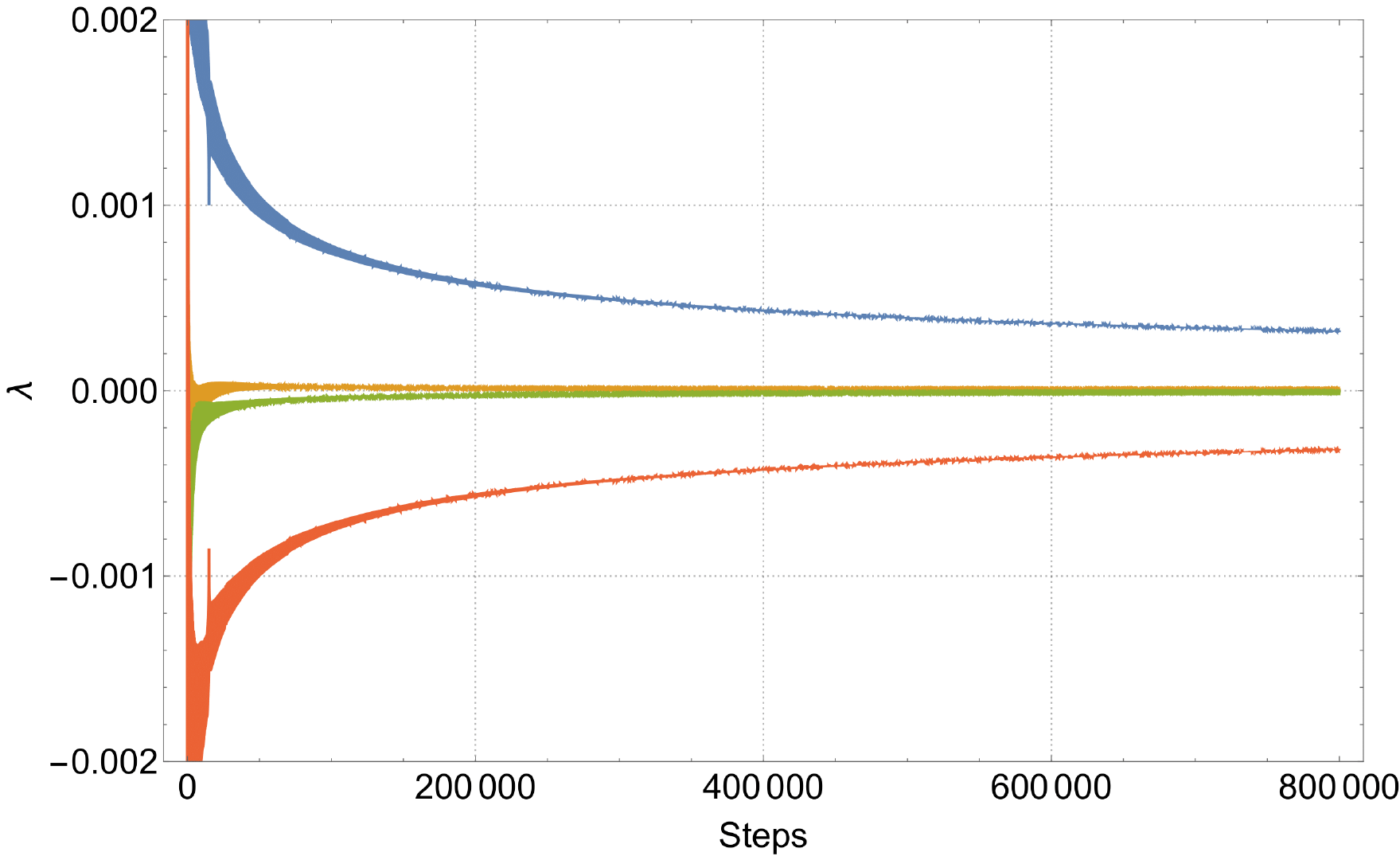} \\
		\includegraphics[width=0.48\linewidth]{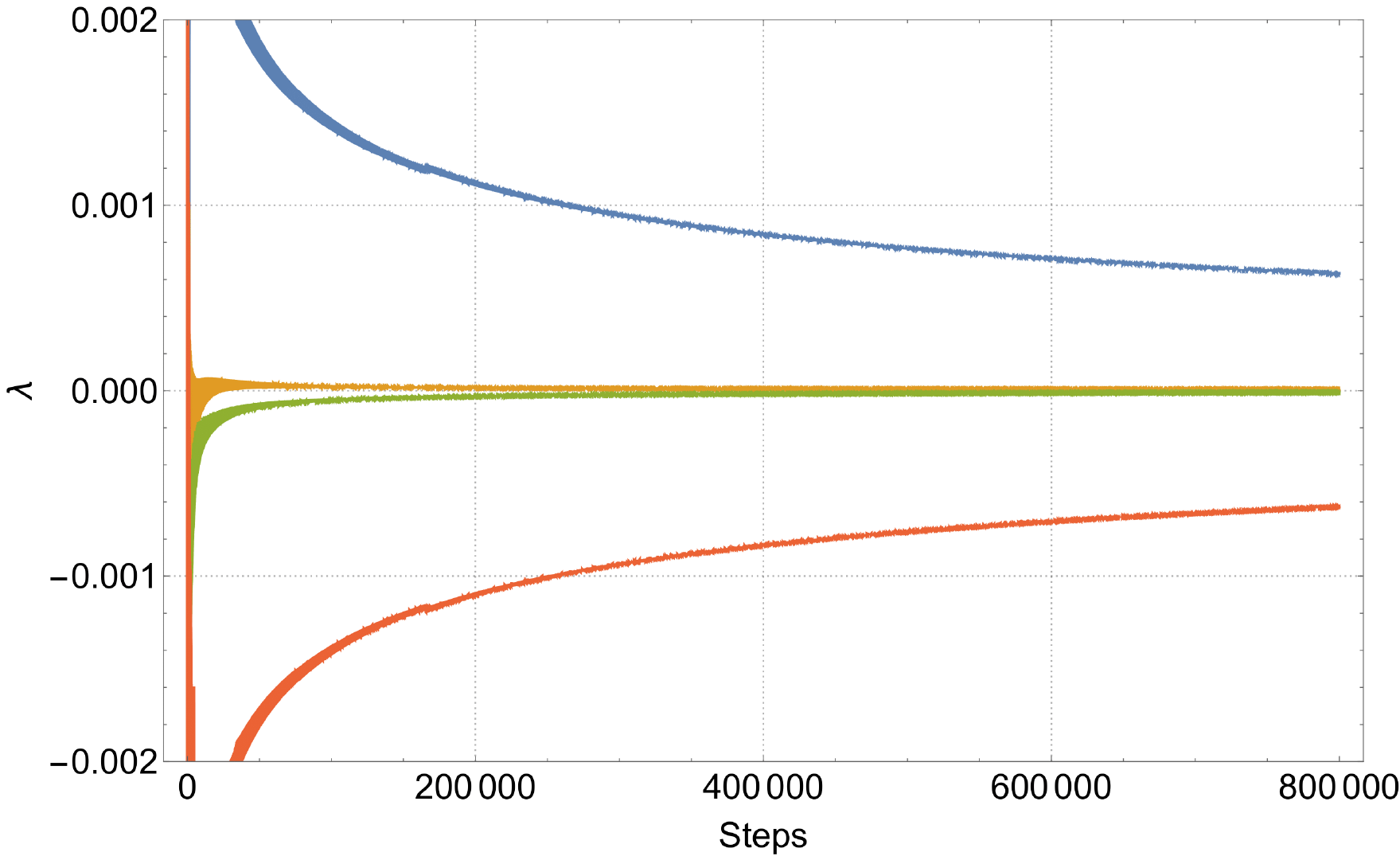} & \includegraphics[width=0.48\linewidth]{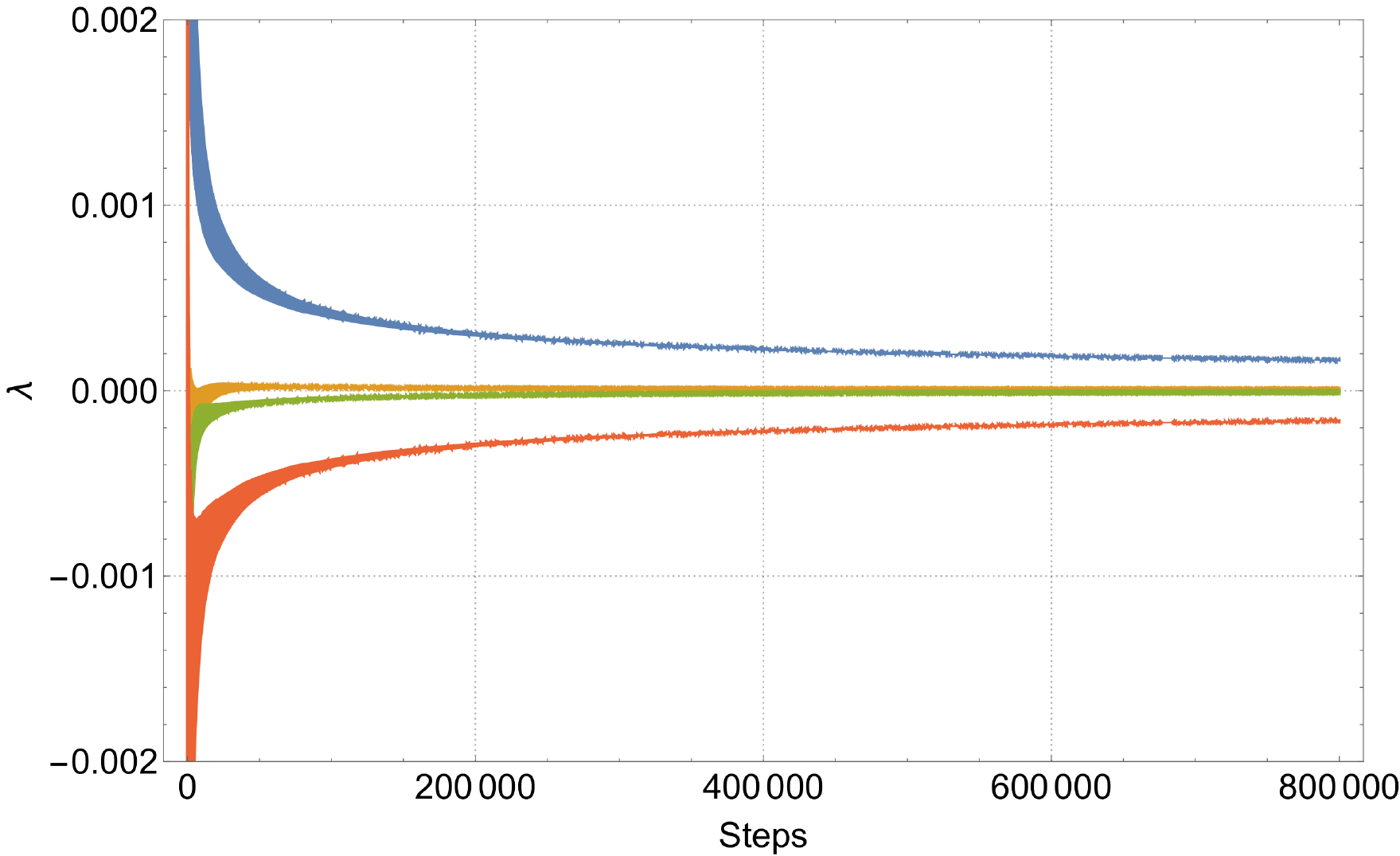} \\
		\includegraphics[width=0.48\linewidth]{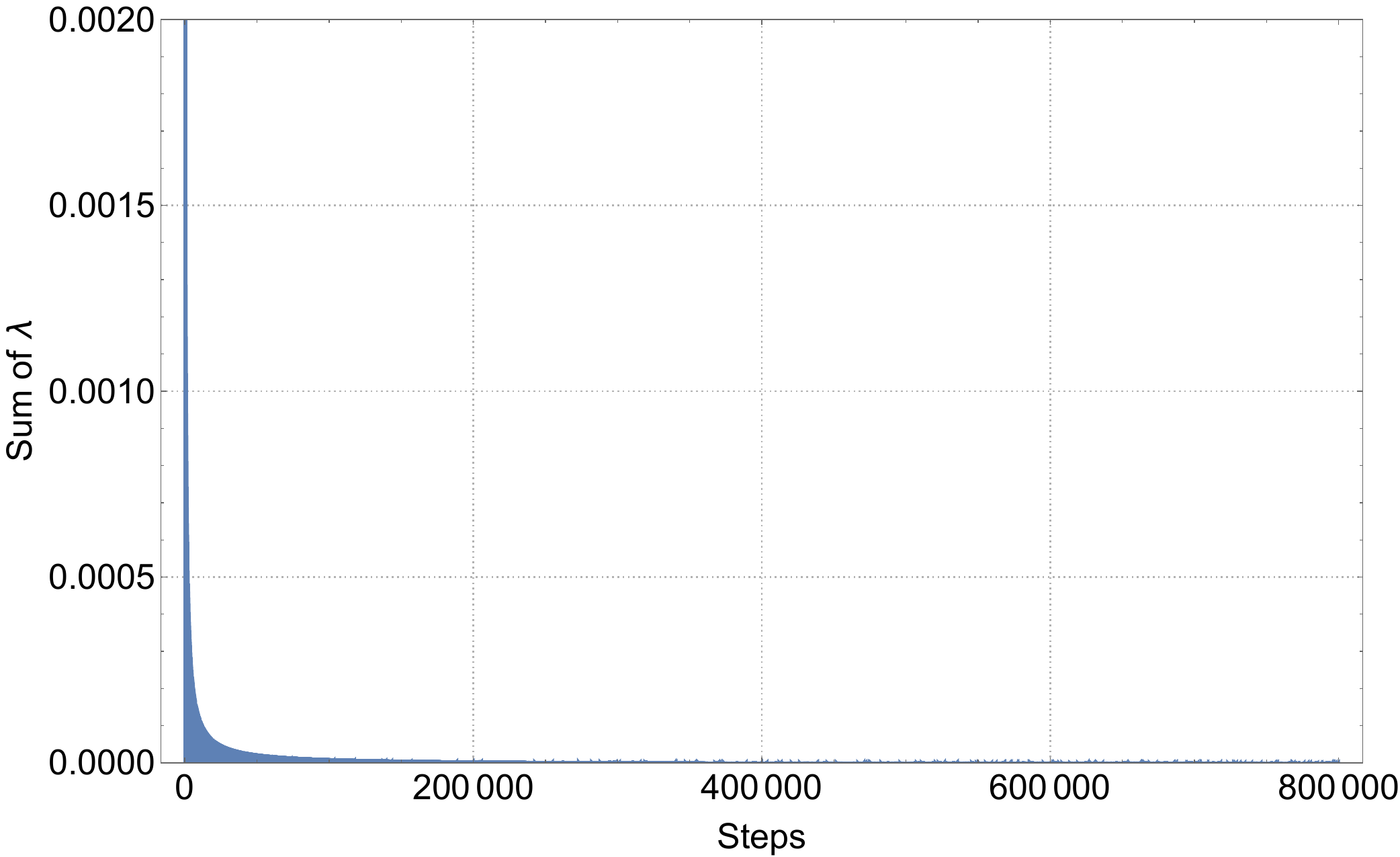} & \includegraphics[width=0.48\linewidth]{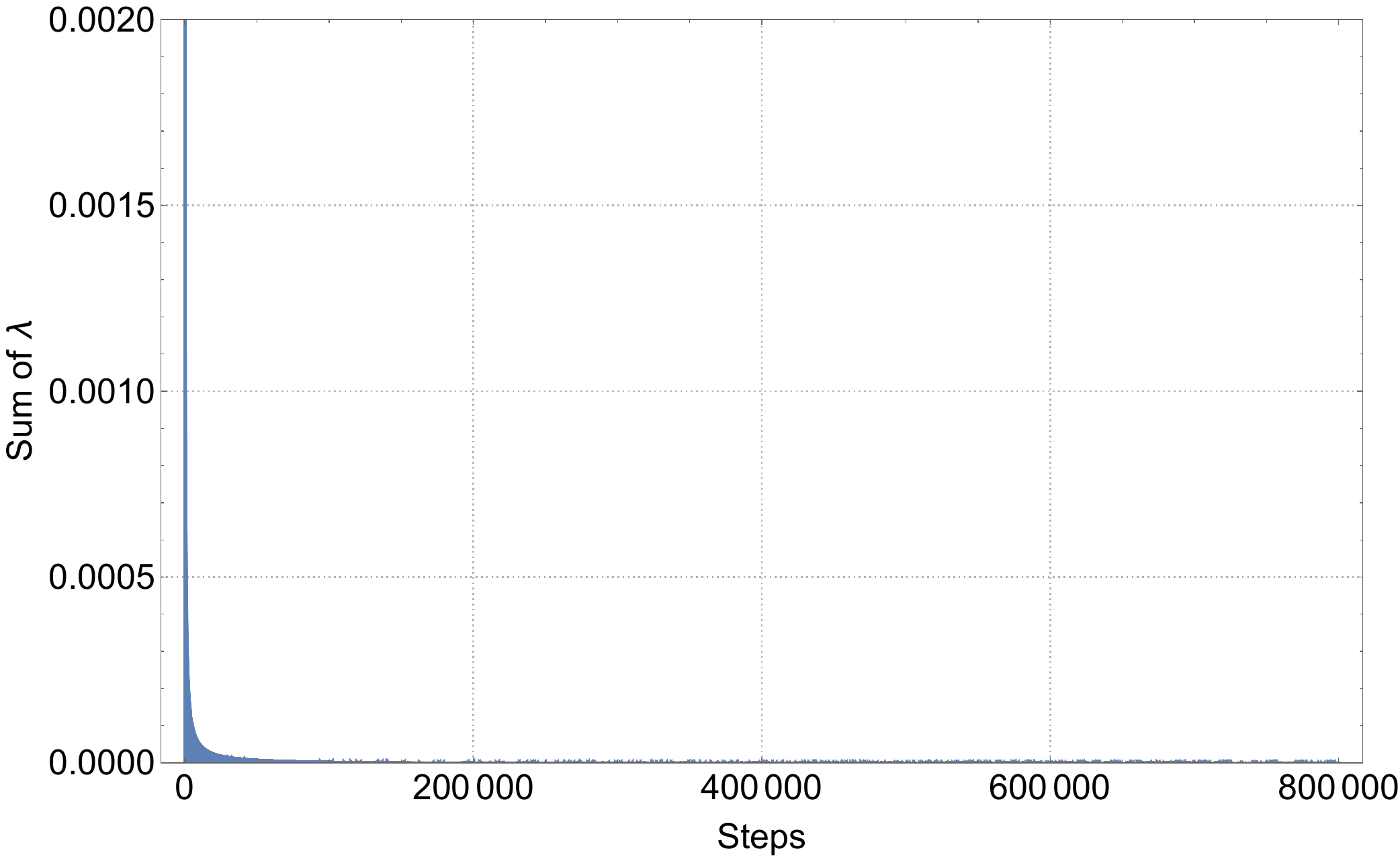} \\
	\end{tabular}
	\caption{Convergence plots of the four Lyapunov exponents for a string orientated in the parallel (left column) and perpendicular (right column) directions with respect to the magnetic field. Here we have
taken $8 \times 10^{5}$ time steps. For the initial conditions, the energy is set to $E=10^{-5}$ and along with $\tilde{c_{0}}=-0.0002$, $\dot{\tilde{c}}_{0}=0.00001$, and $\tilde{c_{1}}=0.0007$. Here $L=0.75$ is fixed and $B$ is varied from $B=0.1$ (first row) to $B=0.3$ (second row) and $B=0.5$ (third row). The sum of Lyapunov exponents is shown in the last row for $B=0.3$. In units of GeV.}
	\label{fig12}
\end{figure}
Like in the string frame case, we can similarly compute the Lyapunov exponents in the four-dimensional ($\tilde{c_{0}}$, $\tilde{c_{1}}$) phase-space with the help of the same numerical methods as for the Einstein frame. Our numerical results for the convergency of the four Lyapunov exponents, along with their sum, are shown in Fig.~\ref{fig12} for the  parallel and perpendicular configurations. From these plots we observe that the convergence is again similar to a damped oscillator. The system remains conservative since the sum of Lyapunov exponents converges to zero
as can be seen in the last row of the Fig.~\ref{fig12}. This is true for all values of the magnetic field, irrespective of its orientation relative to the string.

\begin{figure}[t]
	\centering
	\includegraphics[width=0.5\linewidth]{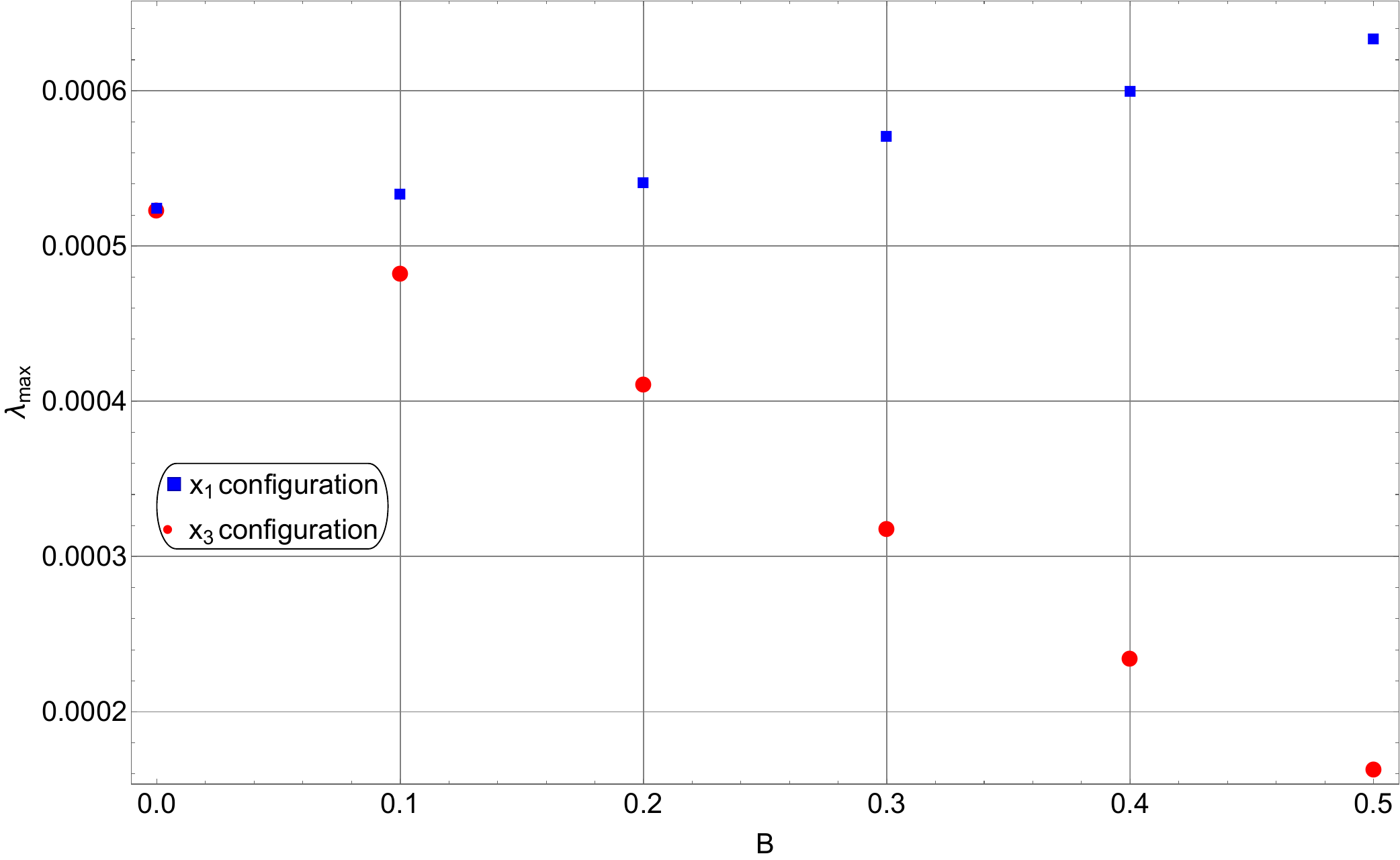}
	\caption{The largest Lyapunov exponent $\lambda_{max}$ of Fig.~\ref{fig12} for different values of $B$ at $L = 0.75$ for two different orientations of the string. In units of GeV.}
	\label{fig13}
\end{figure}

The analysis of the Poincar\'{e} plots tells us that chaos is produced near the black hole horizon, and that the dynamics of the string becomes more chaotic when we increase the magnetic field in the parallel direction and the reverse happens in the perpendicular direction. This is confirmed by the largest Lyapunov exponents $\lambda_{max}$ for both the directions, as shown in Fig.~\ref{fig13}. For the same string length $L$, increasing the value of magnetic field makes $\lambda_{max}$ larger in the parallel direction. This implies that the magnetic field has a destabilizing effect on the string dynamics in this direction. This result should be contrasted with the String frame result where $\lambda_{max}$ was found to be decreasing with the magnetic field. For the perpendicular case, the behaviour of $\lambda_{max}$ is quite similar to the string frame case. In particular, higher magnetic field values make $\lambda_{max}$ smaller, suggesting that the magnetic field has a stabilizing effect on the string dynamics in this direction. Notice also from Fig.~\ref{fig13} that $\lambda_{max}$ inside the potential trap is around three orders of magnitude smaller than the MSS bound for all $B$. This implies that the MSS bound is always satisfied in the Einstein frame as well.

\subsection{Analysis of the saddle point and testing the MSS bound}\label{sec:3.6}
\begin{figure}[b]
	\centering
	\subfloat[Parallel configuration]{\label{saddlepoint_x1_einsteinframe}\includegraphics[width=0.45\textwidth]{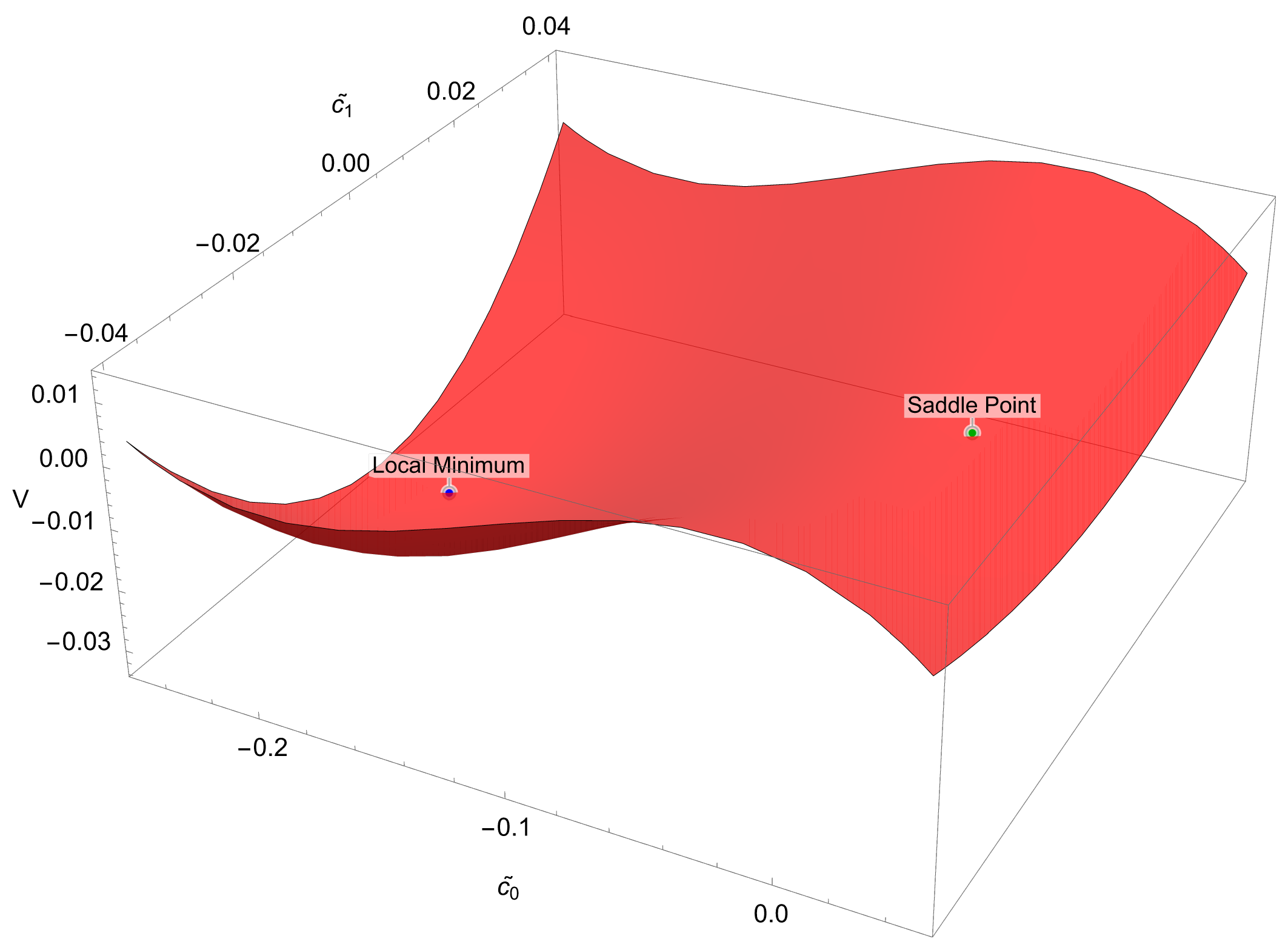}}
	\hfill
	\subfloat[Perpendicular configuration]{\label{saddlepoint_x3_einsteinframe}
		\includegraphics[width=0.45\textwidth]{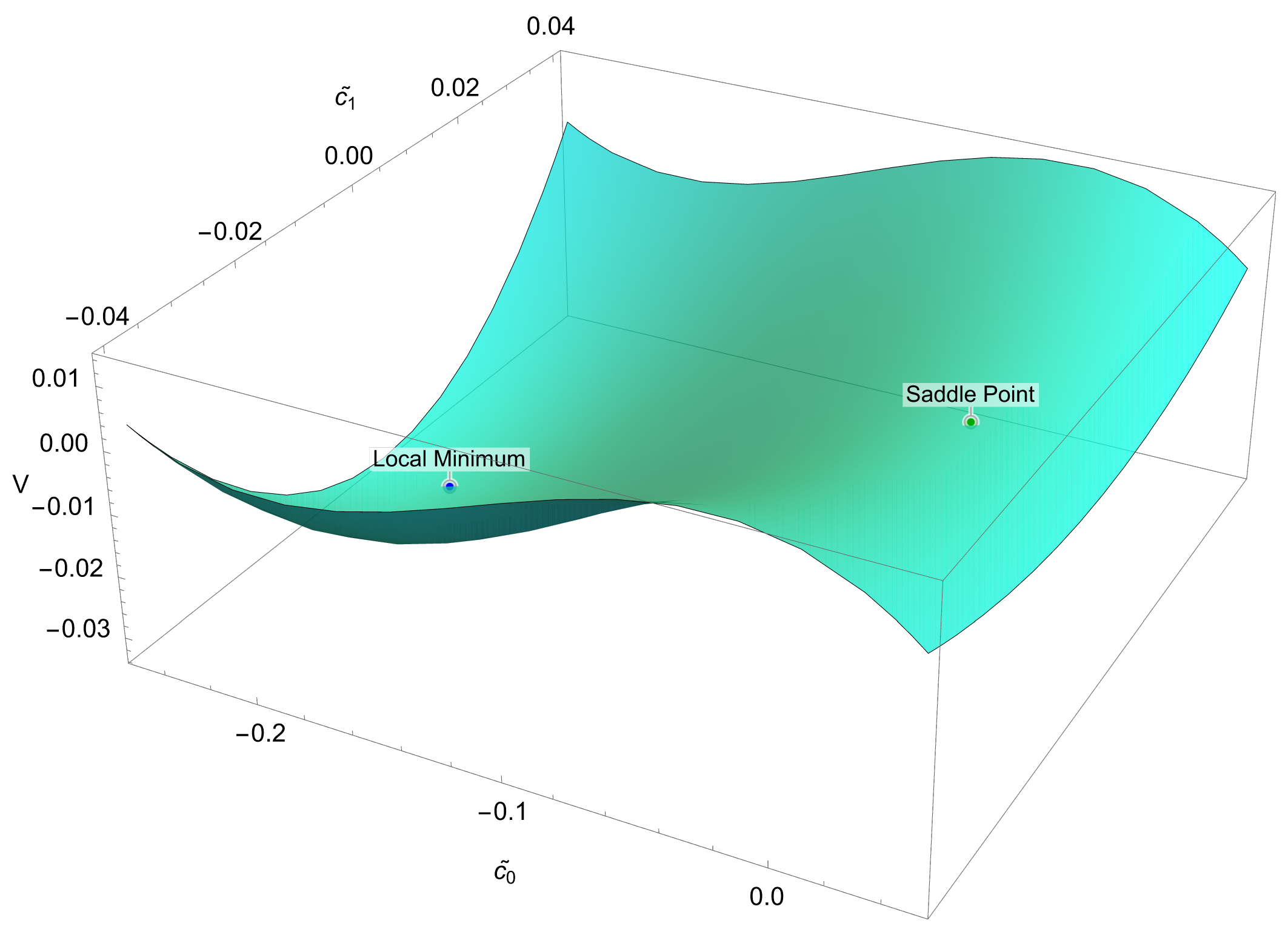}}
	\caption{\label{saddlepoint_einsteinframe}Potentials obtained from Eq.~(\ref{33}) for $B=0.1$. In units of GeV.}
\end{figure}
To complete our discussion on the Lyapunov exponent and its relative comparison with the MSS bound in the Einstein frame, here we discuss the Lyapunov exponent at the unstable fixed points. Notice that, as mentioned
earlier, the potentials obtained from the action (\ref{33}) in the Einstein frame also show a stable and unstable fixed points. These two fixed points differ in terms of their position and appearance for the parallel and perpendicular configurations. This is shown in Fig.~\ref{saddlepoint_einsteinframe}.

\begin{figure}[h]
	\centering
	\includegraphics[width=0.5\linewidth]{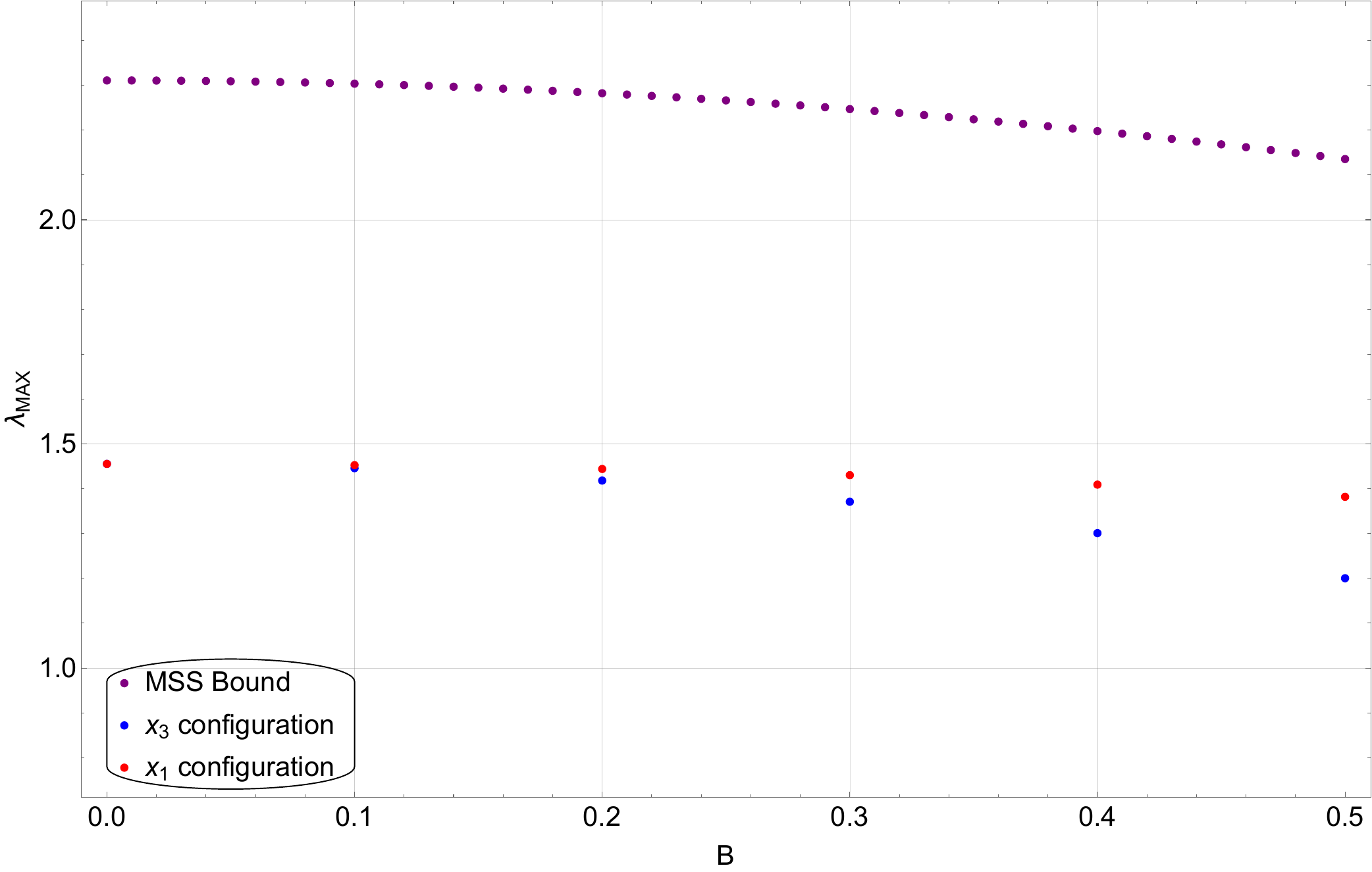}
	\caption{Comparison between MSS bound and largest Lyapunov exponent at the saddle point as a function of $B$ for $L = 0.75$ for two different orientations of the string. In units of GeV.}
	\label{mssbound_einsteinframe}
\end{figure}

For the unstable string configuration with energy $E=0$, the unstable fixed point is located at $\vec{y}=(\tilde{c_0},\dot{\tilde{c_0}},\tilde{c_1},\dot{\tilde{c_1}})=(0,0,0,0)$. The Lyapunov exponents at the unstable fixed point again asymptotically converge to $(\sqrt{-\omega_{0}^{2}},-\sqrt{-\omega_{0}^{2}},0,0)$. The comparison between the largest Lyapunov exponent $\lambda_{max}=\sqrt{-\omega_{0}^{2}}$ at the unstable fixed point and the MSS bound for different magnetic field values is shown in Fig.~\ref{mssbound_einsteinframe}. Notice that, although the largest Lyapunov exponent at the unstable fixed point is three orders of magnitude higher than the stable point, it always remain below the MSS bound. Moreover, at the unstable fixed point, the largest Lyapunov exponent is found to be decreasing with $B$ for both orientations of the magnetic field. This is different from the stable fixed point result, where the largest Lyapunov exponent is found to be increasing/decreasing for parallel/perpendicular magnetic field. This should also be contrasted with the results in the string frame, where the largest Lyapunov exponent exhibited similar behaviour near the stable and unstable fixed points for both parallel and perpendicular magnetic fields. Our analysis therefore provides a curious and intriguing example where the Lyapunov exponent exhibits different structure depending upon the fixed points involved in the system. In particular, the magnetic field tries to soften the chaotic behaviour for both parallel and perpendicular orientation at the unstable fixed point, whereas it increases/decreases the chaotic behaviour for both parallel/perpendicular orientation at the stable fixed point.

Let us also mention that, like in the string frame case, we can calculate the Lyapunov exponents analytically at the fixed point as the real part of the eigenvalues of the Jacobian matrix in the Einstein frame as well. At the unstable fixed point, the Jacobian matrix in the Einstein frame takes the same form as in string frame, i.e. Eq.~(\ref{jacobian}), whose eigenvalues are again given by $(\sqrt{-\omega_{0}^{2}},-\sqrt{-\omega_{0}^{2}},i\sqrt{\omega_{1}^{2}},-i\sqrt{\omega_{1}^{2}})$. Our numerical results above are consistent with this analytic result, and in particular, the largest Lyapunov exponent is given by $\lambda_{max}=\sqrt{-\omega_{0}^{2}}$ at the unstable fixed point.

\section{Conclusions}\label{sec:4}
In this paper, we have analysed the effect of a background magnetic field on the chaotic dynamics of the string. For this purpose, we considered the Einstein-Maxwell-dilaton magnetized gravity model of \cite{Bohra2020Feb}, which captures several lattice supported magnetised QCD features holographically. We considered two closely related gravitational backgrounds, namely string and Einstein frames, and thoroughly analysed the effects of a background magnetic field on the string dynamics. Our investigation confirms the MSS bound both for the parallel and the perpendicular orientation of the magnetic field, and this in both frames. We found that, depending upon the frame under consideration, the magnetic field can introduce non-trivial anisotropic effects in the string dynamics and to its associated chaotic behaviour. Our key results are highlighted as under:
\subsection{String frame}
\begin{itemize}
\item The tip of the unstable string configuration moves further away from the horizon for both parallel and perpendicular magnetic fields, with the tip moving farther in the perpendicular case.
	\item The Poincar\'{e} sections become more structured with a substantial decrease in the number of scattered points when the magnetic field is increased both in the parallel and perpendicular orientations, with the latter being more prominent.
	\item With the increase of the background magnetic field, the largest Lyapunov exponent decreases in both orientations. The Lyapunov exponent is smaller in the perpendicular than parallel direction.
   \item The largest Lyapunov exponent at the unstable fixed point is three orders of magnitude higher than at the stable fixed point. At the unstable fixed points, the largest Lyapunov exponent is again found to be decreasing for both orientations.
\item These results suggests that the effect of background magnetic field is to stabilize the system in both orientations of the string, with a stronger stabilization in the case of perpendicular orientation.
\end{itemize}
\subsection{Einstein frame}
\begin{itemize}
\item The tip of the unstable string configurations moves towards/away from the horizon for parallel/perpendicular case.
	\item The Poincar\'{e} sections become less structured with a substantial increase in the number of scattered points when the magnetic field is increased in the parallel case. On the other hand, the Poincar\'{e} sections become more structured with a substantial decrease in the number of scattered points when the magnetic field is increased in the perpendicular case.
	\item The largest Lyapunov exponent increases/decreases with the magnetic field for parallel/perpendicular cases.
	\item The largest Lyapunov exponent at the unstable fixed point is three orders of magnitude higher than at the stable fixed point. At the unstable fixed points, the largest Lyapunov exponent is again found to be decreasing for both orientations. Therefore, depending upon the fixed point under consideration, the anisotropic effects of the magnetic field are quite distinct in the Einstein frame.
\end{itemize}

To our knowledge, the magnetic field induced anisotropies in the chaotic dynamics have not been well explored in a consistent bottom-up model from a holographic viewpoint, although specific effects of anisotropy have been studied in \cite{Mateos2011Aug,Itsios2019Mar,Giataganas2018Sep,Giataganas2012Jul,Aref'eva2019May,Gursoy:2020kjd,Gursoy:2016ofp}. Holographic methods have been used to study several chaotic properties of a super Yang-Mills theory at temperature $T$ in the presence of a background magnetic field in \cite{Avila2018Sep}. It was found that the system becomes more rigid in presence of magnetic field, in the sense that it increases the mutual information between
the subsystems  \cite{Avila2018Sep,Jain:2022hxl}. Their results highlight that the internal interaction of the system is increased by the magnetic field. The exact physics of what causes such stabilizing effect has not been unravelled to the best of our knowledge. One might speculate that, by enlarging $\frac{B}{T^{2}}>1$, the coupling strength of the QCD increases, which could support the above observations \cite{Ayala2018Aug}.
Moreover, it is still unclear as to why the (de)/stabilizing effect of magnetic field is orientation dependent. This might be related to our previous work \cite{Bohra2020Feb}, where anisotropic confinement was found, expressed by an orientation dependent string tension, see also lattice data of \cite{Bonati:2016kxj,D'Elia2021Dec}. As speculated in \cite{Hashimoto2018Oct}, the chaotic dynamics might also be connected to the (strong) entropy production near deconfinement, another QCD feature we were able to catch already with our model, see \cite{Jena2022Apr}, so it would be interesting to make this conjectured relation more concrete in future work.

\section*{Acknowledgements}
B.S.~would like to thank Pranaya Pratik Das and Siddhi Swarupa Jena for discussions and assistance with some of the coding. The work of S.M.~is supported by the Department of Science and Technology, Government of India under the Grant Agreement number IFA 17-PH207 (INSPIRE Faculty Award).

\appendix

\section{Results for fixed $r_0$ in string and Einstein frames} \label{A}
In our study above, we fixed the string length $L$ and examined the impact of a magnetic field on the (unstable) string's tip location $r_0$. We found that the magnetic field causes $r_0$ to move closer or farther from the black hole horizon, depending on the orientation of the string configuration. This led to substantial changes in the chaotic dynamics of the string. To make contact with the setup and results of \cite{Colangelo2022Apr},  it is also interesting to fix $r_0 = 1.1$ throughout and analyse the chaotic dynamics of the strings. Note that fixing $r_0 = 1.1$ for all values of $B$ makes the quark separation $L$ a $B$-dependent quantity, we already explained in the main text this is a less natural thing to do.

\begin{figure}[h]
 	\centering
 	\subfloat[String frame]{\label{lypmaxvsB_stringframe_r0}	\includegraphics[width=0.45\textwidth]{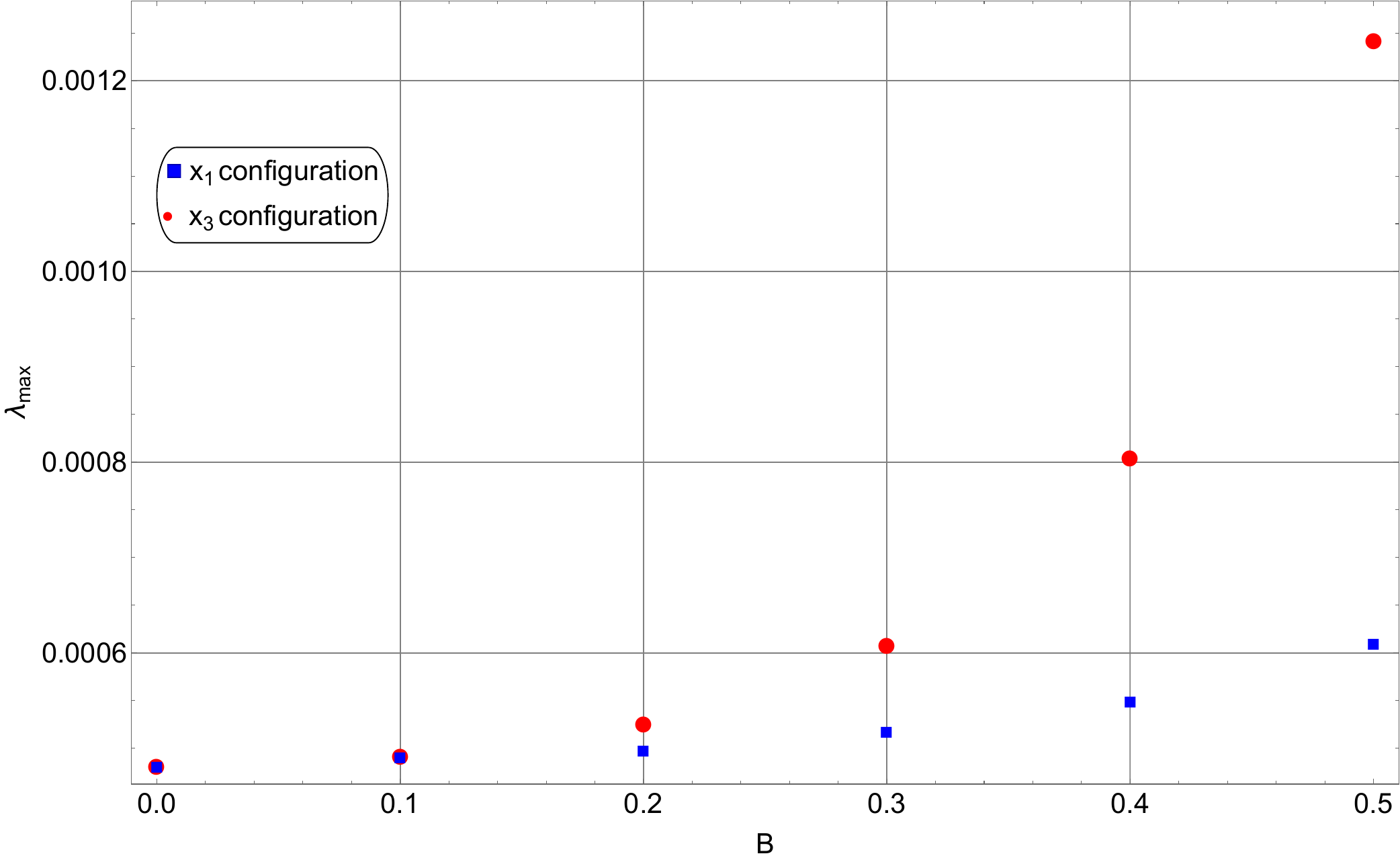}}
 	\hfill
 	\subfloat[Einstein frame]{\label{lypmaxvsB_einsteinframe_r0} \includegraphics[width=0.45\textwidth]{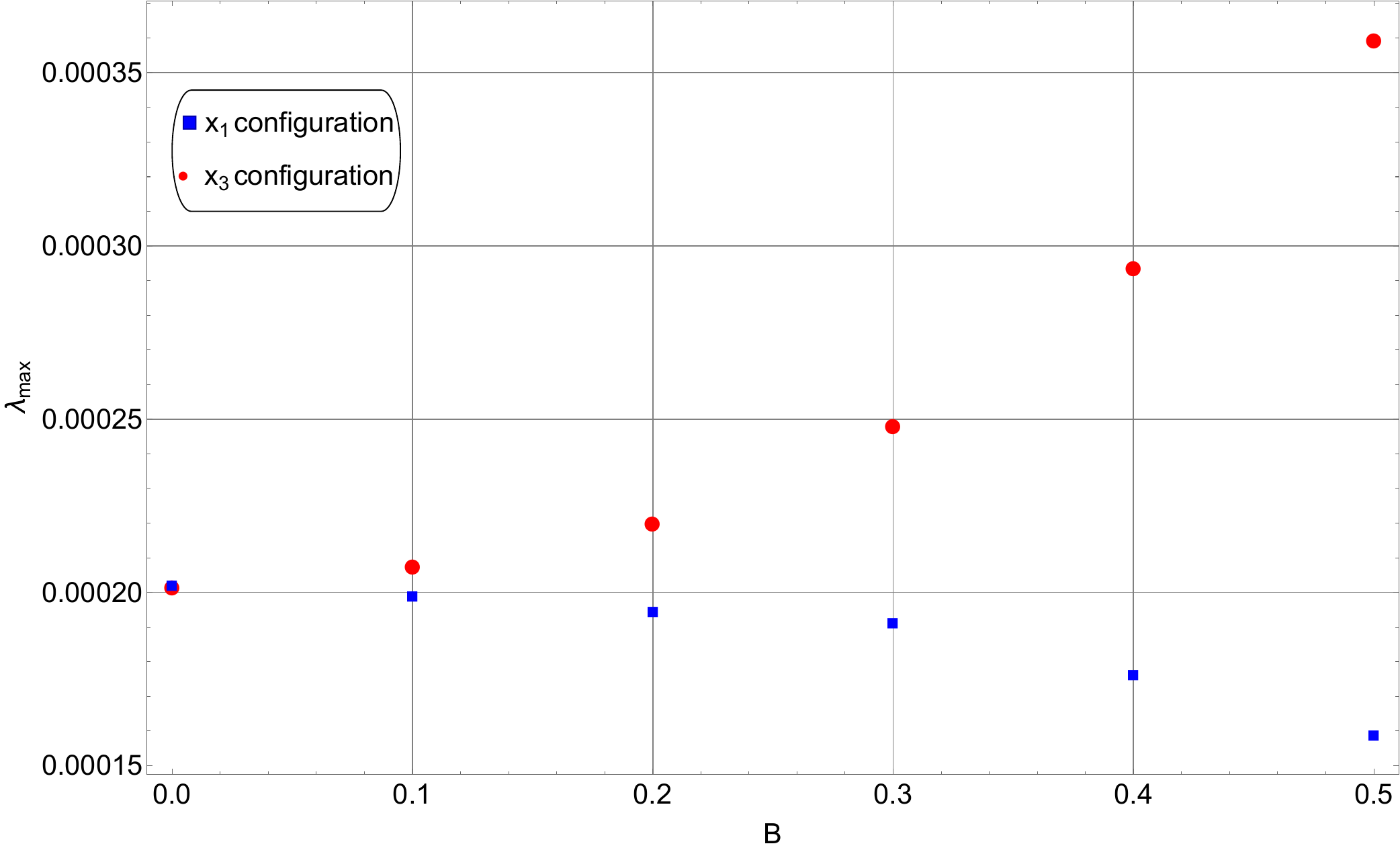}}
 	\caption{\label{lypmaxvsB_r0}Largest Lyapunov exponent $\lambda_{max}$ versus $B$ for $r_0=1.1$ in the string and Einstein frames for the parallel and perpendicular orientations. In the string frame, we used $E = 10^{-5}$, $\tilde{c_{0}}=-0.001$, $\dot{\tilde{c_{0}}}=0$, and $\tilde{c_{1}}=0.00002$ while in the Einstein frame, we used $E = 10^{-5}$, $\tilde{c_{0}}=-0.002$, $\dot{\tilde{c_{0}}}=0$, and $\tilde{c_{1}}=0.001$. In units of GeV. As before, these values were chosen for illustrative purposes.}
 \end{figure}

Our results for the largest Lyapunov exponent in the string and Einstein frames for various values of parallel and perpendicular magnetic field are shown in Figs.~\ref{lypmaxvsB_r0}. We observe that increasing the magnetic field also increases $\lambda_{max}$ for both orientations of the magnetic field in the String frame case. In contrast, in the Einstein frame, increasing the magnetic field decreases/increases $\lambda_{max}$ for the parallel/perpendicular orientation of the magnetic field. For completeness, we also mention that the $\lambda_{max}$ was found to be decreasing for both parallel and perpendicular magnetic field in \cite{Colangelo2022Apr}. These results clearly suggest that substantial differences arise in the chaotic structure of the string not only due to the running coupling constants (cf.~dilaton) but also due to the different frame choice. Overall, we can say that there is a clear anisotropy present in the chaotic dynamics of the string.

\bibliography{mybib}

\end{document}